\documentclass[aps,prc,showpacs,floatfix,nofootinbib,10pt,twocolumn]{revtex4}

\newcommand{\ifproofpre}[2]{#1}

\usepackage{graphicx}
\usepackage{amsmath}
\usepackage{amssymb}
\usepackage{mathtools}
\usepackage{isotope}
\usepackage{dsfont} 

\usepackage{liealg}
\usepackage{wignert}
\usepackage{mcmath}



\newcommand{\LaguerreL}[2]{{L_{#1}^{#2}}}
\newcommand{\LaguerreLhat}[2]{{\hat{L}_{#1}^{#2}}}

\newcommand{\Phat}{{\hat{P}}}
\newcommand{\Lambdat}{{\tilde{\Lambda}}}

\newcommand{\Rt}{{\tilde{R}}}
\newcommand{\St}{{\tilde{S}}}
\newcommand{\Nt}{{\tilde{N}}}
\newcommand{\Nmax}{{N_\text{max}}}
\newcommand{\Ntot}{{N_\text{tot}}}

\newcommand{\Hin}{{H_\text{in}}}
\newcommand{\Jin}{{J_\text{in}}}
\newcommand{\Ecm}{{E_\text{c.m.}}}
\newcommand{\Ein}{{E_\text{in}}}
\newcommand{\Tcm}{{T_\text{c.m.}}}
\newcommand{\Trel}{{T_\text{rel}}}
\newcommand{\Ucm}[1][]{{U_\text{c.m.}^{#1}}}  
\newcommand{\Urel}[1][]{{U_\text{rel}^{#1}}}  
\newcommand{\Hcm}[1][]{{H_\text{c.m.}^{#1}}}  
\newcommand{\bcm}{b_\text{c.m.}}   
\newcommand{\bho}{b_\text{HO}}   
\newcommand{\Ncm}[1][]{{N_\text{c.m.}^{#1}}}  
\newcommand{\Nin}[1][]{{N_\text{in}^{#1}}}  
\newcommand{\lcm}{{l_\text{c.m.}}}
\newcommand{\krel}{k_\text{rel}} 
\newcommand{\prel}{p_\text{rel}} 
\newcommand{\rrel}{r_\text{rel}} 
\newcommand{\Omegat}{{\tilde{\Omega}}}
\newcommand{\OmegaL}{\Omega_L}
\newcommand{\unity}{\mathds{1}}

\newcommand{\rmax}{{r_\text{max}}}

\newcommand{\pn}{{pn}}
\newcommand{\as}{{\text{AS}}}
\newcommand{\nas}{{\text{NAS}}}
\newcommand{\Ncut}{{N_\text{cut}}}  

\newcommand{\wfgen}{\psi} 
\newcommand{\wfgencm}[1][]{\psi_{\text{c.m.}#1}}
\newcommand{\wfgenin}[1][]{\psi_{\text{in}#1}}

\newcommand{\wfbasiscm}[1][]{\phi_{\text{c.m.}#1}}
\newcommand{\wfbasisin}[1][]{\phi_{\text{in}#1}}

\newcommand{\wfho}{\Psi}
\newcommand{\wfhot}{\tilde{\wfho}}
\newcommand{\wfcsx}{\Phi}

\newcommand{\cd}{c^\dagger} 

\newcommand{\sH}[1][]{{\mathcal{H}^{#1}}}  
\newcommand{\sHcm}[1][]{{\mathcal{H}_\text{c.m.}^{#1}}}  
\newcommand{\sHin}[1][]{{\mathcal{H}_\text{in}^{#1}}}  

\newcommand{\MeV}{{\mathrm{MeV}}}
\newcommand{\fm}{{\mathrm{fm}}}

\begin{document}


\title{Coulomb-Sturmian basis for the nuclear many-body problem}

\author{M. A. Caprio}
\affiliation{Department of Physics, University of Notre Dame, Notre Dame, Indiana 46556-5670, USA}

\author{P. Maris}
\affiliation{Department of Physics and Astronomy, Iowa State University, Ames, Iowa 50011-3160, USA}

\author{J. P. Vary}
\affiliation{Department of Physics and Astronomy, Iowa State University, Ames, Iowa 50011-3160, USA}

\date{\today}

\begin{abstract}
Calculations in \textit{ab initio} no-core configuration interaction
(NCCI) approaches, such as the no-core shell model or no-core
full configuration methods, have
conventionally been carried out using the harmonic-oscillator
many-body basis.  However, the rapid falloff (Gaussian asymptotics) of
the oscillator functions at large radius makes them poorly suited for
the description of the asymptotic properties of the nuclear
wave function.  We establish the foundations for carrying out NCCI
calculations with an alternative many-body basis built from
Coulomb-Sturmian functions.  These provide a complete, discrete set of
functions with a realistic exponential falloff.  We present
illustrative NCCI calculations for $\isotope[6]{Li}$ with a
Coulomb-Sturmian basis and investigate the center-of-mass separation
and spurious excitations.
\end{abstract}


\maketitle

\section{Introduction}
\label{sec-intro}

The combination of powerful theoretical frameworks with modern
computing capabilities is making possible significant
advances towards one of the basic goals of nuclear theory, namely, an
\textit{ab inito} understanding of the nucleus directly as a system of
interacting protons and neutrons with realistic interactions.  Nuclear interactions motivated by
quantum chromodynamics are being developed, via effective
field theory
methods~\cite{entem2003:chiral-nn-potl,epelbaum2009:nuclear-forces}, to provide an
underlying Hamiltonian for the problem.  It is then necessary to solve
the nuclear many-body problem for this Hamiltonian, obtaining nuclear
eigenstates and predictions for observables.  In a no-core
configuration interaction (NCCI) approach, such as the no-core shell model (NCSM)~\cite{navratil2000:12c-ncsm-COMBO}, 
the eigenproblem is formulated as a matrix diagonalization problem, in
which the Hamiltonian matrix is represented with respect to a basis of
antisymmetrized products of single-particle states.  The
nuclear eigenproblem is then solved for the full $A$-body system of
nucleons, \textit{i.e.}, there is no assumption of an inert core.

In practice, NCCI calculations have been based almost exclusively on a
harmonic oscillator basis.  In this article, we consider instead an
alternative basis for the NCCI approach, built from Coulomb-Sturmian
functions~\cite{rotenberg1962:sturmian-scatt,weniger1985:fourier-plane-wave}.
These functions have previously been applied to few-body problems in
atomic~\cite{hylleraas1928:helium-sturmian,loewdin1956:natural-orbital,rotenberg1962:sturmian-scatt,rotenberg1970:sturmian-scatt}
and
hadronic~\cite{jacobs1986:heavy-quark-sturmian,fulcher1993:quarkonium-sturmian,keister1997:on-basis,pervin2005:diss}
physics.  The Coulomb-Sturmian functions have the distinctive property
of constituting a complete, discrete set of square-integrable functions,
while also possessing realistic exponential asymptotics appropriate to
the nuclear problem. In the present work, the foundations
for carrying out nuclear many-body calculations with the
Coulomb-Sturmian basis are established.  Then, illustrative
calculations for the nucleus $\isotope[6]{Li}$ are carried out with
the NCCI approach in a Coulomb-Sturmian basis.\footnote{The
Coulomb-Sturmian single-particle states used in the present
calculations arise as solutions to a general Sturm-Liouville equation (Sec.~\ref{sec-cs-fcn}),
rather than a Schr\"odinger equation or Hartree-Fock
problem.  They consequently do not physically
   correspond to ``shells'' in the conventional sense, \textit{i.e.},
   orbitals for independent-particle motion in some mean-field
   potential describing the zeroth-order dynamics of the system.
   Therefore, we use the more inclusive term \textit{configuration
   interaction}, rather than specifically \textit{shell model},
   throughout the present work.}
Many of the considerations
addressed here specifically in the context of the Coulomb-Sturmian
basis are more broadly applicable to alternative
single-particle bases for the nuclear problem.

Actual NCCI calculations must be carried out in a finite, truncated
space.  Progress in expanding the domain of applicability of the
method is hampered by a combinatorial scale explosion in the dimension
of the problem, with increasing size of the included space of
single-particle states and with the number of nucleons in the system.
The challenge is to reach a reasonable approximation of the converged
results which would be achieved in the full, untruncated space for
the many-body system.  The success of the calculation is determined by
the rate of convergence of calculated observables (energies, charge or
mass radii, electromagnetic moments and transition rates, \textit{etc.}) with
increasing basis size and the ability to reliably extrapolate these
results for finite spaces to the full many-body
space~\cite{forssen2008:ncsm-sequences,maris2009:ncfc,coon2012:nscm-ho-regulator}.  Convergence rates may be expected to be
sensitive to the choice of single-particle states from
which the NCCI many-body basis is constructed, as well as the
truncation scheme used for the many-body basis.

Before considering alternative bases, it is worth noting that the
oscillator functions present significant advantages as a basis for the
nuclear problem, which require further assessment in moving to another basis:

(1) An exact factorization of center-of-mass and intrinsic
wave functions is obtained in many-body calculations when the
oscillator basis is used in conjunction with the $\Nmax$ truncation
scheme (see Sec.~\ref{sec-nmax}), which is based on the total number of
oscillator quanta.  Thus, the oscillator basis with this truncation
allows precise removal of or correction for spurious center-of-mass
contributions to the dynamics.

(2) Matrix elements of the nucleon-nucleon two-body interaction are
naturally formulated in the relative oscillator basis, of functions
$\wfho_{nl}(\vec{r}_1-\vec{r}_2)$ (see Sec.~\ref{sec-ho}).  These matrix elements can easily
be transformed to the two-body oscillator basis, of functions
$\wfho_{n_1l_1}(\vec{r}_1)\wfho_{n_2l_2}(\vec{r}_2)$, by the Moshinsky
transformation~\cite{moshinsky1996:oscillator}.  The simplicity of this
transformation is lost with any other single-particle basis.  This is
a fundamental concern, since the starting point of the many-body
calculation is evaluation of the two-body matrix elements.  (Similar
comments apply for three-body or higher-body interactions.)

(3) The oscillator functions constitute a complete \textit{discrete}
basis for square-integrable functions.  Many alternative bases do not
provide this convenience.  For instance, the bound state
eigenfunctions of the Schr\"odinger equation for 
\textit{finite-depth} potentials, such as the Woods-Saxon potential,
are typically finite in number and in general do not constitute a
complete set of square-integrable functions, without inclusion of the
unbound continuum Schr\"odinger equation solutions as well.

Nonetheless, there are also strong reasons to consider moving beyond
the oscillator basis.  The classic and long-recognized (\textit{e.g.},
Ref.~\cite{davies1966:hartree-fock}) physical limitation of the
oscillator basis, for application to the nuclear problem, lies in the
Gaussian falloff ($\propto e^{-\alpha r^2}$) at large distance $r$,
which is a consequence of the quadratic confining harmonic oscillator
potential.  In contrast, for particles bound by a finite-range force,
the actual asymptotics are exponential ($\propto e^{-\beta r}$).  This
mismatch in asymptotics, \textit{i.e.}, the wave function tails,
between the expansion basis and the physical system imposes a serious
handicap on the convergence of calculations with increasing basis
size.  The problem is especially significant for observables, such as
the root-mean-square radius or $E2$ strengths, which are sensitive to
the large-$r$ properties of the nuclear wave functions.

To adapt the Coulomb-Sturmian basis to the nuclear many-body problem,
we must overcome the aforementioned technical challenges of moving
away from the oscillator basis.  The Coulomb-Sturmian functions, as
already noted, are complete and offer the convenience of being a
discrete set.  The remaining challenges~--- transformation of matrix
elements and center-of-mass factorization or spuriosity~--- are found
to be tractable.  First, we review the relevant aspects of the
NCCI approach as conventionally implemented, including the oscillator single-particle basis
(Sec.~\ref{sec-ho}), the Hamiltonian
(Sec.~\ref{sec-hamil}), and the $\Nmax$ many-body truncation scheme
(Sec.~\ref{sec-nmax}).  Then, procedures and results are
established for using the Coulomb-Sturmian basis for nuclear many-body
calculations.  The Coulomb-Sturmian functions are defined
(Sec.~\ref{sec-cs-fcn}), practicalities related to the radial length
parameter are considered (Sec.~\ref{sec-cs-scale}), the transformation
of interaction two-body matrix elements from the oscillator basis to
the Coulomb-Sturmian basis is addressed (Sec.~\ref{sec-xform}), and it
is shown how the two-body matrix elements of the relative kinetic
energy (and certain other operators) can be evaluated via separability
(Sec.~\ref{sec-sep}).  Finally, illustrative NCCI calculations for
$\isotope[6]{Li}$ with the Coulomb-Sturmian basis
(Sec.~\ref{sec-calc-over}) are compared with oscillator-basis
calculations of the same dimensionality.  The convergence of energies
(Sec.~\ref{sec-calc-en}) and the root-mean-square radius
(Sec.~\ref{sec-calc-r}) is examined, and issues of center-of-mass
factorization and spurious states are explored in detail
(Sec.~\ref{sec-calc-cm}).  Preliminary results were reported in Ref.~\cite{caprio2012:csbasis-hites12}.

\section{Background: No-core shell model}
\label{sec-ncsm}

\subsection{Harmonic-oscillator basis}
\label{sec-ho}

The basis
states conventionally used in the NCCI approach are antisymmetrized products of
single-particle harmonic oscillator states.  These single-particle
states are
eigenstates of the
Hamiltonian
\begin{equation}
\label{eqn-h-sp}
h^\Omega=\frac{p^2}{2m_N}+\frac{m_N\Omega^2r^2}{2},
\end{equation}
where $\Omega$ denotes the oscillator frequency and $m_N$ the nucleon
mass, and $\vec{r}$ and $\vec{p}$ are the single-particle coordinates
and momenta.  For the spatial part of the solution, we have the usual
three-dimensional oscillator wave functions
\begin{equation}
\label{eqn-ho-wf}
\wfho_{nlm}(\vec{r})= N_{nl}(r/b)^{l}
L_n^{l+1/2}[(r/b)^2]
e^{-(r/b)^2/2}
\,Y_{lm}(\uvec{r}),
\end{equation}
with normalization factor
\begin{equation}
\label{eqn-ho-wf-N}
N_{nl}=
\frac{1}{b^{3/2}}
\biggl[\frac{2\,n!}{(l+n+1/2)!}\biggr]^{1/2},
\end{equation}
where the $L_n^\alpha$ are generalized Laguerre polynomials, the $Y_{lm}$ are
spherical harmonics, $n$ is the radial quantum number, $l$ and $m$ are
the orbital angular momentum and $z$-projection, and $b$ is the
oscillator length, given by $b=[\hbar/(m_N\Omega)]^{1/2}$.  We use
factorial notation [$x!\equiv\Gamma(x+1)$] uniformly, for both integer
and half-integer arguments.  Letting
$\wfho_{nlm}(\vec{r})=r^{-1}R_{nl}(r)Y_{lm}(\uvec{r})$, the radial
wave function is thus
\begin{equation}
\label{eqn-ho-R}
R_{nl}(r)=bN_{nl}(r/b)^{l+1}
L_n^{l+1/2}[(r/b)^2]
e^{-(r/b)^2/2}.
\end{equation}
The functions $R_{nl}$ form an orthonormal set, with
$\int_0^\infty dr\,R_{n'l}(r) R_{nl}(r)=\delta_{n'n}$.   
The full single-particle states $\tket{nljm}$, including spatial and
spin degrees of freedom, are defined as usual for the
nuclear shell model, by coupling the orbital and spin-$\tfrac12$ angular momenta
to good total angular momentum $j$, with its $z$-projection again denoted by $m$.

The many-body basis states, for calculations in a space of fixed
total many-body angular momentum projection $M$ (\textit{$M$-scheme basis}), are then
\begin{equation}
\label{eqn-m-scheme-state}
\wfgen=\mathcal{A}\tket{n_1 l_1 j_1 m_1}\tket{n_2 l_2 j_2 m_2}\cdots\tket{n_A l_A j_A m_A},
\end{equation}
where the operator $\mathcal{A}$ represents antisymmetrization, over
protons and neutrons separately.  The many-body basis states are thus
eigenstates of the Hamiltonian for noninteracting particles in a
harmonic oscillator potential, $H^\Omega=\sum_ih_i^\Omega$.  These
states may be classified according to the total number of oscillator
quanta $\Ntot=\sum_i N_i=\sum_i (2n_i+l_i)$ and have energy eigenvalue
$E=(\Ntot+\tfrac32)A\hbar\Omega$ with respect to $H^\Omega$.  Thus, truncations by
$\Ntot$, as considered in the following section, are energy
truncations under this noninteracting Hamiltonian.

Since we will later need to consider momentum-space wave functions, note that these
are obtained as the Fourier transform
\begin{equation}
\label{eqn-ho-ft}
\wfhot_{nlm}(\vec{k})\equiv(2\pi)^{-3/2}\int d^3\vec{r}\,e^{-i\vec{k}\cdot\vec{r}}\wfho_{nlm}(\vec{r}).
\end{equation}
The radial wave function $\Rt_{nl}$ in momentum space is
defined by
\begin{equation}
\label{eqn-ho-wf-tilde-generic}
\wfhot_{nlm}(\vec{k})=(-i)^l\frac{\Rt_{nl}(k)}{k}Y_{lm}(\uvec{k})
\end{equation}
and is obtained as the
Fourier-Bessel transform~\cite{fluegge1971:qm}
\begin{equation}
\label{eqn-ho-fbt}
\Rt_{nl}(k)=(2/\pi)^{1/2}\int_0^\infty
dr\,krj_l(kr) R_{nl}(r).
\end{equation}
For the oscillator, $\Rt_{nl}$ has the same functional form as the coordinate-space oscillator wave function $R_{nl}$,
with~\cite{weniger1985:fourier-plane-wave}
\begin{equation}
\label{eqn-ho-Rt}
\Rt_{nl}(k)=(-)^n\frac{1}{b}  \Nt_{nl}(bk)^{l+1} L_n^{l+1/2}[(bk)^2]
e^{-(bk)^2/2},
\end{equation}
where
\begin{equation}
\label{eqn-ho-wf-Nt}
\Nt_{nl}=
b^{3/2} 
\biggl[\frac{2\,n!}{(l+n+1/2)!}\biggr]^{1/2}.
\end{equation}
The $\Rt_{nl}$  form an orthonormal set, with
$\int_0^\infty dk\,\Rt_{n'l}(k) \Rt_{nl}(k)=\delta_{n'n}$.

\subsection{Hamiltonian}
\label{sec-hamil}

We now review the properties of the nuclear Hamiltonian which are most
relevant to understanding the solution method based on the
Coulomb-Sturmian basis (Sec.~\ref{sec-cs}) and the results from
applying this method (Sec.~\ref{sec-calc}).  The
NCCI approach is based upon a nonrelativistic nuclear many-body Hamiltonian of
the form
\begin{equation}
\label{eqn-H}
H=T+V,
\end{equation}
where $T$ is the one-body kinetic energy operator and $V$ 
represents the interaction of the nucleons.  Commonly, the isoscalar kinetic energy
\begin{equation}
\label{eqn-T}
T=\frac{1}{2m_N}\sum_i p_i^2
\end{equation}
is used, that is, protons and neutrons are treated equivalently as
having the average nucleon mass $m_N$, and the summation index $i$
runs over all $A$ nucleons. The potential $V$ is a
Galilean-invariant operator involving two-body and possibly higher
many-body terms. 

The Hamiltonian~(\ref{eqn-H}) has the essential property that it may be separated into
center-of-mass and intrinsic (Galilean-invariant) contributions. The kinetic energy
operator separates into a term
\begin{equation}
\label{eqn-Tcm}
\Tcm=\frac{1}{2Am_N}\Bigl( \sum_i \vec{p}_i\Bigr)^2=\frac{P^2}{2Am_N}
\end{equation}
representing the center-of-mass kinetic energy and a term
\begin{equation}
\label{eqn-Trel}
\Trel=\frac{1}{4Am_N}{\sumprime_{ij}}(\vec{p}_i-\vec{p}_j)^2=\frac{\prel^2}{2Am_N}
\end{equation}
representing the kinetic energy of relative motion of the nucleons,
where the prime on the summation ${\tsumprime_{ij}}$ indicates $i\neq j$. 
The decomposition of both the $r^2$ and $p^2$ operators into center-of-mass and relative contributions is
summarized in 
Appendix~\ref{app-r2k2}, which also serves to define a uniform notation
for the present work.
The operator $\Trel$ depends only upon relative momenta
$\vec{p}_i-\vec{p}_j$ and is therefore Galilean invariant.
Thus, the full nuclear Hamiltonian~(\ref{eqn-H}) may be separated as
$H=\Tcm+\Hin$, where 
\begin{equation}
\label{eqn-Hin}
\Hin=\Trel+V
\end{equation}
is the Galilean-invariant \textit{intrinsic Hamiltonian}.
As a consequence of the separability of $H$, a complete set of
eigenstates may be found with coordinate-space wave functions which
have the factorized form
\begin{equation}
\label{eqn-wf-factorized}
\wfgen(\vec{r}_i;\vec{\sigma}_i)=\wfgencm(\vec{R})\wfgenin[,k](\vec{r}_{ij};\vec{\sigma}_i).
\end{equation}
The factor $\wfgencm(\vec{R})$ depends only on the center-of-mass
coordinate, and the factor $\wfgenin(\vec{r}_{ij};\vec{\sigma}_i)$
depends only on relative coordinates
$\vec{r}_{ij}=\vec{r}_i-\vec{r}_j$ and intrinsic spin degrees of
freedom, indicated schematically here by the arguments
$\vec{\sigma}_i$.  For each intrinsic excitation, with wave function
$\wfgenin$, an infinite set of eigenstates sharing this same intrinsic
structure but different center-of-mass excitations $\wfgencm$ is
obtained.  The corresponding energy eigenvalue separates into
eigenvalues of $\Tcm$ and $\Hin$, as $E=\Ecm+\Ein$.

The ``interesting'' many-body spectroscopy of the nucleus resides in
the intrinsic wave functions $\wfgenin$ and eigenvalues $\Ein$, but the
``uninteresting'' center-of-mass motion remains as an unavoidable and
potentially obfuscating element of the solution. 
In principle, the
center-of-mass motion may be completely eliminated from the problem,
by explicitly changing variables to relative coordinates.  However,
the nuclear many-body state must be antisymmetrized, and this process rapidly becomes intractable with
increasing nucleon number.  On the other hand, if we instead solve the
nuclear eigenproblem in a many-body basis constructed from
antisymmetrized products of single-particle states, antisymmetrization is
straightforward, but we are consigned to simultaneously solving for
center-of-mass and intrinsic excitations.

Before we consider the specifics of formulating the eigenproblem with
respect to a basis, it is worth considering the solutions in the full
coordinate space further.  First, it is convenient to remove the
complication of the center-of-mass kinetic energy operator, by
considering the eigenproblem not for the full Hamiltonian $H$
of~(\ref{eqn-H}) but rather for the intrinsic Hamiltonian $\Hin$
of~(\ref{eqn-Hin}).  The full spectroscopic information of the
original problem is maintained, since the eigenstates still have wave functions of the form
$\wfgen(\vec{r}_i;\vec{\sigma}_i)=\wfgencm(\vec{R})\wfgenin(\vec{r}_{ij};\vec{\sigma}_i)$,
but these are now simply associated with eigenvalues $E=\Ein$.  Thus, for each
intrinsic wave function $\wfgenin$, an infinite set of eigenstates
sharing the same intrinsic structure but different center-of-mass
excitations $\wfgencm$ is still obtained, and these are now strictly degenerate with each other.

Since $\Tcm$ has
been eliminated from the Hamiltonian, we are free to consider any
complete set of wave functions to span the degenerate space of center-of-mass wave functions.  For instance, suppose plane wave solutions
$\wfgencm(\vec{R})=e^{-i\vec{K}\cdot\vec{R}}$ are taken for the center
of mass.  Then, for each intrinsic excitation $\wfgenin[,k]$, with
intrinsic eigenvalue $E_k$, a continuum of eigenstates will be
obtained, having wave functions
$\wfgen(\vec{r}_i;\vec{\sigma}_i)=e^{-i\vec{K}\cdot\vec{R}}\wfgenin[,k](\vec{r}_{ij};\vec{\sigma}_i)$.
Under the full Hamiltonian $H$, these states form a continuum, with
$E=\hbar^2K^2/(2Am_N)+E_k$, but, under $\Hin$, these states are
infinitely degenerate, all with $E=E_k$.

Although they provide the simplest illustration, plane wave
center-of-mass wave functions do not naturally occur in our actual
solutions to the eigenproblem, which are obtained in terms of
\textit{spatially localized}
single-particle basis wave functions.  The choice of basis for
center-of-mass wave functions with direct practical significance in
oscillator-basis calculations consists instead of three-dimensional
harmonic oscillator wave functions,
$\wfgencm(\vec{R})=\wfho_{nlm}(\vec{R})$.  The $\wfho_{nlm}(\vec{R})$
are eigenfunctions of the center-of-mass harmonic oscillator
Hamiltonian $\Hcm[\Omega]$, defined with oscillator frequency $\Omega$
and mass $Am_N$, \textit{i.e.},
\begin{equation}
\label{eqn-Hcm}
\Hcm[\Omega]=\Tcm+\frac{Am_N\Omega^2R^2}{2}.
\end{equation}
The center-of-mass excitation is thus characterized by the
number $\Ncm=2n+l$ of oscillator quanta.
This particular choice of center-of-mass wave functions is 
enforced for the eigenstates of the Hamiltonian, if the degeneracy of center-of-mass
states is broken by introducing a \textit{Lawson term}~\cite{gloeckner1974:spurious-com} proportional to
$\Hcm[\Omega]$.  It is both conventional and convenient to subtract
the zero-point energy of center-of-mass motion with respect to this
term, so the Lawson term has the form $\lambda
(\Hcm[\Omega]-\tfrac32\hbar\Omega)$, with $\lambda$ positive, or, more transparently,
$a\Ncm[\Omega]$, where $\Ncm[\Omega]=(\Hcm[\Omega]-\tfrac32\hbar\Omega)/(\hbar\Omega)$ is the number operator
associated with~(\ref{eqn-Hcm}).  The Hamiltonian thus becomes
\begin{equation}
\label{eqn-HLawson}
H=\Trel+V+a\Ncm[\Omega].
\end{equation}
The factorized eigenstates
have coordinate space wave functions
$\wfgen(\vec{r}_i;\vec{\sigma}_i)=\wfho_{nlm}(\vec{R})\wfgenin(\vec{r}_{ij};\vec{\sigma}_i)$,
and the eigenvalues are now
$E=E_k+a\Ncm$.  Thus, the eigenvalues for
states with $\Ncm=0$ are unchanged by the Lawson term, still simply
the intrinsic energies $E_k$, while the eigenvalues of spurious
states, with $\Ncm>0$, are raised out of the low-lying spectrum, to an
excitation energy of at least $a$.

\subsection{\boldmath Many-body $\Nmax$ truncation}
\label{sec-nmax}

The factorization of the wave function just described is possible in
the full space of the many-body system.  However, in practice,
diagonalization of the Hamiltonian must be carried out in a
finite-dimensional subspace spanned by some truncated basis.  In
general, one cannot expect to be able to construct center-of-mass
factorized states in such a subspace.  The separation
$\wfgen(\vec{r}_i;\vec{\sigma}_i)=\wfgencm(\vec{R})\wfgenin(\vec{r}_{ij};\vec{\sigma}_i)$
will be lost, and it will not be possible to divide the set of
eigenstates into ``nonspurious'' states, consisting of a simple
product of a $0s$ center-of-mass wave function with a single intrinsic
excitation, and ``spurious'' states, involving center-of-mass
excitations.  However, there is an important special case in which
factorization occurs, namely, for a harmonic-oscillator many-body
basis in the so-called $\Nmax$ truncation scheme, which is based on
the total number of oscillator quanta for the many-body state.  This
truncation is commonly used in NCCI calculations.  In this section, we
briefly examine the structure of the $\Nmax$-truncated space, both to
understand what changes as we go to a general single-particle basis
and as a prerequisite to understanding the spurious state spectrum
observed for NCCI calculations with the Coulomb-Sturmian basis in
Sec.~\ref{sec-calc-cm}.

Factorization is to be expected if the truncated space $\sH$ for the calculation
has a simple product structure, before antisymmetrization,\footnote{If the truncated space has
the form $\sHcm\otimes\sHin$, it is in principle
possible to choose a basis consisting of product functions
$\wfbasiscm[,i]\wfbasisin[,j]$.  Since
$\Hin$ acts only on intrinsic degrees of freedom, it does not connect
basis states involving different $\wfbasiscm[,i]$.  Therefore, the 
Hamiltonian matrix with respect to this basis is block diagonal, with
each block simply consisting of the matrix representation of $\Hin$
on the basis of intrinsic states $\wfbasisin[,j]$.}
\begin{equation}
\label{eqn-Sfactor}
\sH=\sHcm\otimes\sHin.
\end{equation}
Most simply, if
all nucleons are restricted to occupy a filled core plus valence orbitals taken from a \textit{single}
major oscillator shell, the many-body space does factorize in the form~(\ref{eqn-Sfactor}), with
pure $0s$ motion for the center of mass, as shown by Elliott and Skyrme~\cite{elliott1955:com-shell}.
The essential reason is that the total number $\Ntot$ of
harmonic oscillator quanta is identical whether calculated as the sum
of single particle oscillator quanta, $\Ntot=\sum_i N_i$, or as the sum
of center-of-mass and intrinsic quanta $\Ntot=
\Ncm+\Nin$~\cite{elliott1955:com-shell}.  The equivalence may be seen from the decomposition of the
one-body number operator
$N=(\hbar\Omega)^{-1}[p^2/(2m_N)+(m_N\Omega^2/2)r^2-3\hbar\Omega/2]$ into
center-of-mass and intrinsic parts (which follows from
Appendix~\ref{app-r2k2}).  Thus, the space for this situation is $\sH[0]=\sHcm[0]\otimes\sHin[0]$, where
$\sHcm[0]$ is the one-dimensional space containing the $0s$ oscillator
function, and $\sHin[0]$ is the space of intrinsic functions with no
excitations above the valence shell.

The $\Nmax$ truncation scheme is a generalization, which likewise yields factorized eigenstates.
Consider a space spanned by product states subject to the truncation
\begin{equation}
\label{eqn-Nmax-trunc}
\Ntot=\sum_i N_i \leq N_0+\Nmax,
\end{equation}
where $N_0$ is the minimal number of oscillator quanta for the given
number of protons and neutrons, if all
nucleons occupy the lowest permitted shells.  (The Elliott and Skyrme
space described above is obtained for $\Nmax=0$.)  The $\Nmax$-truncated space may be
decomposed as a direct sum of product spaces, before antisymmetrization,\footnote{Since the $\Nmax$-truncated space has
the form~(\ref{eqn-space-sum}), it is  in principle possible to obtain a basis for
$\sH[\Nmax]$ consisting of products of the form
$\wfbasiscm[,i]^{\Ncm}\wfbasisin[,j]^{\Nmax-\Ncm}$.  (The actual basis used in NCCI calculations
need not be, and generally is not, of this form.)  Since
$\Hin$ does not connect
basis states involving different center-of-mass wave functions, the 
Hamiltonian matrix with respect to this basis is block diagonal, with
each block, corresponding to a given $\wfbasiscm[,i]^{\Ncm}$, simply consisting of the matrix representation of $\Hin$
on the basis of intrinsic states for $\sHin[\Nmax-\Ncm]$.}
\begin{multline}
\label{eqn-space-sum}
\sH[\Nmax]=\sHcm[0]\otimes\sHin[\Nmax]
+\sHcm[1]\otimes\sHin[\Nmax-1]
\ifproofpre{\\}{}
+\sHcm[2]\otimes\sHin[\Nmax-2]
+\cdots
+\sHcm[\Nmax]\otimes\sHin[0],
\end{multline}
where $\sHcm[N]$ is the space of center-of-mass functions with exactly
$N$ oscillator quanta, and $\sHin[N]$ is the space of intrinsic
functions with $N$ or fewer intrinsic excitation quanta above $N_0$.
Consequently, factorization is maintained,
but, in the solution to the many-body problem in an $\Nmax$-truncated
space, several approximate copies of the intrinsic spectroscopy are obtained, each
in a more highly-truncated space.  The $\sHcm[0]\otimes\sHin[\Nmax]$
block yields the ``interesting'' solutions, or nonspurious states, consisting of a $0s$
center-of-mass function multiplied by the solutions in the
least-truncated intrinsic space $\sHin[\Nmax]$.  Then the
$\sHcm[1]\otimes\sHin[\Nmax-1]$ block yields a $0p$ center-of-mass
function multiplied by the solutions of the intrinsic
problem in the $\sHin[\Nmax-1]$ space,
the $\sHcm[2]\otimes\sHin[\Nmax-2]$ block yields $1s$ and $0d$
center-of-mass functions multiplied by the solutions of the intrinsic
problem in the
$\sHin[\Nmax-2]$ space,
\textit{etc.} In actual calculations, these
``uninteresting'' solutions, or spurious states, may be identified by
evaluating the expectation value $\tbracket{\Ncm[\Omega]}$.  

The presence of such spurious states in the low-lying calculated
spectrum has considerable practical implications.  Although these
states are clearly identifiable, as noted, diagonalization of such
large matrices as encountered in NCCI calculations typically relies
upon methods such as the Lanczos
algorithm~\cite{lanczos1950:algorithm}, which efficiently extract
a selected set of energy eigenvalues (and corresponding eigenvectors),
namely, those lowest in the energy spectrum.  With increasing $\Nmax$,
the low-energy spectrum would be increasingly cluttered with spurious
states (as illustrated more concretely in Sec.~\ref{sec-calc-cm}),
limiting the ability of the Lanczos diagonalization to access the low-lying
intrinsic excited states.  The spurious states are therefore, in practice,
typically shifted to higher
energy by inclusion of a Lawson term (Sec.~\ref{sec-hamil}) in the Hamiltonian, so that they
do not interfere with the low-lying spectrum obtained by
diagonalization.  

As a final practical matter, it is necessary to note that, for
calculations with parity-conserving nuclear interactions, the $\Nmax$
truncation of~(\ref{eqn-Nmax-trunc}) is further restricted either to
$\Ntot$ even or to $\Ntot$ odd.  If, \textit{e.g.}, even $\Ntot$ are
taken, so $\sH[\Nmax]$ is the even-parity space for the nucleus, then
the $\sHcm[0]\otimes\sHin[\Nmax]$ subspace yields only even-parity
intrinsic excitations, the $\sHcm[1]\otimes\sHin[\Nmax-1]$ subspace
yields the odd-parity $0p$ center-of-mass function multiplied by
odd-parity intrinsic excitations, the $\sHcm[2]\otimes\sHin[\Nmax-2]$
subspace yields the even-parity intrinsic excitations again but
evaluated in the smaller $\Nmax-2$ intrinsic space, \textit{etc.}

\section{The Coulomb-Sturmian basis}
\label{sec-cs}

\subsection{Coulomb-Sturmian functions}
\label{sec-cs-fcn}

The harmonic oscillator functions have the desirable properties, as
basis functions for an eigenfunction expansion, that these form a
complete \textit{discrete} set (of square-integrable functions on
$\bbR^3$), \textit{i.e.}, without a continuum.  However,
the oscillator functions are obtained from an infinitely bound
potential and decay with Gaussian ($e^{-\alpha r^2}$) asymptotics,
\textit{i.e.}, they satisfy an undesirable boundary condition for
problems involving finite binding.  Conversely, the Schr\"odinger
equation for the Coulomb potential yields a set of eigenfunctions
which have exponentially decaying asymptotics ($e^{-\beta r}$), as
desired, but which do not form a complete set (of square-integrable
functions on $\bbR^3$) unless the positive-energy continuum Coulomb wave functions are
included.  However, a closely related set of functions, the
\textit{Coulomb-Sturmian}
functions~\cite{hylleraas1928:helium-sturmian,rotenberg1962:sturmian-scatt,rotenberg1970:sturmian-scatt,weniger1985:fourier-plane-wave,keister1997:on-basis},
can be obtained as the solutions to a Sturm-Liouville problem
associated with the Coulomb potential.  These functions retain the
exponential asymptotics of the Coulomb problem while also forming, in
the final form in which we will write
them, a complete and
\textit{discrete} set of square-integrable functions on $\bbR^3$.
The Coulomb-Sturmian functions thus combine favorable attributes of
both the oscillator and Coulomb functions, as an expansion
basis for three-dimensional Schr\"odinger
problems.

To begin with, let us recall the Schr\"odinger equation solutions for
the Coulomb potential.  The functions
\begin{multline}
\label{eqn-W}
W_{nlm}(\vec{r})=N_{nl}
\biggl(\frac{2r}{n+l+1}\biggr)^l
\ifproofpre{\\\times}{}
L_n^{2l+1}\biggl(\frac{2r}{n+l+1}\biggr)
e^{-r/(n+l+1)}
Y_{lm}(\uvec{r}),
\end{multline}
with
\begin{equation}
\label{eqn-W-N}
N_{nl}=
\biggl(\frac{2}{n+l+1}\biggr)^{3/2}
\biggl[\frac{n!}{2(n+l+1)(n+2l+1)!}\biggr]^{1/2},
\end{equation}
satisfy the Schr\"odinger equation 
\begin{equation}
\label{eqn-W-se}
\biggl(-\delsqr -\frac{2}{r} -2E_{nl}\biggr)W(\vec{r})=0,
\end{equation}
with energy eigenvalue
\begin{equation}
\label{eqn-W-E}
E_{nl}=-\frac{1}{2(n+l+1)^2}.
\end{equation}
This is the Schr\"odinger equation, written in dimensionless form
($\hbar^2/m=1$),  for the potential
$V(r)=1/r$.
The functions $W$ are orthonormal with respect to the standard inner
product on $\bbR^3$, that is,
\begin{equation}
\label{eqn-W-on}
\int d^3\vec{r}\, W_{n'l'm'}^*(\vec{r}) W_{nlm}(\vec{r})=\delta_{(n'l'm')(nlm)}.
\end{equation}

Observe that $r$ always appears in the
usual Coulomb functions divided by a scale $n+l+1$,
which depends
upon the quantum numbers $n$ and
$l$.\footnote{The combination $n+l+1$ is in fact the
\textit{principal}, or energy, quantum number, which enters into the energy eigenvalue
$E_{nl}$ in~(\ref{eqn-W-E}).  In comparing with the literature, it should be borne in mind that,
traditionally, the principal quantum number for the Coulomb problem is
denoted by $n$~\cite{messiah1999:qm}, and this notation propagates to
some discussions of the Coulomb-Sturmian functions (\textit{e.g.},
Refs.~\cite{rotenberg1962:sturmian-scatt,weniger1985:fourier-plane-wave}). However,
consistency with conventional notation for the oscillator
problem~\cite{moshinsky1996:oscillator} and nuclear shell
model~\cite{suhonen2007:nucleons-nucleus} is strongly desirable in the
present context.  Hence, we reserve the symbol $n$ for the radial
quantum number ($n=0$, $1$, $\ldots$).}  The Coulomb-Sturmian
functions are obtained by replacing $(n+l+1)\rightarrow b$
in~(\ref{eqn-W}), that is, by carrying out a radial change of variable
on each function so as to obtain a constant length
scale $b$, yielding
\begin{equation}
\label{eqn-cs-Sigma}
\wfcsx_{nlm}(\vec{r})=N_{nl}
(2r/b)^l
L_n^{2l+1}(2r/b)
e^{-r/b}
Y_{lm}(\uvec{r}),
\end{equation}
with
\begin{equation}
\label{eqn-cs-Sigma-N}
N_{nl}=
\Bigl(\frac{2}{b}\Bigr)^{3/2}
\biggl[\frac{n!}{2(n+l+1)(n+2l+1)!}\biggr]^{1/2}.
\end{equation}
By making the same change of variable
in~(\ref{eqn-W-se}), it is seen that the functions $\wfcsx$ satisfy
\begin{equation}
\label{eqn-cs-Sigma-sle}
\biggl(-\delsqr +\frac{1}{b^2} - \alpha_{nl}\frac{2}{r} \biggr)\wfcsx(\vec{r})=0,
\end{equation}
with eigenvalue $\alpha_{nl}=(n+l+1)/b$.  They are thus solutions to a
Sturm-Liouville eigenproblem, with the Coulomb potential as weighting
function.\footnote{More precisely, the one-dimensional radial equation
associated with~(\ref{eqn-cs-Sigma-sle}), 
\begin{displaymath}
\biggl[-\frac{d^2}{dr^2}+\biggl(\frac{l(l+1)}{r^2}+\frac{1}{b^2}\biggr) -\alpha_{nl}\frac{2}{r} \biggr]\varphi(r)=0,
\end{displaymath}
obtained by setting
$\wfcsx(\vec{r})=r^{-1}\varphi(r)Y_{lm}(\uvec{r})$, has the form of a
Sturm-Liouville equation $[(d/dr)\,p(r)\,(d/dr)+q(r)+\lambda
w(r)]u(r)=0$~\cite{arfken1995:mathmethods}, with weight function
$w(r)\propto1/r$.}
The solutions
$\wfcsx(\vec{r})$ consequently are orthogonal, with respect to the
same weighting function.  In particular,
\begin{equation}
\label{eqn-cs-Sigma-on}
\int d^3\vec{r}\, \wfcsx_{n'l'm'}^*(\vec{r}) \, \frac{1}{r} 
\, \wfcsx_{nlm}(\vec{r})=\frac{1 }{b(n+l+1)}\delta_{(n'l'm')(nlm)}.
\end{equation}

Since~(\ref{eqn-cs-Sigma-sle}) is
obtained from the Schr\"odinger equation simply by a change of
variable, the solutions $\wfcsx_{nlm}$ may also be considered~\cite{rotenberg1962:sturmian-scatt} as
a set of solutions to the Schr\"odinger equation.  However, by
comparison of~(\ref{eqn-W-se}) with~(\ref{eqn-cs-Sigma-sle}), it is seen then that the 
\textit{scale}, or depth, of the potential must be taken to vary with each
solution, as $\alpha_{nl}$, so the solutions to the problem 
share a constant
\textit{energy} $E_0= -1/(2b^2)$, equal to the ground state energy
$E_{00}$ of the associated Schr\"odinger equation,
from~(\ref{eqn-W-E}), after the substitution $(n+l+1)\rightarrow b$.

For use as an expansion basis in quantum mechanical problems, it is desirable to
obtain a set of functions which are orthonormal with respect to the standard integration
metric.  This may be accomplished by absorbing the integration weight $1/r$
and norm $1/[b(n+l+1)]$ appearing in~(\ref{eqn-cs-Sigma-on}) into the Coulomb-Sturmian function itself,
\textit{i.e.}, multiplying the function $\wfcsx_{nlm}$ of~(\ref{eqn-cs-Sigma}) by $[b(n+l+1)/r]^{1/2}$.  However,
the radial dependence of the resulting functions involves a
half-integral power of $r$, $\wfcsx\sim r^{l-1/2}$, for
$r\rightarrow0$.  In contrast, the harmonic oscillator
functions~(\ref{eqn-ho-wf}) have dependence $\wfho\sim r^{l}$ for
$r\rightarrow0$.  We can recover this relation between the
$r\rightarrow0$ asymptotics and the angular momentum by furthermore
shifting $l\rightarrow l+1/2$ in the radial part of the
Coulomb-Sturmian functions, yielding new
functions~\cite{filter1980:sturmian-translation,weniger1985:fourier-plane-wave}
\begin{equation}
\label{eqn-cs-Lambda}
\Lambda_{nlm}(\vec{r})=N_{nl}
(2r/b)^l
L_n^{2l+2}(2r/b)
e^{-r/b}
Y_{lm}(\uvec{r}),
\end{equation}
where now 
\begin{equation}
\label{eqn-cs-Lambda-N}
N_{nl}=
\Bigl(\frac{2}{b}\Bigr)^{3/2}
\biggl[\frac{n!}{(n+2l+2)!}\biggr]^{1/2}.
\end{equation}
Although both $\wfcsx_{nlm}$ and $\Lambda_{nlm}$ are defined in terms
of generalized Laguerre polynomials $L_n^\alpha$, the polynomials
appearing in the $\wfcsx_{nlm}$ have \textit{odd} $\alpha=2l+1$, while
those appearing in the $\Lambda_{nlm}$ have \textit{even}
$\alpha=2l+2$.  The functions $\Lambda(\vec{r})$ are orthogonal with
respect to the standard inner product, \textit{i.e.},
\begin{equation}
\label{eqn-cs-Lambda-on}
\int d^3\vec{r}\, \Lambda_{n'l'm'}^*(\vec{r}) \Lambda_{nlm}(\vec{r})=\delta_{(n'l'm')(nlm)}.
\end{equation}
Moreover, they can be shown to form a complete set on the space of
square-integrable functions on $\bbR^3$~\cite{klahn1977:rayleigh-convergence-part2-special,weniger1985:fourier-plane-wave}.
Letting $\Lambda_{nlm}(r)=r^{-1}
S_{nl}(r)Y_{lm}(\uvec{r})$, the
radial wave function for our Coulomb-Sturmian expansion basis is thus
\begin{equation}
\label{eqn-cs-S}
S_{nl}(r)=(2/b)^{-1}N_{nl}(2r/b)^{l+1}
L_n^{2l+2}(2r/b)
e^{-r/b}.
\end{equation}
The $S_{nl}$ form an orthonormal set, with $\int_0^\infty dr\,S_{n'l}(r)
S_{nl}(r)=\delta_{n'n}$.  

The momentum-space
representation of the Coulomb-Sturmian functions (as
for Coulomb functions in general) may be evaluated
analytically~\cite{weniger1985:fourier-plane-wave,keister1997:on-basis}.
This property of the basis is particularly useful, in the
present application, for evaluation of matrix elements of the kinetic energy
operators.  The momentum-space wave function, defined as
in~(\ref{eqn-ho-ft})--(\ref{eqn-ho-fbt}), 
is simply
expressed in terms of Jacobi polynomials.  If we let 
$\Lambdat_{nlm}(\vec{k})=k^{-1}(-i)^l\St_{nl}(k)Y_{lm}(\uvec{k})$,
then
\begin{equation}
\label{eqn-cs-St}
\St_{nl}(k)=\frac{1}{b}  \Nt_{nl}
\frac{(bk)^{l+1}}{[(bk)^2+1]^{l+2}}
P_n^{(l+3/2,l+1/2)}\biggl[\frac{(bk)^2-1}{(bk)^2+1}\biggr],
\end{equation}
with normalization factor
\begin{equation}
\label{eqn-cs-Lambdat-Nt}
\Nt_{nl}=
2b^{3/2}
\frac{[n!(n+2l+2)!]^{1/2}}{(n+l+\tfrac12)!}.
\end{equation}
The  $\St_{nl}$ form an orthonormal set, with $\int_0^\infty dk\,\St_{n'l}(k)
\St_{nl}(k)=\delta_{n'n}$.  

\subsection{Length parameter}
\label{sec-cs-scale}
\begin{figure*}
\begin{center}
\includegraphics*[width=\ifproofpre{0.75}{0.95}\hsize]{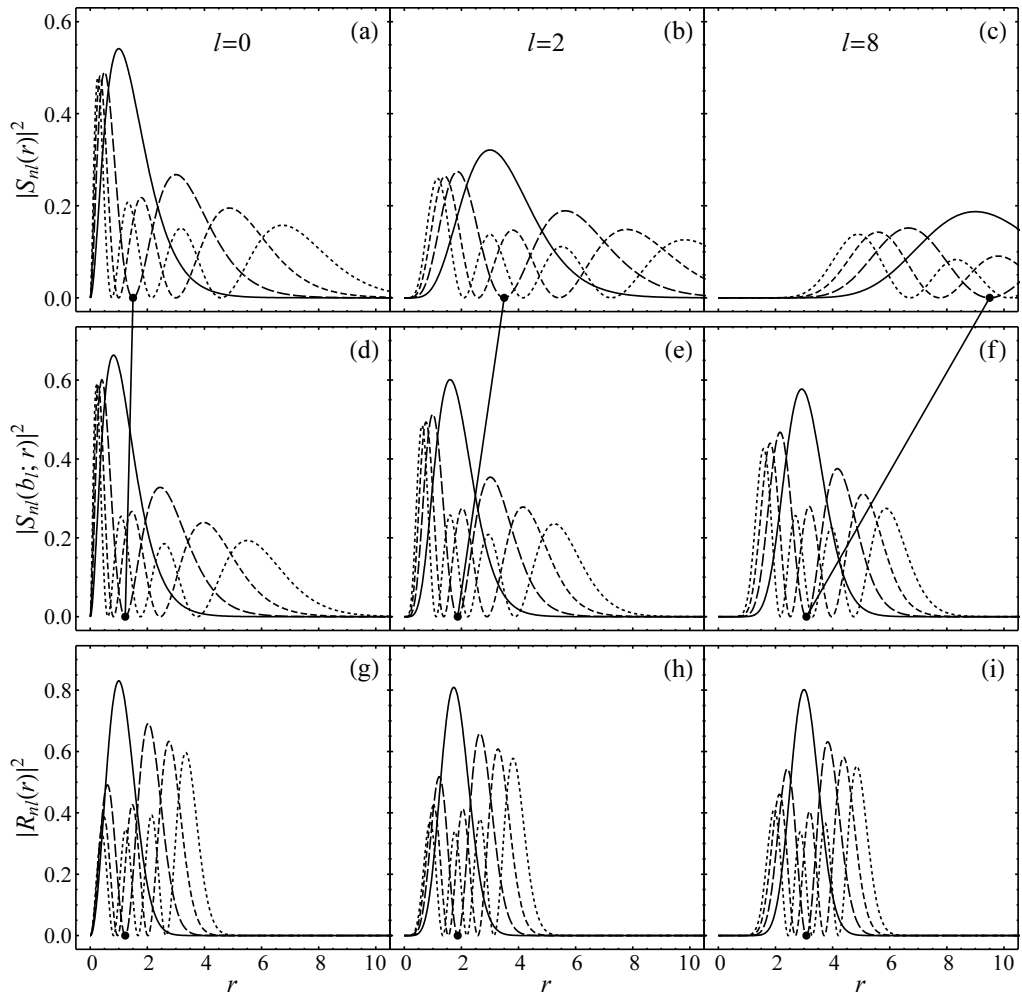}
\end{center}
\caption{Radial basis functions (shown as squared amplitudes), with $0\leq n \leq 3$, for $l=0$
(left), $l=2$ (center),
and $l=8$ (right): (a--c)~Coulomb-Sturmian functions $S_{nl}(r)$, with
fixed length parameter $b=1$, (d--f)~rescaled Coulomb-Sturmian functions
$S_{nl}(b_l;r)$, with $l$-dependent length parameter $b_l$ given by
the 
prescription~(\ref{eqn-bl-formula}), and (g--i)~harmonic oscillator functions $R_{nl}(r)$, with
fixed length parameter $b=1\,[\equiv \bho]$.  The dots
mark the location of the node of the $n=1$ function in each panel, and
the connector lines highlight the shift in this node between the
 top
and middle rows.
}
\label{fig-functions}
\end{figure*}

For any given value of $l$, the radial wave functions $S_{nl}(r)$,
with $n=0$, $1$, $\ldots$, constitute a complete and orthogonal set 
on $\bbR^+$,
regardless of the choice of length scale parameter $b$
in~(\ref{eqn-cs-S}).  For the full wave functions
$\Lambda_{nlm}(\vec{r})$ on $\bbR^3$, orthogonality of functions with
\textit{different} $l$ quantum numbers is enforced by the
$Y_{lm}(\uvec{r})$ factor, regardless of the radial wave function.
Therefore, the choice of length parameter $b$ may be made
independently for each $l$-space, and orthogonality of the basis of
single-particle states on $\bbR^3$ will still be preserved.\footnote{In fact, when
spin is introduced in the single-particle basis, a distinct value
$b_{lj}$ may be chosen for the length parameter independently for each
$lj$-space, much as different sets of radial wave functions are obtained for
each $lj$ value in the shell model Woods-Saxon
basis~\cite{suhonen2007:nucleons-nucleus}.  Different values may also
be chosen for the proton and neutron spaces.}  

The freedom to define distinct $b_l$, for different values of $l$,
appears to be crucial to the present use of
a Coulomb-Sturmian basis for the nuclear problem.  A many-body basis
built from oscillator wave functions has had considerable past success
in providing a reasonable first approximation to the central portion
of the wave functions in the nuclear problem and also clearly
enjoys the advantage of complete separability of
center-of-mass motion.  As we introduce the
Coulomb-Sturmian basis, we wish to retain the successes enjoyed by the
oscillator basis, to the extent possible, while also now providing for exponential asymptotics in the
tail, or large $r$, region.  

If $b$ is simply taken independent of $l$, Coulomb-Sturmian radial
functions $S_{nl}(r)$ are obtained as shown in
Fig.~\ref{fig-functions} (top).  For illustration, we use the
dimensionless value $b=1$ for the length parameter.  The first four
radial functions ($0\leq n \leq 3$) are shown as probability
distributions $\abs{S_{nl}(r)}^2$, for $l=0$
[Fig.~\ref{fig-functions}(a)], $l=2$ [Fig.~\ref{fig-functions}(b)],
and $l=8$ [Fig.~\ref{fig-functions}(c)].  These functions may be
compared with the corresponding radial functions $R_{nl}(r)$ for the
harmonic oscillator, as shown in Fig.~\ref{fig-functions} (bottom),
again taking the dimensionless value $b=1$ for the length parameter.
For the $S_{nl}(r)$, it may be observed that the radial probability
distribution migrates rapidly to large $r$ as $l$ increases.  By
$l=8$, the $n=0$ function [Fig.~\ref{fig-functions}(c)] shares virtually no overlap with the several
lowest-$n$ oscillator functions [Fig.~\ref{fig-functions}(i)].  Physically, it is reasonable to
expect that the success of the oscillator basis in describing the
central portion of the nuclear wave function may be lost in such a
basis.  Convergence of the description of center-of-mass motion may
also be compromised.  Computationally, there is a purely pragmatic
difficulty which effectively precludes calculations with such a basis.
It will be seen in Sec.~\ref{sec-xform} that significant overlaps
between the low-$n$ members of the Coulomb-Sturmian and oscillator
bases are required, to carry out a change-of-basis transformation on
the interaction matrix elements with reasonable accuracy.

We therefore seek an alternative prescription for $b_l$, which
provides a closer alignment of the low-$n$ Coulomb-Sturmian basis functions
with the harmonic oscillator basis functions.  A straightforward,
though certainly not unique, solution is to choose $b_l$  so as to align the node of the
$n=1$ Coulomb-Sturmian function, for the given value of $l$, with the node of the $n=1$ oscillator
function, for this same value of $l$.  It is convenient to work in
this fashion, with nodes rather
than, say, maxima, since the nodes are given by the zeros of
generalized Laguerre
polynomials~\cite{olver2010:handbook}.  Let $x^\alpha_{n,s}$ denote the
$s$th zero of the generalized Laguerre polynomial $L_n^\alpha(x)$.
The condition obtained for $b_l$, relative to the oscillator length $\bho$, is
\begin{equation}
\label{eqn-bl-x}
\frac{b_l}{\bho}=\frac{2(x_{1,1}^{l+1/2})^{1/2}}{x_{1,1}^{2l+2}},
\end{equation}
which yields the simple analytic result
\begin{equation}
\label{eqn-bl-formula}
\frac{b_l}{\bho}=\sqrt{\frac{2}{2l+3}}.
\end{equation}
Thus, \textit{e.g.}, $b_0/\bho\approx0.8165$,
$b_1/\bho\approx0.6325$, and $b_2/\bho\approx0.5345$.  The
nodes under consideration are marked by dots in
Fig.~\ref{fig-functions}.  Selecting $b_l/\bho$ according
to~(\ref{eqn-bl-formula}) yields radially rescaled Coulomb-Sturmian
functions as in Fig.~\ref{fig-functions} (middle).  These functions
are seen to provide a much closer match to the oscillator functions of
Fig.~\ref{fig-functions} (bottom) in the small-$r$ central region,
than do the unscaled functions of Fig.~\ref{fig-functions} (top),
while still retaining greater support than the oscillator functions in
the large-$r$ tail region.

The optimal approach to choosing the $b_l$ may be expected to depend
upon the problem at hand~--- nucleus, interaction, states of interest,
observables of interest, and many-body truncation scheme in use~---
and warrants thorough investigation.  The
prescription~(\ref{eqn-bl-formula}) would appear to be a reasonable
starting point and is therefore used in the example NCCI calculations
of Sec.~\ref{sec-calc}.  However, it remains to be determined what
prescription for $b_l$ might ultimately yield the most rapid
convergence in the many-body problem.  Under some circumstances, it
may even be appropriate to choose the $b_l$ separately for the proton
and neutron spaces, for instance, for neutron halo nuclei.

\subsection{Transformation of matrix elements}
\label{sec-xform}

For the many-body problem, we now consider a basis built up from the
Coulomb-Sturmian functions $\Lambda_{nlm}$, combined with spin to give
$nlj$ states as usual.  The angular and spin dependence is thus the
same as for the harmonic oscillator single-particle states, but with
the harmonic oscillator radial wave functions $R_{nl}$ replaced by the
$S_{nl}$. Many-body basis states may be built as antisymmetrized
products of these single-particle states exactly as before,
\textit{i.e.}, according to~(\ref{eqn-m-scheme-state}).  For the many-body
calculation, it is necessary for one to evaluate the matrix elements
of the Hamiltonian with respect to the many-body basis
states.  However, the specific choice of single-particle basis enters
into the problem only through the two-body matrix elements of this
Hamiltonian, if the interaction is limited to two-body
contributions, or three-body matrix elements if a three-body
interaction is considered, \textit{etc.}  Here we consider
specifically two-body interactions and matrix elements, but the
discussion readily generalizes to higher-body interactions.

If the two-body matrix elements of an interaction are known with
respect to the oscillator basis, matrix elements with respect to the
Coulomb-Sturmian basis may then be obtained by a straightforward sum
over two-body states.  Strong practical considerations suggest first
generating the nuclear interaction two-body matrix elements in the
oscillator representation.  By Galilean invariance, the interaction
itself is a function only of the relative $\vec{r}_2-\vec{r}_1$ degree
of freedom and the intrinsic spins.  Conventionally, for NCCI
calculations, the two-body interaction is first represented via its
matrix elements in a basis of harmonic oscillator states in the
relative spatial degree of freedom, coupled to the spins,
\textit{i.e.}, $\tket{nl;SJ}$.  The transformation from a relative
oscillator basis to a single-particle oscillator basis, \textit{i.e.},
to product states $\tket{n_al_aj_a,n_bl_bj_b;J}$ for the two-particle
system, can then be carried out through the well-developed framework
of the Moshinsky transformation~\cite{moshinsky1996:oscillator}.  Such
a convenient means of transformation is not, in general, available for
other bases.\footnote{We note, however, that the weakly-convergent
two-center expansion methods of
Ref.~\cite{weniger1985:fourier-plane-wave} might provide a viable
approach for carrying out such a transformation.}  Therefore, only
after this transformation to single-particle degrees of freedom do we
carry out the transformation to the Coulomb-Sturmian basis.

For purposes of discussing the change of basis, let us label
single-particle orbitals for the oscillator basis by unbarred symbols $a=(n_al_aj_a)$,
$b=(n_bl_bj_b)$, \textit{etc.}, and those for the Coulomb-Sturmian basis
by barred symbols $\bar{a}=(\bar n_a
\bar l_a \bar j_a)$, $\bar{b}=(\bar n_b \bar l_b \bar j_b)$,
\textit{etc.}  Then the two-body matrix elements in the oscillator
basis are of the form  $\tme{cd;J}{V}{ab;J}$, and we wish to
obtain transformed matrix elements
$\tme{\bar{c}\bar{d};J}{V}{\bar{a}\bar{b};J}$.
The basic ingredient is the transformation of single-particle states,
\begin{equation}
\label{eqn-xform-sp}
\tket{\bar{a}}=\sum_{a} \toverlap{a}{\bar{a}} \, \tket{a}.
\end{equation}
The angular functions $Y_{lm}$ and the coupling with spin to yield $j$
are identical for both bases, so
$\toverlap{a}{\bar{a}}=\toverlap{R_{n_al_a}}{S_{\bar{n}_al_a}}\delta_{(l_a
j_a)(\bar l_a \bar j_a)}$, and the sum over orbitals $a$ in fact only
involves a sum over
\textit{radial} quantum numbers $n_a$.   In writing out the overlap
$\toverlap{R_{n_al_a}}{S_{\bar{n}_al_a}}$, it is worthwhile to
explicitly indicate the different choices of length parameter
appearing in
$R_{nl}(r)$ and
$S_{nl}(r)$, for which we adopt the notations $R_{nl}(b;r)$ and
$S_{nl}(b;r)$. Then, the overlap is given by the radial integral\footnote{The oscillator wave functions as defined in~(\ref{eqn-ho-wf}) are positive at the origin,
\textit{i.e.}, as $r\rightarrow0$. The Coulomb-Sturmian functions as
defined in~(\ref{eqn-cs-Lambda}) have this property as well.  It should be noted that a conventional phase factor
$(-)^n$ may be included in the definition of $\wfho_{nlm}$, so
that the functions are instead positive at
infinity, \textit{i.e.}, as $r\rightarrow\infty$.  If so, this sign must be accounted for in
evaluating the transformation bracket~(\ref{eqn-ol-S-R})
for the change of basis.  Alternatively, the phase convention for the
Coulomb-Sturmian basis may be adjusted analogously.} 
\begin{equation}
\label{eqn-ol-S-R}
\toverlap{R_{nl}}{S_{\bar{n}l}} 
= \int_0^\infty dr\,
R_{nl}(\bho;r)
S_{\bar{n}l}(b_l;r) 
.
\end{equation}
Equivalently, the overlaps may be evaluated in momentum space, as
\begin{equation}
\label{eqn-ol-St-Rt}
\toverlap{\tilde{R}_{nl}}{\tilde{S}_{\bar{n}l}} 
= 
\int_0^\infty dk\,
\tilde{R}_{nl}(\bho;k)
\tilde{S}_{\bar{n}l}(b_l;k) 
.
\end{equation}
\begin{figure}
\begin{center}
\includegraphics*[width=\ifproofpre{0.95}{0.5}\hsize]{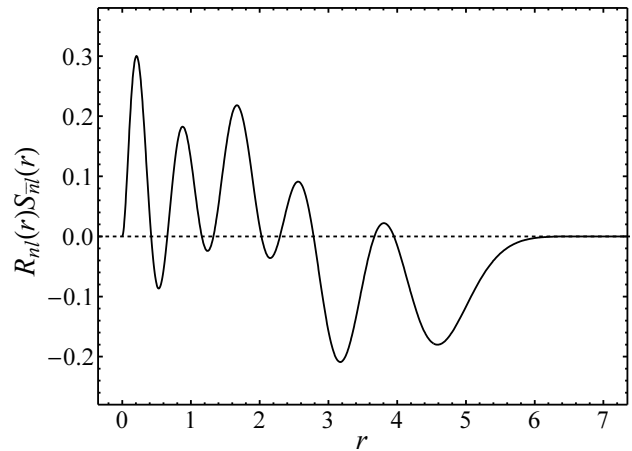}
\end{center}
\caption{Integrand $R_{nl}(\bho;r) S_{\bar{n}l}(b_l;r)$ for
the overlap integral~(\ref{eqn-ol-S-R}), taken for a representative case ($l=0$,
$\bar{n}=5$, and $n=5$).  For this plot, $b_l/\bho$ is given by
the 
prescription~(\ref{eqn-bl-formula}), and $\bho$ is taken to be unity.
}
\label{fig-integrand-overlap}
\end{figure}

When larger values for the radial quantum numbers $\bar{n}$ or $n$ are
considered, the integrand appearing in the overlap
integral~(\ref{eqn-ol-S-R}) or~(\ref{eqn-ol-St-Rt}) is highly
oscillatory, as illustrated in Fig.~\ref{fig-integrand-overlap}~---
as is to be expected for overlap integrals of functions with large
numbers of nodes.  Therefore, care must be taken in evaluating the
overlap integral through numerical quadrature.  Conventional
quadrature formulas are found to be slowly-converging and unreliable.
However, the zeros of the integrand are easily determined, from the
zeros of the generalized Laguerre polynomials or Jacobi polynomials,
in terms of which the radial functions are defined, as summarized in
Appendix~\ref{app-zeros}.  Integration can then be carried out in a
numerically robust fashion if the full integration range $[0,\infty)$
is first broken into intervals between successive zeros.  Within each
interval, the integrand is well-behaved, and conventional numerical
quadrature can be carried out reliably.  The results may then be summed
to give the full integral.  It is found that a $32$-point
Gauss-Legendre quadrature on each interval suffices for present
purposes, yielding numerical errors of $\lesssim10^{-8}$ (and
generally much better) for calculations involving radial wave functions
with $n \lesssim 20$.  Integration in the tail region, between the
last zero of the integrand and infinity, requires special treatment,
since Gauss-Legendre quadrature is only defined on finite intervals.
One can map the tail region onto
a finite interval by a suitable transformation of integration variable.
Alternatively, and most simply, the integration may be truncated at a sufficiently large
cutoff $\rmax$, \textit{e.g.}, $\rmax/b \approx 50$ is found to suffice in the
present calculations.   

For proton-neutron matrix elements, the
two-body states  transform as
\begin{equation}
\label{eqn-tb-xform-pn}
\tket{\bar{a}\bar{b};J}_\pn
=
\sum_{ab} \toverlap{a}{\bar{a}}\toverlap{b}{\bar{b}} \, \tket{ab;J}_\pn,
\end{equation}
and the matrix elements consequently transform as
\begin{multline}
\label{eqn-tbme-xform-pn}
\tme{\bar{c}\bar{d};J}{V}{\bar{a}\bar{b};J}_\pn
\ifproofpre{\\}{}
=
\sum_{abcd} \toverlap{a}{\bar{a}}\toverlap{b}{\bar{b}}
\toverlap{c}{\bar{c}}\toverlap{d}{\bar{d}}
\, \tme{cd;J}{V}{ab;J}_\pn.
\end{multline}
As noted above for~(\ref{eqn-xform-sp}), the sums over orbitals $a$,
$b$, $c$, and $d$ need only traverse the radial quantum numbers $n_a$,
$n_b$, $n_c$, and $n_d$, preserving the same angular quantum
numbers. 

For proton-proton or neutron-neutron
matrix elements, normalization considerations related to
antisymmetrization must be taken into account in carrying out the
transformation.  Since different normalization conventions arise in
the description of two-particle states, the present conventions are
briefly summarized in Appendix~\ref{app-tb}.  It is easiest to state
the transformation rule if the two-body matrix elements are defined in
terms of the \textit{antisymmetrized} (AS) two-particle states $\tket{ab;JM}_\as$
of~(\ref{eqn-state-as}), which are properly
normalized except in the case in which both particles occupy the same
orbital.  Then, it maybe be seen [\textit{e.g.}, by carrying out a
change of basis on the creation operators in~(\ref{eqn-state-as})~\cite{negele1988:many-particle}] that we simply have
\begin{equation}
\label{eqn-tb-xform-as}
\tket{\bar{a}\bar{b};J}_\as=\sum_{ab} \toverlap{a}{\bar{a}}\toverlap{b}{\bar{b}} \, \tket{ab;J}_\as.
\end{equation}
Consequently, for the two-body matrix elements,
\begin{multline}
\label{eqn-tbme-xform-as}
\tme{\bar{c}\bar{d};J}{V}{\bar{a}\bar{b};J}_\as
\ifproofpre{\\}{}
=
\sum_{abcd} \toverlap{a}{\bar{a}}\toverlap{b}{\bar{b}}
\toverlap{c}{\bar{c}}\toverlap{d}{\bar{d}}
\, \tme{cd;J}{V}{ab;J}_\as.
\end{multline}
The corresponding expression for the transformation
in terms of the strictly \textit{normalized antisymmetrized} (NAS) states
$\tket{ab;JM}_\nas$ of~(\ref{eqn-state-nas}) is less transparent, since the case of
identical orbitals must be treated specially within the sum, giving
\ifproofpre{\begin{widetext}}{}
\begin{multline}
\label{eqn-tbme-xform-nas}
\tme{\bar{c}\bar{d};J}{V}{\bar{a}\bar{b};J}_\nas=
(1+\delta_{\bar{a}\bar{b}})^{-1/2}
(1+\delta_{\bar{c}\bar{d}})^{-1/2}
\ifproofpre{}{\\\times}
\sum_{abcd} 
(1+\delta_{ab})^{1/2}
(1+\delta_{cd})^{1/2}
\toverlap{a}{\bar{a}}\toverlap{b}{\bar{b}}
\toverlap{c}{\bar{c}}\toverlap{d}{\bar{d}}
\, \tme{cd;J}{V}{ab;J}_\nas.
\end{multline}
\ifproofpre{\end{widetext}}{}
It is trivial to convert between AS and NAS matrix elements,
and thus to use either relation~(\ref{eqn-tbme-xform-as})
or~(\ref{eqn-tbme-xform-nas}), but
it is important to note the distinction.

For actual calculation of the transformed matrix elements, the
infinite sums over orbitals appearing in the transformation
rule~(\ref{eqn-tbme-xform-pn})
and~(\ref{eqn-tbme-xform-as}) [or (\ref{eqn-tbme-xform-nas})] must be
truncated, limited in practice by the available set of
oscillator-basis matrix elements.  If a shell-based cutoff,
\textit{i.e.}, by number of oscillator quanta, is applied
to the single-particle space, then $N\leq\Ncut$ for the
single-particle states, and the sum $\sum_{abcd}$ appearing
in~(\ref{eqn-tbme-xform-pn}) and~(\ref{eqn-tbme-xform-as}) is
truncated to $\sum_{abcd}^{N_a,N_b,N_c,N_d\leq\Ncut}$.  For example,
the set of oscillator basis two-body matrix elements required for a
transformation with cutoff $\Ncut=13$ ($14$ shells) consists of
$9.2\times10^7$ proton-neutron two-body matrix elements and
$2.3\times10^7$ proton-proton or neutron-neutron matrix
elements.\footnote{These are the possible nonzero two-body matrix
elements (Appendix~\ref{app-tb}), with single-particle states taken
from $14$ shells, for an interaction which is parity-conserving but
with no further assumptions about isospin (or charge) symmetry.
Actual nucleon-nucleon interactions may in fact contain fewer
independent matrix elements.}  The
summations~(\ref{eqn-tbme-xform-pn}) or~(\ref{eqn-tbme-xform-as}) only
involve matrix elements sharing the same angular momentum $J$, parity
$P$, and isospin projection $T_z$ ($pn$, $pp$, or $nn$), and thus in
practice the transformation may be carried out separately for each sector
of matrix elements, characterized by these quantum numbers.
\textit{After} transformation, substantially fewer matrix elements are
required for an $\Nmax$-truncated many-body calculation in the same
number of shells,
\textit{e.g.}, for $p$-shell nuclei, an $\Nmax=12$ calculation
involves $14$ shells but only $4.1\times10^6$ proton-neutron two-body
matrix elements and $1.0\times10^6$ proton-proton or
neutron-neutron matrix elements, due to the further restriction on
$\Ntot$.

The accuracy of the resulting two-body matrix elements 
obtained for the Coulomb-Sturmian basis depends on the inclusion of an
adequate number of oscillator shells.  The effect of truncation may in
general be expected to vary depending on the two-body operator under
consideration.  In practice, the adequacy of the transformation may be
judged by the sensitivity of the final many-body calculation to
$\Ncut$.  Calculations with $\Ncut=9$ ($10$ shells), $\Ncut=11$ ($12$
shells), and $\Ncut=13$ ($14$ shells) are considered in
Sec.~\ref{sec-calc}.
\begin{figure*}
\begin{center}
\includegraphics*[width=\ifproofpre{0.75}{0.95}\hsize]{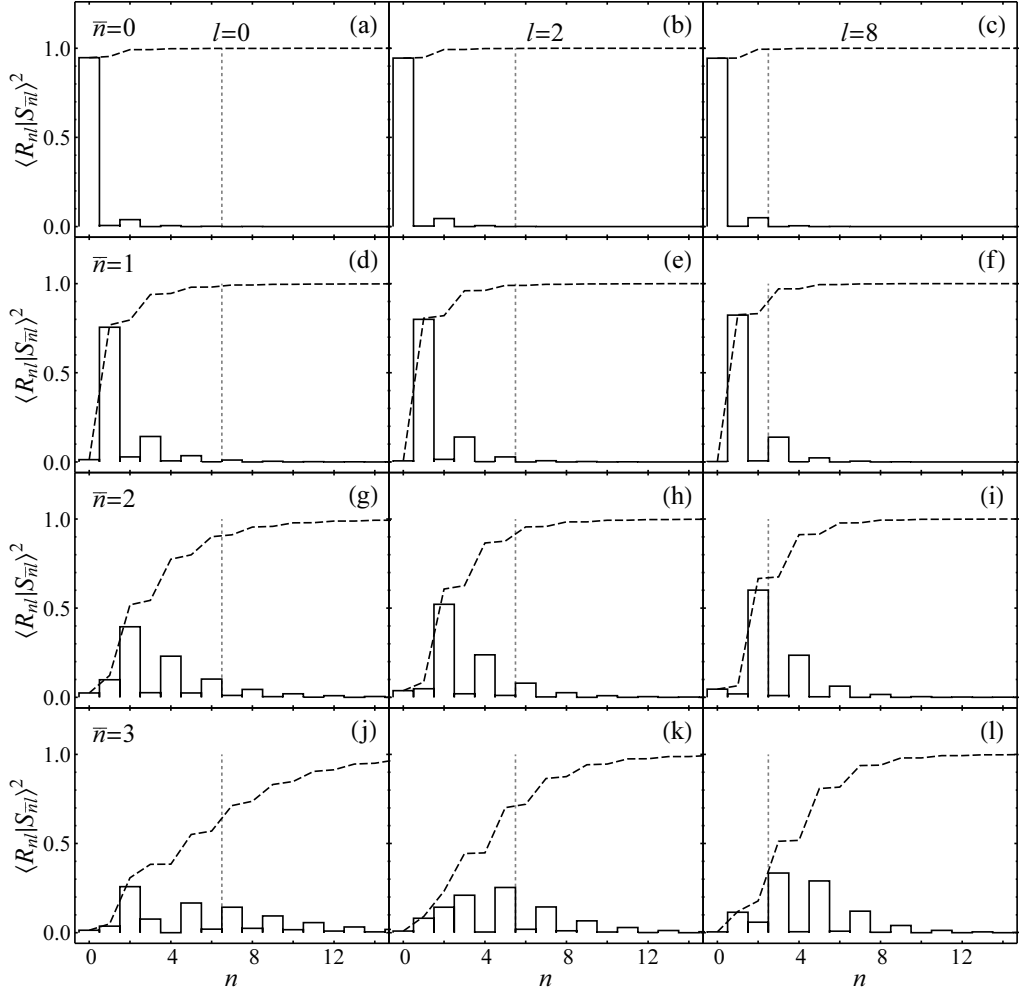}
\end{center}
\caption{Probability decomposition of the Coulomb-Sturmian radial functions
$S_{\bar{n}l}(b_l;r)$ with respect to the basis of harmonic oscillator
radial functions $R_{nl}(\bho;r)$.  Results are shown for
Coulomb-Sturmian functions with $0\leq n \leq 3$ (top to bottom) and
for $l=0$ (left), $l=2$ (center), and $l=8$ (right), with
$b_l/\bho$ given by the node-matching
prescription~(\ref{eqn-bl-formula}).  The histogram bars indicate
squared amplitudes
$\toverlap{R_{nl}(\bho;r)}{S_{\bar{n}l}(b_l;r)}^2$ with respect to
\textit{individual} oscillator basis functions.  The dashed curve
indicates accumulated probability, \textit{i.e.}, for \textit{all} oscillator basis
functions of lesser or equal $n$.  The vertical dotted line indicates
the truncation of the radial basis in effect if the oscillator
functions are limited to $14$ major shells ($\Ncut=13$), as in the
least-truncated calculations of Sec.~\ref{sec-calc}.  }
\label{fig-overlaps}
\end{figure*}

Since the change of basis~(\ref{eqn-xform-sp})  represents a
transformation of radial wave functions, the underlying approximation
in applying a cutoff is that we are effectively representing the
Coulomb-Sturmian radial functions in terms of a truncated set of
oscillator radial functions, as
\begin{equation}
S_{\bar{n}l}(b_l;r)=\sum_{n}^{N\leq\Ncut} \toverlap{R_{nl}}{S_{\bar{n}l}} 
 \,  R_{nl}(\bho;r),
\end{equation}
with $N=2n+l$, so $n\leq(\Ncut-l)/2$.  The decomposition of Coulomb-Sturmian functions in
terms of oscillator functions, shown as squared amplitudes
(probabilities), is given in Fig.~\ref{fig-overlaps}.  Results are
shown for the functions previously plotted in
Fig.\ref{fig-functions}(d--f), that is, with $0\leq n \leq 3$ and for
$l=0$, $2$, and $8$, with the length scales of the functions
determined according to the prescription~(\ref{eqn-bl-formula}).
While the first two or three Coulomb-Sturmian functions for each value of $l$ are
easily expanded in the oscillator basis, the
required number of shells is seen to grow rapidly for higher radial
quantum numbers.  The degree to which the Coulomb-Sturmian radial
function is successfully expanded in a truncated set of  oscillator
radial functions is
seen from the dashed curves in Fig.~\ref{fig-overlaps}, which indicate the
accumulated probability $P_n=\sum_{n'\leq n}
\toverlap{R_{n'l}}{S_{\bar{n}l}}^2$.  The set of oscillator radial
functions retained in the most generous truncation used in
Sec.~\ref{sec-calc}, $\Ncut=13$, can be seen from the vertical
dotted line in each panel of Fig.~\ref{fig-overlaps}.

\subsection{Evaluation of two-body matrix elements for separable radial
and kinetic operators}
\label{sec-sep}

If the two-body matrix elements of the entire Hamiltonian are first
evaluated in the oscillator basis then transformed to the
Coulomb-Sturmian basis, according to the procedure of
Sec.~\ref{sec-xform}, it is found (Sec.~\ref{sec-calc}) that the kinetic
energy term requires an unacceptably large number of oscillator shells
for its expansion.  That is, the $\Ncut$-dependence of the transformed
relative kinetic energy, rather than of the transformed
nucleon-nucleon interaction, dominates the cutoff dependence of the
many-body calculations.

In this section, we therefore instead consider a scheme which permits
the two-body matrix elements of the center-of-mass and relative
components of the $r^2$ and $p^2$ operators~--- $R^2$, $\rrel^2$, $P^2$, and
$\prel^2$~--- to be evaluated directly in the Coulomb-Sturmian
basis.  The approach makes use of separability, together with the
explicitly known form~(\ref{eqn-cs-St}) of the Coulomb-Sturmian radial
wave function in momentum space.  The operators $R^2$, $\rrel^2$,
$P^2$, and $\prel^2$ all appear in the NCCI problem.
Specifically, $\prel^2$ appears through the relative kinetic energy
operator, $\rrel^2$ through the root-mean-square (RMS) radius observable, and $R^2$ and
$P^2$ through the center-of-mass oscillator Hamiltonian appearing in
the Lawson term.  The definitions of and relations among these
operators are summarized for reference in Appendix~\ref{app-r2k2}.

Each of the operators $R^2$, $\rrel^2$, $P^2$, and $\prel^2$ may be
decomposed into one-body terms and separable two-body terms.  In the
following, we let $\vec{p}=\hbar\vec{k}$ and work with $K^2$ and
$\krel^2$ instead of $P^2$ and $\prel^2$.  Then we have (see
Appendix~\ref{app-r2k2}):
\begin{equation}
\label{eqn-r2-k2-decomp}
\begin{aligned}
A^2R^2&=\sum_ir_i^2+{\sumprime_{ij}}\vec{r}_i\cdot\vec{r}_j
\\
A^2 \rrel^2&=(A-1)\sum_ir_i^2-{\sumprime_{ij}} \vec{r}_i\cdot\vec{r}_j
\\
K^2&=\sum_i k_i^2+{\sumprime_{ij}}\vec{k}_i\cdot\vec{k}_j
\\
\krel^2&=(A-1)\sum_i k_i^2-{\sumprime_{ij}} \vec{k}_i\cdot\vec{k}_j.
\end{aligned}
\end{equation}
The terms involving $\sum_i$ are manifestly one-body operators, and
those involving ${\tsumprime_{ij}}$ are manifestly two-body operators.  The
important property of these expressions~(\ref{eqn-r2-k2-decomp}) for
the present approach is that the two-body term in each case~---
${\tsumprime_{ij}}\vec{r}_i\cdot\vec{r}_j$ or ${\tsumprime_{ij}}
\vec{p}_i\cdot\vec{p}_j$~--- has the separable form ${\tsumprime_{ij}}
\vec{T}_i\cdot\vec{T}_j$, where $\vec{T}_k$ is a spherical tensor (in
the present case, rank-$1$ or vector) operator acting on particle $k$
only.  The procedure for calculating two-body matrix elements
therefore reduces to the evaluation of radial integrals (either in
coordinate space or momentum space, for $\vec{r}_i$ or $\vec{k}_i$,
respectively), which are then combined using standard angular momentum
coupling and recoupling results.

First, let us consider the matrix elements of the one-body terms
$\sum_ir_i^2$ and $\sum_ik_i^2$ appearing in~(\ref{eqn-r2-k2-decomp}).
For the Coulomb-Sturmian basis, the one-body matrix
elements of $r^2$ and $k^2$ are
\begin{equation}
\label{eqn-me-r2}
\tme{b}{r^2}{a}=\delta_{l_bl_a}\delta_{j_bj_a}\int_0^\infty dr\,S_{n_bl_b}(b_{l_b};r)\,r^2\,S_{n_al_a}(b_{l_a};r)
\end{equation}
and
\begin{equation}
\label{eqn-me-k2}
\tme{b}{k^2}{a}=\delta_{l_bl_a}\delta_{j_bj_a}\int_0^\infty dk\,\St_{n_bl_b}(b_{l_b};k)\,k^2\,\St_{n_al_a}(b_{l_a};k).
\end{equation}
The radial integrals appearing in these expressions may be evaluated
by numerical quadrature.  Since the integrands are highly oscillatory,
the comments and methods of Sec.~\ref{sec-xform} apply to this
integration.  The integrals are again evaluated piecewise between
zeros of the integrands, through Gauss-Legendre quadrature.

Although $\sum_ir_i^2$ and $\sum_ik_i^2$ are one-body operators, they
are being considered here as contributions to the two-body operators
$R^2$, $\rrel^2$, $P^2$, and $\prel^2$,
through~(\ref{eqn-r2-k2-decomp}), for which \textit{two-body} matrix
elements are therefore required as input to the many-body calculation.
The appropriate two-body matrix elements are readily
obtained from the one-body matrix elements $\tme{b}{r^2}{a}$ and
$\tme{b}{k^2}{a}$ considered in~(\ref{eqn-me-r2})
and~(\ref{eqn-me-k2}).  In general, corresponding to any one-body operator
$U=\sum_i u_i$, we may define a two-body operator $V_U$ via
$V_U=\tfrac12{\tsumprime_{ij}}v_{ij}$, where $v_{ij}=u_i+u_j$.  By comparing
the sums appearing in the definitions of $U$ and $V_U$, it may be seen
that these operators  are identical, except for an $A$-dependent
normalization.  Specifically, the operators are
related by 
\begin{equation}
\label{eqn-vt-t}
U=\frac{1}{A-1}V_U,
\end{equation}
when acting
on the many-body states of an $A$-particle system.  

We therefore consider two-body matrix elements of $V_U$.  For the proton-neutron matrix elements, 
\begin{equation}
\label{eqn-tbme-obme-pn}
\tme{c d;J}{V_U}{a b;J}_\pn=
\tme{c}{U}{a}\delta_{db}
+\tme{d}{U}{b}\delta_{ca}
.
\end{equation}
For the proton-proton or neutron-neutron matrix
elements,  the antisymmetrized matrix element may be evaluated by first
reexpressing it in terms of
unsymmetrized matrix elements,
as
\begin{multline}
\label{eqn-tbme-obme-as-prod}
\tme{cd;J}{V_U}{ab;J}_\as=\tpme{cd;J}{v_{12}}{ab;J}
\ifproofpre{\\}{}
-(-)^{J-j_a-j_b}\tpme{cd;J}{v_{12}}{ba;J},
\end{multline}
with $v_{ij}$ as defined above.
It follows that
\begin{multline}
\label{eqn-tbme-obme-as}
\tme{cd;J}{V_U}{ab;J}_\as
=\tme{c}{U}{a}\delta_{db}
+\tme{d}{U}{b}\delta_{ca}
\ifproofpre{\\}{}
-(-)^{J-j_a-j_b}\tme{c}{U}{b}\delta_{da}
-(-)^{J-j_a-j_b}\tme{d}{U}{a}\delta_{cb}.
\end{multline}
Thus, the matrix elements of interest for the one-body terms appearing
in~(\ref{eqn-r2-k2-decomp}) are obtained by setting $U=r^2$ or $k^2$
and using one-body matrix elements~(\ref{eqn-me-r2})
or~(\ref{eqn-me-k2}), respectively, in~(\ref{eqn-tbme-obme-pn}) and~(\ref{eqn-tbme-obme-as}).

Now let us consider the matrix elements of the two-body terms $\tfrac12{\tsumprime_{ij}}
\vec{r}_i\cdot\vec{r}_j$ and
$\tfrac12{\tsumprime_{ij}} \vec{k}_i\cdot\vec{k}_j$ appearing
in~(\ref{eqn-r2-k2-decomp}).  We include a factor of $1/2$ in these
expressions to bring them into the standard form for two-body
operators, namely, $V=\tfrac12{\tsumprime_{ij}} v_{ij}$, with
$v_{ij}=v_{ji}$.  The operator defined by the sum, in either
case, is of the separable form
$V_{\vec{T}_1\cdot\vec{T}_2}=\tfrac12{\tsumprime_{ij}}\vec{T}_i\cdot\vec{T}_j$,
where $\vec{T}_i$ is a vector operator acting on particle $i$.  
Since the summand is a
spherical tensor product of operators acting on two different
subsystems (namely, particles
$i$ and $j$), it is possible to evaluate the matrix elements by Racah's
reduction formula~\cite{edmonds1960:am}.  
For the proton-neutron matrix elements, 
\begin{multline}
\label{eqn-me-sep-pn}
\tme{cd;J}{\vec{T}_1\cdot\vec{T}_2}{ab;J}_\pn
\ifproofpre{\\}{}
=(-)^{j_d+j_a+J}\smallsixj{j_c}{j_d}{J}{j_b}{j_a}{1}
\trme{c}{\vec{T}}{a}\trme{d}{\vec{T}}{b}.
\end{multline}
For the proton-proton and neutron-neutron matrix elements, it is
important to note that Racah's reduction formula applies to matrix
elements between ordinary, unsymmetrized product states of
distinguishable subsystems.  Thus, the two-body matrix element between
\textit{antisymmetrized} states of two like nucleons must first be
expanded by~(\ref{eqn-state-as}) in terms of unsymmetrized matrix
elements, as
\begin{multline}
\label{eqn-me-sep-as-prod}
\tme{cd;J}{V_{\vec{T}_1\cdot\vec{T}_2}}{ab;J}_\as
=\tpme{cd;J}{\vec{T}_1\cdot\vec{T}_2}{ab;J}
\ifproofpre{\\}{}
-(-)^{J-j_a-j_b}\tpme{cd;J}{\vec{T}_1\cdot\vec{T}_2}{ba;J}.
\end{multline}
Then, each of the two terms may be evaluated separately through Racah's 
reduction formula, much as in~(\ref{eqn-me-sep-pn}), giving
\begin{multline}
\label{eqn-me-sep-prod}
\tpme{cd;J}{\vec{T}_1\cdot\vec{T}_2}{ab;J}
\ifproofpre{\\}{}
=(-)^{j_d+j_a+J}\smallsixj{j_c}{j_d}{J}{j_b}{j_a}{1}
\trme{c}{\vec{T}}{a}\trme{d}{\vec{T}}{b}
\end{multline}
for the first term, and similarly with $b\leftrightarrow a$ for the
second term.  

The one-body reduced matrix elements 
$\trme{b}{\vec{T}}{a}$ appearing in~(\ref{eqn-me-sep-pn})
or~(\ref{eqn-me-sep-prod}) are expressed in terms of radial integrals,
using the general relation $x_m=\sqrt{4\pi/3}xY_{1m}(\uvec{x})$ for
the spherical components of a coordinate vector $\vec{x}$ in terms of $Y_1$~\cite{suhonen2007:nucleons-nucleus}, as
\begin{multline}
\label{eqn-me-r}
\trme{b}{\vec{r}}{a}=\biggl(\frac{4\pi}{3}\biggr)^{1/2}
\ifproofpre{\\\times}{}
\biggl[\int_0^\infty dr\,S_{n_bl_b}(b_{l_b};r) \, r \,
S_{n_al_a}(b_{l_a};r)\biggr]
\ifproofpre{\\\times}{}
\trme{l_bj_b}{Y_1}{l_aj_a}
\end{multline}
and
\begin{multline}
\label{eqn-me-k}
\trme{b}{\vec{k}}{a}=(-)^{(l_b-l_a-1)/2}i\biggl(\frac{4\pi}{3}\biggr)^{1/2}
\ifproofpre{\\\times}{}
\biggl[\int_0^\infty dk\,\St_{n_bl_b}(b_{l_b};k) \, k \,
\St_{n_al_a}(b_{l_a};k)\biggr]
\ifproofpre{\\\times}{}
\trme{l_bj_b}{Y_1}{l_aj_a}.
\end{multline}
Numerical evaluation of these radial integrals is again subject to the considerations for
oscillatory integrands discussed in Sec.~\ref{sec-xform}.  The
angular factor appearing in (\ref{eqn-me-r})~and~(\ref{eqn-me-k}) is
given by~\cite{suhonen2007:nucleons-nucleus}
\begin{equation}
\label{eqn-me-Y1}
\trme{l_bj_b}{Y_1}{l_aj_a}=
\biggl(\frac{3}{4\pi}\biggr)^{1/2}(-)^{j_b-j_a+1}
\tcg{j_a}{\tfrac12}{1}{0}{j_b}{\tfrac12}
\pi(l_a 1 l_b),
\end{equation}
where $\pi(l_1l_2\cdots)\equiv\tfrac12[1+(-)^{l_1+l_2+\cdots}]$.  The
factor $\pi(l_a 1 l_b)$ enforces the parity selection rule for $Y_1$,
namely, $l_b-l_a$ odd.  Since the angular momentum triangle inequality
also applies, the radial matrix elements $\trme{b}{\vec{r}}{a}$ or
$\trme{b}{\vec{k}}{a}$ need only be evaluated for pairs of orbitals
for which $l_b=l_a\pm 1$.  The phase factor $(-)^{(l_b-l_a-1)/2}$
in~(\ref{eqn-me-k}) arises from the phase factor $(-i)^l$ in the
definition~(\ref{eqn-ho-wf-tilde-generic}) of the momentum-space radial wave function, after simplifications are carried out making use of the
constraints on $l$-values imposed by the angular factor~(\ref{eqn-me-Y1}).

In summary, the two-body matrix elements of $R^2$, $\rrel^2$,
$K^2$, or $\krel^2$ are evaluated by calculating the one body contributions according to~(\ref{eqn-tbme-obme-pn})
or~(\ref{eqn-tbme-obme-as}) and combining these with the matrix elements of the
two-body contribution, calculated
according to~(\ref{eqn-me-sep-pn}) or~(\ref{eqn-me-sep-as-prod}), via
the operator relations~(\ref{eqn-r2-k2-decomp}).  Collecting the various contributions and normalization factors, we have
\ifproofpre{\begin{widetext}}{}
\begin{equation}
\label{eqn-r2-k2-me}
\begin{aligned}
\tme{cd;J}{A^2R^2}{ab;J}&=\frac{1}{A-1}\tme{cd;J}{V_{r^2}}{ab;J}+2\tme{cd;J}{V_{\vec{r}_1\cdot\vec{r}_2}}{ab;J}
\\
\tme{cd;J}{A^2\rrel^2}{ab;J}&=\tme{cd;J}{V_{r^2}}{ab;J}-2\tme{cd;J}{V_{\vec{r}_1\cdot\vec{r}_2}}{ab;J}
\\
\tme{cd;J}{K^2}{ab;J}&=\frac{1}{A-1}\tme{cd;J}{V_{k^2}}{ab;J}+2\tme{cd;J}{V_{\vec{k}_1\cdot\vec{k}_2}}{ab;J}
\\
\tme{cd;J}{\krel^2}{ab;J}&=\tme{cd;J}{V_{k^2}}{ab;J}-2\tme{cd;J}{V_{\vec{k}_1\cdot\vec{k}_2}}{ab;J}.
\end{aligned}
\end{equation}
\ifproofpre{\end{widetext}}{}
Further practical aspects of evaluating these matrix elements are
considered in Appendix~\ref{app-rescaling}.

Although the separable method described here for evaluating two-body
matrix elements of $R^2$, $\rrel^2$, $K^2$, and $\krel^2$ has been
presented in the context of the Coulomb-Sturmian basis, this approach
is applicable to a general radial basis, so long as both the coordinate-space and  
momentum-space radial wave function can be accurately evaluated and
integrated.  The only basis dependence lies in evaluating the radial
integrals~(\ref{eqn-me-r2}), (\ref{eqn-me-k2}), (\ref{eqn-me-r}),
and~(\ref{eqn-me-k}).  For instance, the separable method can be used with the
oscillator basis, applied to the radial functions $R_{nl}(r)$
of~(\ref{eqn-ho-R}) and $\Rt_{nl}(k)$ of~(\ref{eqn-ho-Rt}), in lieu
of Moshinsky transformation.\footnote{In fact, the
separable method has been used to evaluate the matrix elements of the
$\Trel$, $\Ncm$, and $\rrel^2$ operators for the oscillator-basis NCCI
calculations
shown in Sec.~\ref{sec-calc}.  Comparison against the results obtained with
existing Moshinsky-based oscillator-basis calculations provides a vital means of
validating the present computational framework for general bases.}

\section{\boldmath Coulomb-Sturmian calculations for $\isotope[6]{Li}$}
\label{sec-calc}

\subsection{Overview}
\label{sec-calc-over}

As a basic illustration of the use of the Coulomb-Sturmian basis for
NCCI calculations, we consider the nucleus $\isotope[6]{Li}$.  The
code
MFDn~\cite{sternberg2008:ncsm-mfdn-sc08,vary2009:ncsm-mfdn-scidac09,maris2010:ncsm-mfdn-iccs10}
is used for the many-body calculations, taking as its input
Hamiltonian two-body matrix elements obtained according to the
procedures developed in Secs.~\ref{sec-xform} and~\ref{sec-sep}.
Calculations are carried out with respect to a proton-neutron
$M$-scheme basis.

The question arises as to how to truncate a many-body basis built from
Coulomb-Sturmian functions.  For the present calculation, we
\textit{formally} carry over the $\Nmax$ truncation scheme to the
Coulomb-Sturmian basis.  That is, for each Coulomb-Sturmian
single-particle state, we define $N=2n+l$.  Then, as for the oscillator
basis, we label the many-body states by the sum $\Ntot=\sum_i N_i$ and
apply the $\Nmax$ truncation as defined in~(\ref{eqn-Nmax-trunc}).
Since $n$ is now the radial quantum number for the Coulomb-Sturmian
functions, the label $N$ no longer has any direct significance in
terms of oscillator quanta.  Furthermore, when applied to the
Coulomb-Sturmian basis, the $\Nmax$ truncation does \textit{not} imply
the exact separation properties described in Sec.~\ref{sec-nmax}, nor
can it any longer be interpreted as an ``energy'' truncation, with
respect to some noninteracting Hamiltonian.  Nonetheless, as one of
many conceivable truncation schemes, the $\Nmax$ scheme provides a
reasonable starting point for further exploration, and it is
particularly convenient for use with existing NCCI many-body codes.
Furthermore, using an $\Nmax$ truncation facilitates comparison of
convergence rates obtained using the oscillator
and Coulomb-Sturmian bases, since the dimensions of the many-body
spaces are then the same in both cases.

The result for any given observable has a twofold dependence on the
basis used: on the truncation and on the length parameter.  In the
existing literature on the NCCI approach with the oscillator basis, the
oscillator length $b$ for the basis is commonly not stated directly,
but rather the oscillator energy $\hbar\Omega$ is given, in terms of
which we recall $b=[\hbar/(m_N\Omega)]^{1/2}$.  For consistency, we
therefore adopt the same convention for the Coulomb-Sturmian basis.
However, it must be borne in mind that the $\hbar\Omega$ value quoted
for the Coulomb-Sturmian basis is simply the $\hbar\Omega$ of the
\textit{reference} oscillator length $\bho$, from which the actual
$l$-dependent length parameters $b_l$ are derived by the node-matching
prescription of Sec.~\ref{sec-cs-scale}.  It therefore has no
direct significance as an energy scale for the problem.  When
comparing calculations in the harmonic oscillator basis and in the
Coulomb-Sturmian basis, the relationship of $\hbar\Omega$ values
between the two calculations should therefore also not be viewed as one of strict
physical equivalence,
\textit{e.g.}, it is not necessarily most appropriate to compare an
$\hbar\Omega=20\,\MeV$ oscillator basis calculation with 
an $\hbar\Omega=20\,\MeV$ Coulomb-Sturmian basis calculation.  Rather,
a set of calculations for each basis, spanning a range of $\hbar\Omega$
values, should be considered, and best convergence may be obtained for
different $\hbar\Omega$ values in each of the two bases.  However, for either basis, the same
proportionality $b\propto (\hbar\Omega)^{-1/2}$ holds, \textit{e.g.},
a doubling in $\hbar\Omega$ corresponds to a factor of $\sqrt{2}$
contraction of the length scale.

The present $\isotope[6]{Li}$ calculations are carried out for the
JISP16 interaction~\cite{shirokov2007:nn-jisp16}, which is a two-body
interaction derived from neutron-proton scattering data and adjusted
via a phase-shift equivalent transformation to describe light nuclei
without explicit three-body interactions.  All calculations shown here
are for the positive-parity space, spanned by states with even values
$\Ntot = N_0$, $N_0+2$, $\ldots$, $N_0+\Nmax$.  Although isospin is
not strictly conserved by the Hamiltonian, due to the Coulomb
interaction, the isospin $T$ is essentially a good quantum number for
the states in the present calculations.  Therefore, for simplicity, we
restrict attention to the $T=0$ spectrum.  Calculations are carried
out in several truncated spaces with $\Nmax\leq10$, to provide an
initial investigation into convergence.  

The nucleus $\isotope[6]{Li}$
provides a useful case for benchmarking, since calculations with
comparatively large values of $\Nmax$ are feasible with the most-powerful
presently-available computational resources, and detailed
extrapolation studies have recently been carried using the
conventional oscillator basis in such large spaces, specifically,
$\Nmax\leq16$, with the same interaction as used
here~\cite{cockrell2012:li-ncfc}.  These results provide estimates for
the true values of observables, against which the present
Coulomb-Sturmian calculations in smaller spaces can be compared.

\subsection{Energies}
\label{sec-calc-en}
\begin{figure*}
\begin{center}
\includegraphics*[width=\ifproofpre{0.86}{0.75}\hsize]{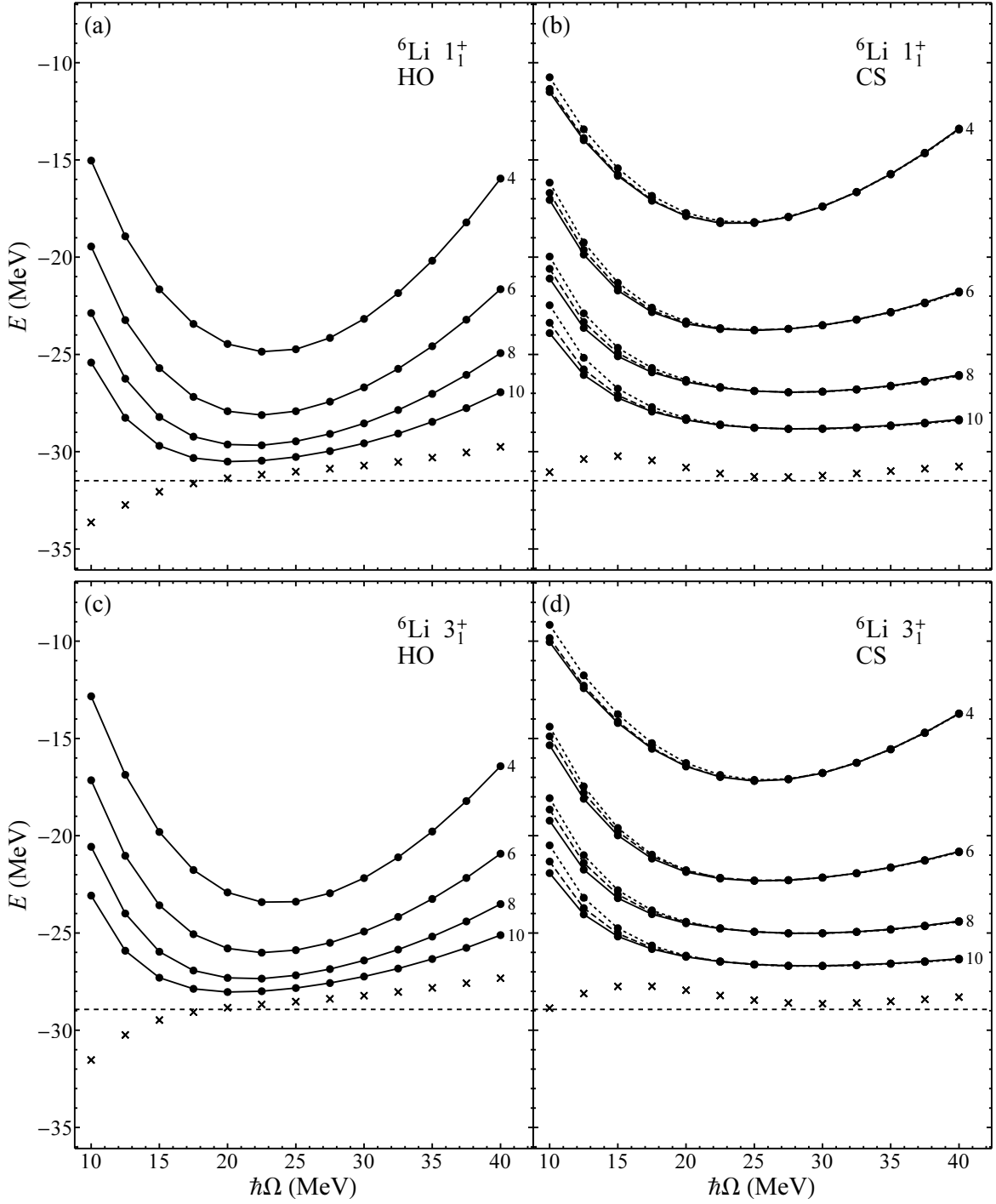}
\end{center}
\caption{The $\isotope[6]{Li}$ $1^+$ ground
state energy~(top) and $3^+$ excited
state energy~(bottom), calculated using the conventional harmonic
oscillator basis~(left) and the Coulomb-Sturmian basis~(right).  Calculated energies are
plotted as a function of the basis $\hbar
\Omega$ parameter, for $\Nmax=4$, $6$, $8$, and $10$ (successive
curves, as labeled).  For the Coulomb-Sturmian basis, calculations are
shown variously for truncations $\Ncut=9$ (dotted curves), $\Ncut=11$
(dashed curves), and $\Ncut=13$ (solid curves) in the change-of-basis
transformation of two-body matrix elements.  Exponentially
extrapolated values (based on the $\Ncut=13$ calculations in the case
of the Coulomb-Sturmian basis) are indicated by crosses ($\times$).
The best extrapolated values from the large-basis calculations of
Ref.~\cite{cockrell2012:li-ncfc} are shown as horizontal dashed lines.
}
\label{fig-conv-e}
\end{figure*}
\begin{figure}
\begin{center}
\includegraphics*[width=\ifproofpre{0.95}{0.5}\hsize]{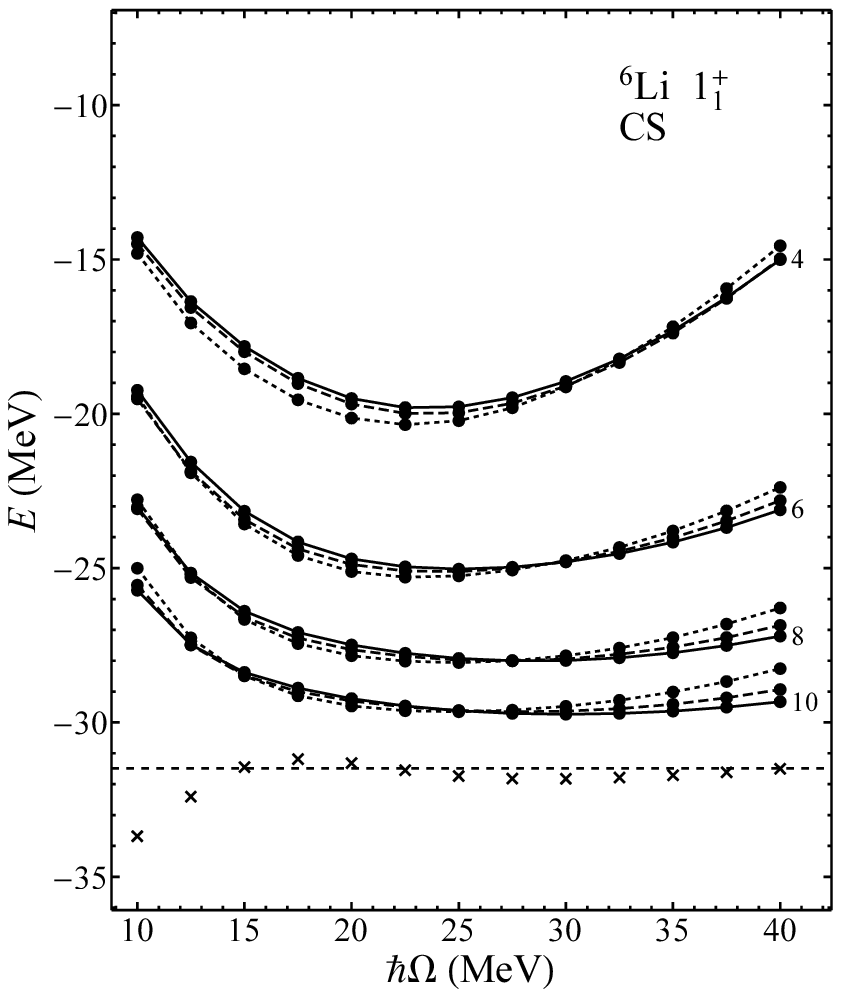}
\end{center}
\caption{The $\isotope[6]{Li}$ $1^+$ ground
state energy, calculated using the Coulomb-Sturmian basis, but
\textit{without} making use of the separable method of
Sec.~\ref{sec-sep} for the two-body
matrix
elements of the $\Trel$ operator, for comparison with
Fig.~\ref{fig-conv-e}(b).  That is, the entire Hamiltonian, including
$\Trel$, is transformed from the oscillator basis following the
approach of
Sec.~\ref{sec-xform}. See the caption to
Fig.~\ref{fig-conv-e} for further
explanation of curves and symbols.
}
\label{fig-conv-e-xform}
\end{figure}

We begin by comparing the ground state energy obtained in NCCI
calculations with the conventional oscillator basis and with the
Coulomb-Sturmian basis.  The calculated energies of the $1^+$ ground
state of $\isotope[6]{Li}$ are shown for the oscillator basis in
Fig.~\ref{fig-conv-e}(a) and for the Coulomb-Sturmian basis in
Fig.~\ref{fig-conv-e}(b).  In each case, the calculations span a range
of $\hbar\Omega$ values from $10\,\MeV$ to $40\,\MeV$ and are carried
out for $\Nmax=4$, $6$, $8$, and $10$.  These $\Nmax$ values
correspond to the highest to lowest curves, respectively, in the
figure.  The bare Hamiltonian has been used, without renormalization
to the finite space, so the variational principle is in effect, and
energies (for the lowest state with each set of conserved quantum
numbers) approach the full-space value monotonically from above with increasing
$\Nmax$.

The goal is not for any single NCCI calculation to actually \textit{reach} a
converged value, but rather to obtain the most reliable
\textit{extrapolation} from a series of NCCI calculations,
to the converged value which would be obtained in the full,
untruncated space for the many-body
problem~\cite{forssen2008:ncsm-sequences,maris2009:ncfc,coon2012:nscm-ho-regulator}.
It is thus first necessary to examine the dependence of the results on
the basis parameters $\hbar\Omega$ and $\Nmax$ as just
described.  Extrapolation schemes are still largely empirical in their
justification, and different prescriptions, varying in their details,
might be used.  However, for energies, at least, the basic procedure
explored in, \textit{e.g.},
Refs.~\cite{forssen2008:ncsm-sequences,bogner2008:ncsm-converg-2N,maris2009:ncfc},
consists of an exponential extrapolation.  The no-core full
configuration (NCFC) approach~\cite{maris2009:ncfc}, in particular,
is based on exponential extrapolations of results
of calculations obtained with an unrenormalized interaction
appropriate to the infinite, untruncated space, so that energies
approach the full-space values monotonically, as noted above. One first finds the
variational minimum with respect to $\hbar\Omega$, for the highest
available $\Nmax$-truncated space.  Then one extrapolates with respect
to $\Nmax$, at this $\hbar\Omega$, to the full-space result
($\Nmax\rightarrow\infty$) by assuming an exponential approach to the
asymptotic value $E_\infty$,
\begin{equation}
\label{eqn-exp}
E(\Nmax)=E_\infty+ae^{-c\Nmax},
\end{equation}
where $E_\infty$, $a$, and $c$ are taken as parameters.

As the baseline for comparison, the calculations of the ground state
energy with the oscillator basis are shown in
Fig.~\ref{fig-conv-e}(a).  The variational minimum with respect to
$\hbar\Omega$ occurs at $\sim20\,\MeV$, for $\Nmax=10$, moving
gradually lower with increasing $\Nmax$.  For each value of
$\hbar\Omega$ at which calculations have been carried out, an
exponential extrapolation of the $\Nmax=4$--$10$ calculations is shown
(indicated by a cross).  The best estimate of the ground state energy
from Ref.~\cite{cockrell2012:li-ncfc}, $E=-31.49(3)\,\MeV$, is
indicated by the dashed horizontal line.  The extrapolated values pass
through this estimate at $\hbar\Omega\approx20\,\MeV$, that is,
roughly the location of the variational minimum.

The calculations of the ground state energy with the Coulomb-Sturmian basis
are shown in Fig.~\ref{fig-conv-e}(b).  The variational minimum with respect to
$\hbar\Omega$ occurs at $\sim30\,\MeV$, for $\Nmax=10$, and moves
\textit{higher} with increasing $\Nmax$.  Notice that at each
$\Nmax$ the variational minimum energy obtained with the
Coulomb-Sturmian basis is substantially higher than that obtained with
the oscillator basis (by $\sim2\,\MeV$ for $\Nmax=10$).  However, the energies obtained with the
Coulomb-Sturmian basis are also falling significantly more rapidly
with increasing $\Nmax$.  (In general, a higher \textit{starting} energy
for the convergence, at low $\Nmax$, need not imply a lower
\textit{rate} of convergence.)  Therefore, let us compare the
exponential fit parameters [see~(\ref{eqn-exp})] near the variational
minimum.  For the oscillator basis at $\hbar\Omega=20\,\MeV$, the
convergence rate is $c\approx0.35$, with an extrapolated ground state
energy of $-31.3\,\MeV$.  For the Coulomb-Sturmian basis at
$\hbar\Omega=30\,\MeV$, the convergence rate is comparable, albeit
marginally lower, at $c\approx0.29$, with an extrapolated ground state
energy which is also comparable, at $-31.2\,\MeV$.  Interestingly, the
extrapolations for the Coulomb-Sturmian basis have a qualitatively
different dependence on $\hbar\Omega$ than those for the oscillator
basis.  Rather than varying monotonically (increasing with increasing
$\hbar\Omega$), they have a minimum, at an $\hbar\Omega$ approximately
equal to that of the variational minimum.

The one significant numerical approximation which is entailed in
setting up the Coulomb-Sturmian calculations, as discussed in
Sec.~\ref{sec-xform}, is in the transformation of the two-body matrix
elements of the nucleon-nucleon interaction from the oscillator basis
to the Coulomb-Sturmian basis.  The transformation is necessarily
carried out in a truncated oscillator basis.  It is therefore
imperative to establish the numerical stability of the results with
respect to the shell truncation $\Ncut$ in the sum over oscillator
states.  Calculations based on two-body matrix elements obtained with
$\Ncut=9$ ($10$ shells), $\Ncut=11$ ($12$ shells), and $\Ncut=13$
($14$ shells) are overlaid in Fig.~\ref{fig-conv-e}(b), as well as in
all subsequent plots of Coulomb-Sturmian calculations.  Calculations
of the ground state energy for $\hbar\Omega\gtrsim20\,\MeV$ are highly
stable with respect to this cutoff, in the present calculations.  This
range safely covers the variational minimum.  However, the
calculations are not stable with respect to this cutoff for
$\hbar\Omega\lesssim20\,\MeV$, and higher cutoffs would therefore be
required for accurate results at these $\hbar\Omega$ values.  The
instability with respect to oscillator basis cutoff appears to
increase with increasing $\Nmax$.  Such a dependence is reasonable,
since higher-$\Nmax$ calculations increasingly probe higher-$n$
Coulomb-Sturmian single-particle basis functions, which in turn
require a higher $\Ncut$ for accurate expansion in an oscillator
basis, as illustrated in Fig.~\ref{fig-overlaps}.

For the calculations shown in Fig.~\ref{fig-conv-e}, the kinetic energy matrix
elements have been calculated by the separable
method of Sec.~\ref{sec-sep}.  It is interesting at this point to
investigate how essential it is to use the separable approach, rather
than simply transforming the kinetic energy matrix elements from the
oscillator basis.  For comparison, we therefore repeat the
calculations for the ground state energy in the Coulomb-Sturmian
basis, but transforming the two-body matrix elements of the entire
Hamiltonian from the oscillator basis, yielding the results shown in
Fig.~\ref{fig-conv-e-xform}.  It is seen that, without the separable
calculation, the results are unstable with respect to $\Ncut$
throughout the entire range of $\hbar\Omega$ values, including the
vicinity of the variational minimum.  Thus, the separable method plays
a major role in obtaining numerically accurate calculations.  It would
otherwise be necessary to start from oscillator two-body matrix
elements in a significantly larger number of oscillator shells,
possibly prohibitively so.

Calculations for the energy of the $3^+$ first excited state for the
oscillator basis are shown in Fig.~\ref{fig-conv-e}(c) and for the
Coulomb-Sturmian basis in Fig.~\ref{fig-conv-e}(d).  The results are
very similar in nature to those for the ground state, so little
additional discussion is required.  The best extrapolation from Ref.~\cite{cockrell2012:li-ncfc} places
this state at $2.56(2)\,\MeV$ excitation energy, corresponding to
$E\approx-28.93\,\MeV$.  For the oscillator basis at
$\hbar\Omega=20\,\MeV$, the convergence rate is $c\approx0.34$, with
an extrapolated ground state energy of $-28.8\,\MeV$.  For the
Coulomb-Sturmian basis at $\hbar\Omega=30\,\MeV$, the convergence rate
is again marginally lower, at $c\approx0.30$, with an extrapolated
energy of $-28.6\,\MeV$, apparently erring on the high side relative
to Ref.~\cite{cockrell2012:li-ncfc}.

From these exploratory calculations for $\isotope[6]{Li}$, it
would appear that convergence properties for energies with the
Coulomb-Sturmian basis are comparable, \textit{i.e.}, not
markedly inferior, to those of the
oscillator basis, with some qualitative differences in the
$\hbar\Omega$ dependence.  We note that these exploratory results have
not yet probed the variational freedoms available with the
Coulomb-Sturmian basis, both in the choice of length parameters
(Sec.~\ref{sec-cs-scale}) and in truncation schemes, as described above.  The convergence rate
alone does not provide conclusive information on the robustness which
can be expected from large-$\Nmax$ extrapolation or on the best
extrapolation procedure.  Some questions regarding extrapolation may be elucidated by extending the calculations
to higher $\Nmax$.  Furthermore, the rates of convergence of
calculations with the oscillator and Coulomb-Sturmain bases will depend on the physical
properties of the nucleus (and particular state) under consideration.
For instance, the asymptotic properties of the single-particle basis
may well play a larger role for halo nuclei or for states involving
clusters with significant spatial separation.

\subsection{Root-mean-square radius}
\label{sec-calc-r}
\begin{figure*}
\begin{center}
\includegraphics*[width=\ifproofpre{0.86}{0.75}\hsize]{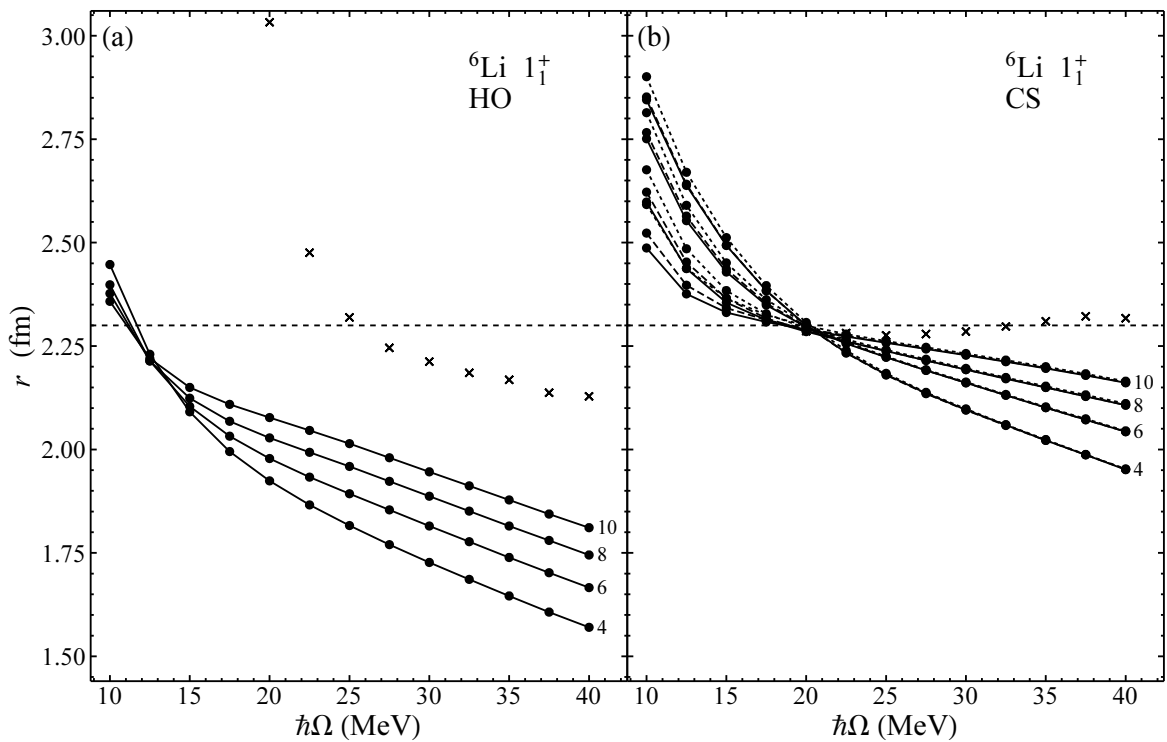}
\end{center}
\caption{
The $\isotope[6]{Li}$ $1^+$ ground state RMS
radius, calculated using the conventional harmonic oscillator
basis~(left) and the Coulomb-Sturmian basis~(right).  Calculated
energies are plotted as a function of the basis $\hbar
\Omega$ parameter, for $\Nmax=4$, $6$, $8$, and $10$ (successive
curves, as labeled).  For the Coulomb-Sturmian basis, calculations
are shown variously for truncations $\Ncut=9$ (dotted curves), $\Ncut=11$ (dashed curves), and $\Ncut=13$ (solid curves) in the change-of-basis
transformation of two-body matrix elements.  Exponentially
extrapolated values (based on the $\Ncut=13$ calculations in the case
of the Coulomb-Sturmian basis) are indicated by crosses ($\times$).  The best
estimated value from the large-basis calculations of
Ref.~\cite{cockrell2012:li-ncfc} is shown as a horizontal dashed line.
}
\label{fig-conv-r}
\end{figure*}

The root-mean-square radius presents challenges for convergence
in NCCI calculations with the conventional oscillator
basis~\cite{bogner2008:ncsm-converg-2N}.  Here we consider the
intrinsic, point-nucleon RMS radius for the ground state, defined by
$\sqrt{\tbracket{\rrel^2}}$ (see Appendix~\ref{app-r2k2}), from which
the center-of-mass contribution has been removed by construction.
Evaluation of the expectation value $\tbracket{\rrel^2}$ in a
many-body state requires that one first calculate the two-body matrix
elements of the $\rrel^2$ operator.  These are obtained for the
Coulomb-Sturmian basis by the separable method of Sec.~\ref{sec-sep}.

The oscillator basis results for the RMS radius in
Fig.~\ref{fig-conv-r}(a) are shown for the same range of calculations
($\Nmax=4$, $6$, $8$, and $10$, with $\hbar\Omega$ values from
$10\,\MeV$ to $40\,\MeV$) as for the energies in
Sec.~\ref{sec-calc-r}.  (The curves proceed from greatest to least slope with increasing $\Nmax$ in the figure.)  Exponential
extrapolations to infinite $\Nmax$ are shown as well.  The
extrapolated values vary strongly with $\hbar\Omega$ and converge very
slowly with $\Nmax$.  For instance, taking $\hbar\Omega$ at the
variational minimum for the energy, \textit{i.e.},
$\hbar\Omega\approx20$, the exponential convergence rate for the RMS
radius with respect to $\Nmax$ is only $c\approx0.024$, and the
extrapolated radius lies $\sim1\,\fm$ above the calculated values.
Alternatively, the value at the crossover point of the curves obtained
for different $\Nmax$ has also been proposed as an estimate of the
full-space value~\cite{bogner2008:ncsm-converg-2N}.  This crossover occurs at $\hbar\Omega\approx12\,\MeV$ in the
present calculations and lies in the vicinity of $2.2\,\fm$.  The best estimate from
Ref.~\cite{cockrell2012:li-ncfc}, similarly obtained from the crossover point, for
calculations with $\Nmax\leq16$, is $\sim2.3\,\fm$, indicated by the
dashed horizontal line in Fig.~\ref{fig-conv-r}.

Examining the calculations for the RMS radius using the
Coulomb-Sturmian basis, as shown in Fig.~\ref{fig-conv-r}(b), the
gross features are similar.  The crossover point for the curves
obtained with $\Nmax$ lies at $\hbar\Omega\approx20$.  The value of
$\sim2.3\,\fm$ is consistent with the estimate of
Ref.~\cite{cockrell2012:li-ncfc} and $\sim0.1\,\fm$ higher than the
crossover for the curves obtained with the oscillator basis, for the
same $\Nmax$, in Fig.~\ref{fig-conv-r}(a).  Moreover, it is seen that
exponential extrapolation may be a viable approach to estimating the
full-space value for the radius.  The extrapolated values obtained for
$\hbar\Omega\gtrsim20\,\MeV$, \textit{i.e.}, above the crossover
point, are reasonably insensitive to $\hbar\Omega$ and are consistent
with the best estimate from Ref.~\cite{cockrell2012:li-ncfc}.  For
instance, taking $\hbar\Omega\approx30$, \textit{i.e.}, at the
variational minimum, the exponential convergence rate for the RMS
radius is $c\approx0.19$, and the extrapolated radius is
$\sim2.28\,\fm$.  Results are stable with respect to the shell cutoff
in the transformation of matrix elements from the oscillator basis,
for $\hbar\Omega\gtrsim20\,\MeV$, as observed above for the energies.

It would thus appear that the rate of convergence of the RMS radius
obtained with the Coulomb-Sturmian basis is superior to that obtained
with the conventional oscillator basis.  However, further systematic
investigation is required, especially into the stability of
extrapolations with increasing $\Nmax$, before general conclusions may
be drawn.

\subsection{Center-of-mass dynamics}
\label{sec-calc-cm}
\begin{figure}
\begin{center}
\includegraphics*[width=\ifproofpre{0.95}{0.75}\hsize]{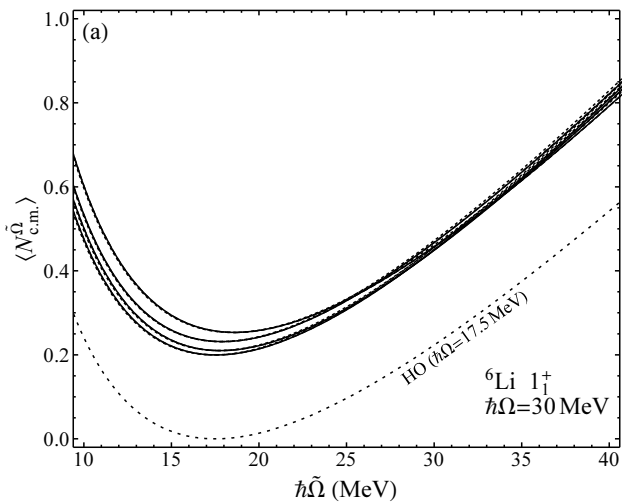}
\end{center}
\caption{Expectation value of the number operator $\Ncm[\Omegat]$ for
center-of-mass oscillator quanta, as a function of oscillator energy
$\hbar\Omegat$.  These calculations are
for the $\isotope[6]{Li}$ $1^+$ ground
state, using the Coulomb-Sturmian basis, 
with $\hbar\Omega=30\,\mathrm{MeV}$.
Calculations
are shown for $\Nmax=4$, $6$, $8$, and $10$ (successive
curves, top to bottom), and for truncations $\Ncut=9$ (dotted curves), $\Ncut=11$
(dashed curves), and $\Ncut=13$ (solid curves) in the change-of-basis
transformation of two-body matrix elements.  The analogous curve
expected for a pure harmonic oscillator $0s$ function, with
$\hbar\Omega=17.5\,\MeV$, is also
shown for comparison (dotted curve, labeled).
}
\label{fig-Ncm-distrib}
\end{figure}
\begin{figure*}
\begin{center}
\includegraphics*[width=\ifproofpre{0.75}{0.75}\hsize]{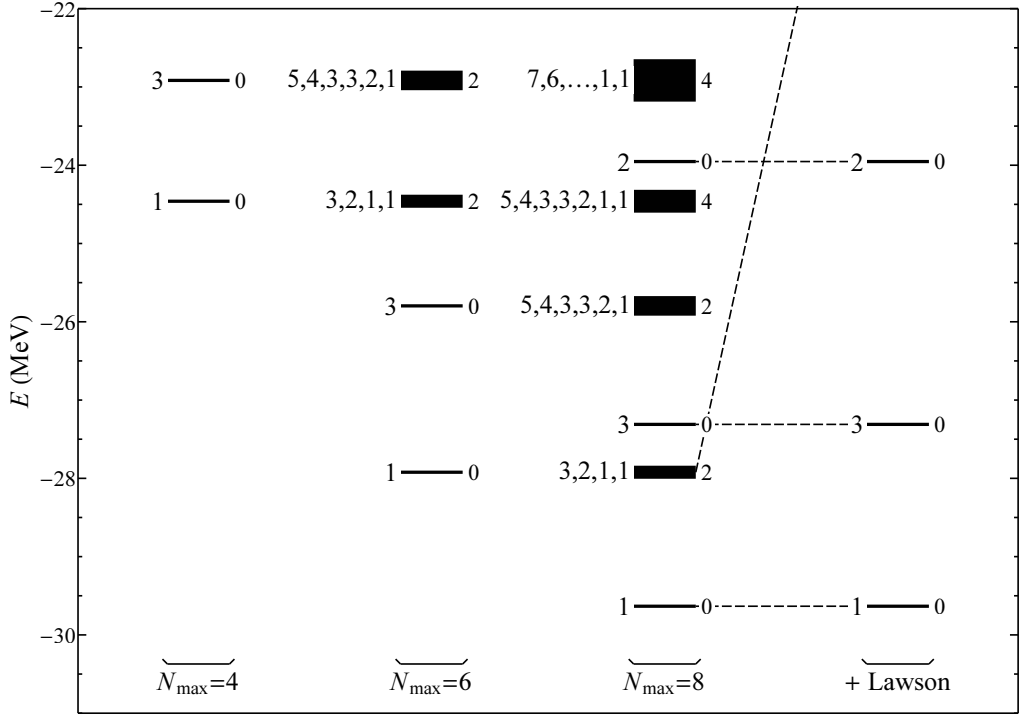}
\end{center}
\caption{Level spectrum for $\isotope[6]{Li}$, including spurious
states, calculated using the
conventional harmonic oscillator basis.  Calculations (left to right)
are for $\Nmax=4$,
$6$, and $8$, and then shown again for $\Nmax=8$ with
addition of a Lawson term, of sufficient strength to shift the spurious
states above the energy range displayed in this plot.  For each level, the
angular momentum $J$ is indicated at left, and $\tbracket{\Ncm}$ is
indicated at right.  For degenerate multiplets of spurious states, the
thickness of the line is proportional to the number of states.   These calculation are for
$\hbar\Omega=20\,\mathrm{MeV}$. 
 }
\label{fig-ho-scheme}
\end{figure*}
\begin{figure*}
\begin{center}
\includegraphics*[width=\ifproofpre{0.75}{0.75}\hsize]{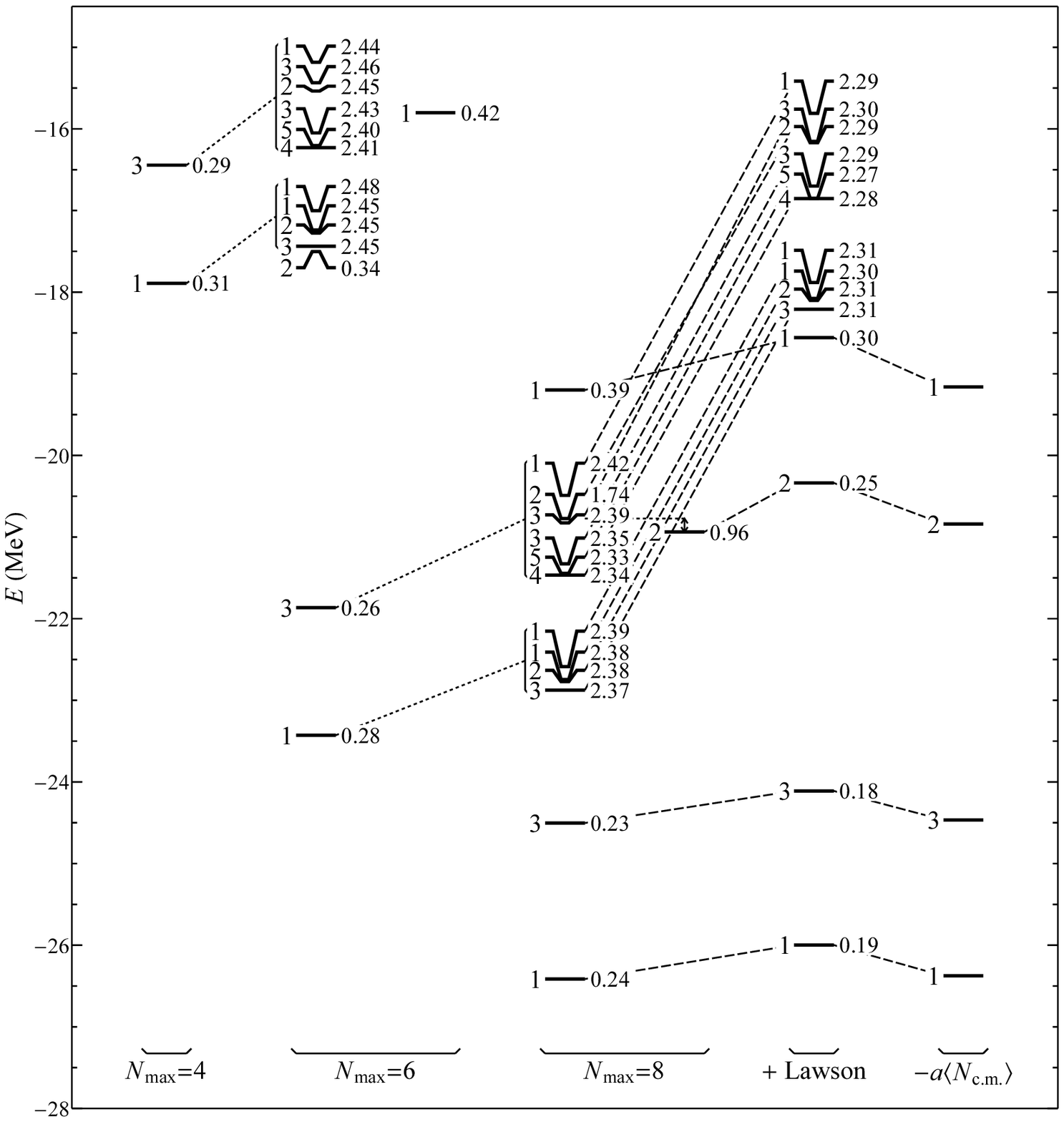}
\end{center}
\caption{Level spectrum for $\isotope[6]{Li}$, calculated using the
Coulomb-Sturmian basis.  Calculations (left to right) are for $\Nmax=4$, $6$, and $8$, and then again for $\Nmax=8$ with addition of a
Lawson term $a\Ncm[\OmegaL]$ of strength $a=2\,\mathrm{MeV}$.  Energies corrected by
$-a\tbracket{\Ncm[\OmegaL]}$ are shown at far right.  For each level, the
angular momentum $J$ is indicated at left, and $\tbracket{\Ncm[\Omegat]}$ is
indicated at right, as a measure of the number
of center-of-mass oscillator quanta.  Approximately degenerate multiplets of spurious
states are marked by brackets and connected to the state at lower
$\Nmax$ to which they are approximately related by
coupling to two center-of-mass quanta.  The dashed lines trace the
change in level energy induced by the Lawson term.  For $\Nmax=8$, an
arrow connects the two $J=2$ levels which may be described (see text) as
admixtures of a nonspurious and spurious level.  These calculation are
for $\hbar\Omega=20\,\mathrm{MeV}$, with $\Ncut=13$.  The quantity
$\tbracket{\Ncm[\Omegat]}$ is evaluated for
$\hbar\Omegat=20\,\mathrm{MeV}$, and the Lawson term is defined for
$\hbar\OmegaL=20\,\mathrm{MeV}$ as well. }
\label{fig-cs-scheme}
\end{figure*}
\begin{figure}
\begin{center}
\includegraphics*[width=\ifproofpre{0.95}{0.50}\hsize]{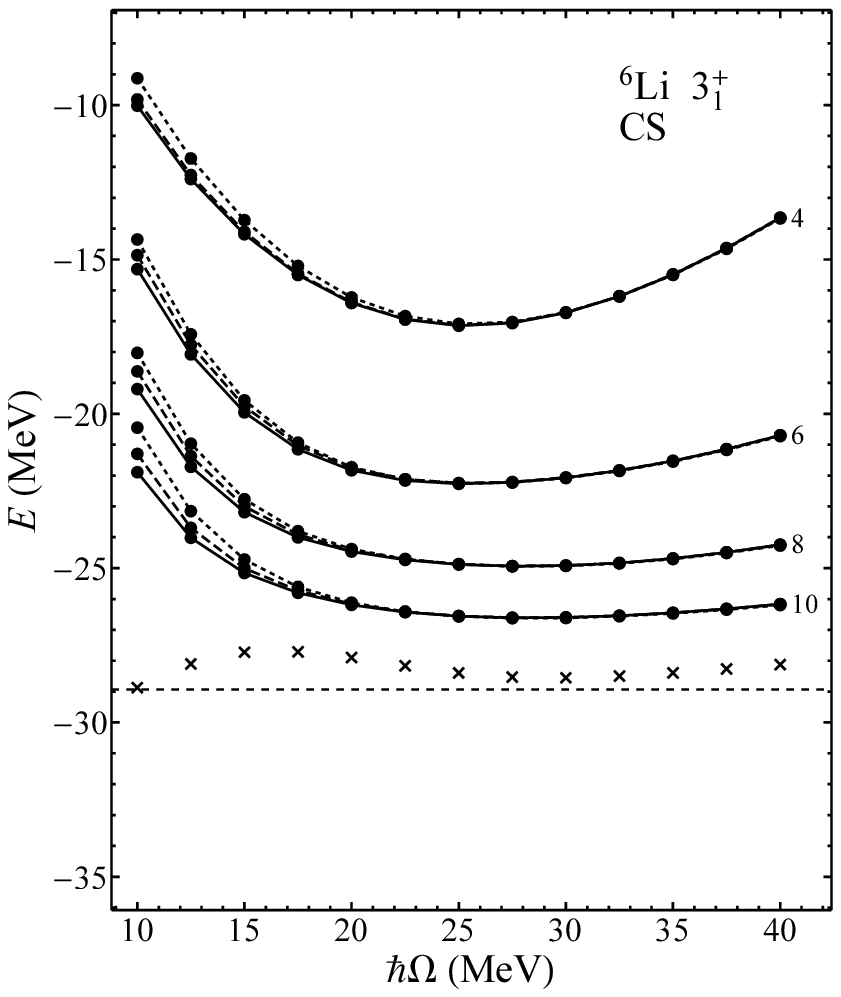}
\end{center}
\caption{The $\isotope[6]{Li}$ $3^+$ excited
state energy, calculated using the Coulomb-Sturmian basis, as in
Fig.~\ref{fig-conv-e}(d), but now including a Lawson term, with
strength $a=2\,\mathrm{MeV}$ and Lawson term oscillator energy
$\hbar\Omega_L$ chosen equal to the basis $\hbar\Omega$.
Energies are corrected by subtracting $a\tbracket{\Ncm[\Omega_L]}$.
See the caption to Fig.~\ref{fig-conv-e} for further explanation of
curves and symbols.  }
\label{fig-conv-e-lawson}
\end{figure}

We now focus on the dominant concern in using any basis other than the
harmonic oscillator basis with $\Nmax$ truncation for nuclear
many-body calculations, namely, incomplete separation of
center-of-mass and intrinsic dynamics.  There are several aspects to
consider: the natural degree of separation arising in calculations
using the Coulomb-Sturmian basis, the spurious state spectrum obtained
in such calculations, and the extent to which a Lawson term can be
used to influence spurious excitations.

The problem of correcting for, or eliminating, spurious contributions
for calculations with a general truncated basis is
unresolved~\cite{lipkin1958:com-shell}.  Nonetheless, it is still
possible that factorized wave functions might approximately be
obtained in a truncated space.  In the full space, factorization is
obtained due to
the separable Hamiltonian, albeit with degeneracies in the 
center-of-mass wave functions multiplying each intrinsic state (Sec.~\ref{sec-hamil}).
Therefore, as larger truncated spaces are taken,
approaching this full space, the structure of the eigenstates may be
expected to converge towards such factorized structure.  For instance, a high degree of factorization has been
reported for coupled-cluster calculations in light
nuclei~\cite{hagen2009:coupled-cluster-com-COMBO}.  Furthermore, introducing a
Lawson term to the Hamiltonian, as in~(\ref{eqn-HLawson}), may serve
to ``purify'' the eigenstates so that the motion more closely
approximates $0s$ center-of-mass motion, as proposed by Gloeckner and
Lawson~\cite{gloeckner1974:spurious-com}.  This Lawson term also
pushes eigenstates dominated by other center-of-mass excitations
higher in the spectrum.  However, caution must be exercised in such
use of the Lawson term, since any improved (or, at least, more
oscillator-like) description of center-of-mass motion may be obtained
at the expense of the quality with which the intrinsic wave function
is approximated~\cite{mcgrory1975:spurious-com}.

A first indication of the degree of separation in the many-body
eigenstate is provided by the expectation
value of the $\Ncm$ operator.  This operator is defined, for an arbitrary center-of-mass harmonic oscillator
energy $\hbar\Omegat$, by
\begin{equation}
\label{eqn-ncm}
\Ncm[\Omegat]
\equiv
\frac{1}{\hbar\Omegat}\biggl(
\frac{P^2}{2Am_N}+\frac{Am_N\Omegat^2R^2}{2}-\frac{3\hbar\Omegat}{2}
\biggr),
\end{equation}
where the tilde serves to distinguish $\hbar\Omegat$ from the basis
$\hbar\Omega$ parameter.  As noted by Hagen~\textit{et
al.}~\cite{hagen2009:coupled-cluster-com-COMBO},
\textit{if} separation occurs, as
$\wfgen(\vec{r}_i;\vec{\sigma}_i)=\wfgencm(\vec{R})\wfgenin(\vec{r}_{ij};\vec{\sigma}_i)$,
and \textit{if} $\wfgencm(\vec{R})$ happens to be an oscillator $0s$
function, corresponding to some oscillator energy $\hbar\Omegat$, then
the many-body wave function will have $\tbracket{\Ncm[\Omegat]}=0$.
Evaluation of the expectation value $\tbracket{\Ncm[\Omegat]}$
requires that one first calculate the two-body matrix elements of
$P^2$ and $R^2$, and thence of $\Ncm[\Omegat]$.  These are readily
obtained for the Coulomb-Sturmian basis by the separable method of
Sec.~\ref{sec-sep}, so evaluation is straightforward.

The expectation value $\tbracket{\Ncm[\Omegat]}$ is shown as a
function of $\hbar\Omegat$ for the $\isotope[6]{Li}$ $1^+$ ground
state in Fig.~\ref{fig-Ncm-distrib}, for the Coulomb-Sturmian basis
calculation with basis $\hbar\Omega=30\,\mathrm{MeV}$ and no Lawson term.  The minimum
value of $\tbracket{\Ncm[\Omegat]}$ is obtained at
$\hbar\Omegat\approx17.5\,\MeV$, shifting gradually towards lower
$\hbar\Omegat$, which corresponds to larger center-of-mass oscillator
length $\bcm=[\hbar/(Am_N\Omega)]^{1/2}$, with increasing $\Nmax$.
(The location of the minimum also depends modestly upon the choice of
basis $\hbar\Omega$ for the calculation, increasing with
$\hbar\Omega$.)  The minimum value of $\tbracket{\Ncm[\Omegat]}$
decreases with increasing $\Nmax$, but it appears to be converging
towards a \textit{nonzero} value of $\sim0.2$.  The fact that
$\tbracket{\Ncm[\Omegat]}$ values significantly less than unity are
obtained in the calculations indicates that a $0s$ oscillator function
dominates the center-of-mass motion, and that an approximate
separation of center-of-mass and intrinsic functions is spontaneously
arising.  However, the nonzero limit indicates that, as the full space
is approached, the separated center-of-mass function is \textit{not}
strictly taking the form of a $0s$ oscillator function.  For
comparison, the dependence of $\tbracket{\Ncm[\Omegat]}$ on
$\hbar\Omegat$ which would be obtained for a pure oscillator $0s$
function with $\hbar\Omega=17.5\,\MeV$, given by
$\tbracket{\Ncm[\Omegat]}=\tfrac34({\Omega}/{\Omegat}+{\Omegat}/{\Omega} -2)$,
is also shown in
Fig.~\ref{fig-Ncm-distrib}.

In interpreting these results, it must be stressed that calculating
$\tbracket{\Ncm[\Omegat]}$ for an eigenstate provides only a lower limit on the degree of factorization.  That is, a nonzero
$\tbracket{\Ncm[\Omegat]}$ does not preclude factorization but can
simply indicate that the factorized center-of-mass wave function is
not of $0s$ oscillator type.  Extracting the true degree of factorization
is more challenging.  To do so likely requires some form of explicit
transformation to center-of-mass and relative coordinates.  For
instance, an expansion $\wfgen=\sum_i s_i
\wfgencm^{(i)}\wfgenin^{(i)}$ may then be obtained through a singular
value decomposition as proposed in
Ref.~\cite{hagen2009:coupled-cluster-com-COMBO}.

Since factorization arises in the full space, the effects of
convergence of the center-of-mass dynamics were already implicitly
included in the extrapolations to the full-space values of the
observables of interest, as explored in Secs.~\ref{sec-calc-en}
and~\ref{sec-calc-r}.  However, for this extrapolation to be possible,
it is necessary that states involving spurious excitations of the
center-of-mass function can be disentangled and removed from the
low-lying spectrum.  This becomes an increasing concern with
increasing $\Nmax$, as we shall now see from examining the spurious
state spectrum.

It is helpful to first consider the eigenvalue spectrum, including
spurious states, obtained in calculations with an $\Nmax$-truncated
oscillator basis.  This is illustrated for $\isotope[6]{Li}$ in
Fig.~\ref{fig-ho-scheme}, for $\Nmax=4$, $6$, and $8$, with
$\hbar\Omega=20\,\MeV$.  The eigenvalue of $\Ncm$ is indicated to the
right of each level in this figure.  For instance, at $\Nmax=4$, where
the two lowest states in the spectrum are shown, these states have
$\Ncm=0$ and are thus nonspurious, corresponding to the intrinsic 
$1^+$ ground state and $3^+$ first excited state, with $0s$
center-of-mass motion.  

Let us examine the
evolution of the spectrum in Fig.~\ref{fig-ho-scheme} with increasing
$\Nmax$, bearing in mind the direct sum
structure~(\ref{eqn-space-sum}) of the $\Nmax$-truncated space.
Moving from $\Nmax=4$ to $\Nmax=6$, the two additional oscillator
quanta introduced to the system may go towards converging the
intrinsic states.  This yields the new $1^+$ ground state
and $3^+$ excited state at lower energies in the $\Nmax=6$
calculation (in the
$\sHcm[0]\otimes\sHin[6]$ subspace).  Alternatively, the two
additional quanta may go into center-of-mass excitation, yielding
spurious states (in the $\sHcm[2]\otimes\sHin[4]$ subspace).  Since the center-of-mass
excitation gives no contribution to the energy, under the
\textit{intrinsic} Hamiltonian we are using, the resulting spurious
states are degenerate with the nonspurious states obtained at $\Nmax=4$
(in the $\sHcm[0]\otimes\sHin[4]$ subspace).  Then, moving to $\Nmax=8$,
new $\Ncm=2$ spurious states appear degenerate with the $\Nmax=6$
nonspurious states, new $\Ncm=4$ spurious states appear degenerate
with the $\Ncm=2$ spurious states from $\Nmax=6$, \textit{etc.}

To ascertain the angular
momenta expected for the spurious states, we note that angular momentum eigenstates of the full eigenproblem
are obtained from those of the intrinsic eigenproblem via angular
momentum coupling as
$\wfgen^{(J)}=[\wfgencm^{(\lcm)}\times\wfgenin^{(\Jin)}]^{(J)}$.
Thus, the angular momenta expected for the spurious states follow by
the triangle inequality for addition of the center-of-mass angular
momentum $\lcm$ and intrinsic angular momentum $\Jin$. 
Recall that the three-dimensional
oscillator spectrum contains angular momenta $l=0$ for $N=0$, $l=1$ for $N=1$,
$l=(0,2)$ for $N=2$, $l=(1,3)$ for $N=3$,  $l=(0,2,4)$ for $N=4$,
\textit{etc.}   
Spurious states with $\Ncm=1$ lie in the
opposite-parity space and therefore do not appear in Fig.~\ref{fig-ho-scheme}.\footnote{Odd spurious exitations of odd-parity intrinsic states,
\textit{e.g.}, in the $\sHcm[1]\otimes\sHin[5]$ subspace, do indeed
appear in the even-parity spectrum, but in $\isotope[6]{Li}$ these are
at higher energy.}  However,
for $\Ncm=2$, coupling $\lcm=0$ and $2$ to the $\Jin=1$ intrinsic
ground state yields a spurious-state multiplet with angular momenta
$(3,2,1,1)$, as seen in Fig.~\ref{fig-ho-scheme}.  Similarly, coupling
these values of $\lcm$ to the $\Jin=3$ intrinsic excited state yields
a spurious-state multiplet with angular momenta $(5,4,3,3,2,1)$.  The
$\Ncm=4$ spurious multiplets seen at $\Nmax=8$ in
Fig.~\ref{fig-ho-scheme} are obtained similarly by coupling $\lcm=0$,
$2$, and $4$ to the intrinsic state.

It is apparent from Fig.~\ref{fig-ho-scheme} that a high level of contamination
of the low-lying spectrum with spurious states will arise with increasing
$\Nmax$, as the difference in energy between intrinsic ground states
in successive $\Nmax$ spaces decreases.  Already several spurious
states arise below the first excited state at $\Nmax=8$ in this
example.  This would become a prohibitive problem for large $\Nmax$,
as suggested in Sec.~\ref{sec-nmax},
since the Lanczos diagonalization in the many-body problem must
converge the spurious states along with the nonspurious states.
However, for the $\Nmax$-truncated oscillator basis, as noted in
Sec.~\ref{sec-nmax}, inclusion of the Lawson term in the Hamiltonian
pushes the spurious solutions to higher energy, without affecting the
nonspurious states, obviating the problem
[Fig.~\ref{fig-ho-scheme}~(far right)].

With this understanding of the spurious state spectrum for the
oscillator basis, we now have a baseline for interpreting the
eigenvalue spectrum obtained with a Coulomb-Sturmian basis, shown in
Fig.~\ref{fig-cs-scheme} for basis $\hbar\Omega=20\,\MeV$.  It is seen
that the same multiplets of spurious states (marked with brackets in
the figure) arise as in the calculation based on the oscillator basis,
but now the degeneracies~--- with the nonspurious state at lower
$\Nmax$ and between the members of the multiplet itself~--- are only
approximate.  Since we are using the intrinsic Hamiltonian, these
energy differences do not arise from any direct contribution of the
center-of-mass dynamics to the energy.  Rather, to the extent that
factorization occurs, these differences arise from variation in the level of
convergence of the intrinsic wave function associated with the
center-of-mass wave function.  Alternatively, to the extent that
factorization is imperfect, these differences can arise from admixtures of
contributions involving different center-of-mass and intrinsic
excitations.

Although these states in Fig.~\ref{fig-cs-scheme} are not eigenstates
of $\Ncm[\Omegat]$, we can still calculate an average number of
center-of-mass oscillator quanta as $\tbracket{\Ncm[\Omegat]}$.  This
expectation is indicated to the right of each level in
Fig.~\ref{fig-cs-scheme}, where $\hbar\Omegat$ has been chosen simply
equal to the basis $\hbar\Omega$, \textit{i.e.},
$\hbar\Omegat=20\,\MeV$.  It is seen that the
$\tbracket{\Ncm[\Omegat]}$ values clearly reflect the identification
of the states as spurious or nonspurious according to the energy
spectrum noted above.  The nonspurious states share a similar range of
values for $\tbracket{\Ncm[\Omegat]}$~--- at $\Nmax=8$, nearly
identical for the ground state and first excited state ($\sim0.24$)
and somewhat higher ($\sim0.3$--$0.4$) for some of the higher states.
The states analogous to the $\Ncm=2$ spurious states of the
oscillator-basis calculation, in contrast, have
$\tbracket{\Ncm[\Omegat]}$ values which cluster closely around $2.4$.

Note the two $2^+$ states at about $-21\,\MeV$ in the $\Nmax=8$
calculation of Fig.~\ref{fig-cs-scheme}.  With exact factorization,
one of these would be nonspurious and the other spurious.  However,
the $\tbracket{\Ncm[\Omegat]}$ values for these two states ($\sim0.96$
and $\sim1.74$) indicate that the spurious and nonspurious states are
strongly mixed.  This mixing provide an illustration of the challenge
associated with contamination of the low-lying spectrum with spurious
states.  As the density of spurious states increases with $\Nmax$, the
close proximity of spurious and nonspurious states may be expected to
lead to extensive mixing and consequently a breakdown of center-of-mass
factorization for even the lowest-lying states.
Therefore, it is even more important that the spurious states be
eliminated from the low-lying spectrum than it is for conventional
oscillator basis calculations.  

With this in mind, we explore the efficacy of the Lawson term when
used with the Coulomb-Sturmian basis.  At right in
Fig.~\ref{fig-cs-scheme}, the effect of introducing a Lawson term
$a\Ncm[\OmegaL]$ to the Hamiltonian is shown.  For simplicity in this illustration, we
choose $\hbar\OmegaL=20\,\MeV$, corresponding to the basis
$\hbar\Omega$ and the $\hbar\Omegat$ for the
$\tbracket{\Ncm[\Omegat]}$ values indicated.  This choice is arbitrary,\footnote{When used with the
oscillator basis, the
center-of-mass oscillator energy $\hbar\OmegaL$ in the Lawson operator is
generally chosen equal to the basis $\hbar\Omega$, to preserve factorization.  However, when used with a
general, non-oscillator basis, there is no such requisite
pairing, and $\hbar\OmegaL$ may be chosen freely, so as to obtain the
most effective removal of spurious dynamics.} and
another value, such as that at which
$\tbracket{\Ncm[\OmegaL]}$ is minimized for the ground state, might
well be profitably used.   A Lawson strength $a=2\,\MeV$ has been
adopted, as sufficiently large to expunge spurious states from the lowest few
$\MeV$ of the spectrum, but not so large as to place undue weight on
coercing the
center-of-mass wave function into a pure $0s$ oscillator state, at the
possible expense of compromising convergence of the intrinsic state.
The change of energies with introduction of the Lawson term is traced
by dashed lines in Fig.~\ref{fig-cs-scheme}.  Notice that the mixing
of the nonspurious and spurious $2^+$ states discussed above (now
at energies of about $-20\,\MeV$ and $-16\,\MeV$, respectively) has been eliminated.

The Lawson term is also seen, from the expectation values indicated in
Fig.~\ref{fig-cs-scheme}, to reduce $\tbracket{\Ncm[\OmegaL]}$ for
these states.  It is not yet clear how much of this change reflects
improvement of the center-of-mass factorization and how much simply
relects modification of an already-factorized center-of-mass function
towards oscillator $0s$ form.

Since even the nonspurious states have nonzero values for
$\tbracket{\Ncm[\OmegaL]}$, their energies are raised by introduction
of the Lawson term, by $\sim a
\tbracket{\Ncm[\OmegaL]}$.  This contribution is \textit{not} expected
to vanish in the large $\Nmax$ limit, since $\tbracket{\Ncm[\OmegaL]}$
has already been seen not to approach zero. 
To recover the eigenvalue of
the \textit{intrinsic} Hamiltonian on the \textit{intrinsic} wave function, to the extent that good factorization is obtained, we must
correct the calculated energy for the contribution of the Lawson term acting
on the center-of-mass function, by subtracting $a
\tbracket{\Ncm[\OmegaL]}$ back off.
The energies obtained after this correction, for the nonspurious states,
are shown at far right in Fig.~\ref{fig-cs-scheme}.  After correction,
the original
values for the energies, as obtained before introduction of the Lawson term, are almost (but not quite) recovered.
The corrected energies are still marginally higher, likely reflecting
the compromise in convergence of the
intrinsic state incurred by the Lawson term, and this discrepancy
increases with the Lawson strength $a$ used in the calculation.

The Lawson term thus appears to be a credible means of eliminating spurious states from the
low-lying spectrum, in calculations with the Coulomb-Sturmian basis.  The
essential question, if the Lawson term is to be used in practice, is whether or not
the Lawson term has any significant adverse impact on convergence
properties.  Taking
the energy of the first excited state as an example, we repeat the calculations of
Fig.~\ref{fig-conv-e}(d), but now including a Lawson term of
strength $a=2\,\MeV$, resulting in the energies in Fig.~\ref{fig-conv-e-lawson}.  The $a
\tbracket{\Ncm[\OmegaL]}$ correction to the energies, described above, has
been included.  The results are virtually indistinguishable from those
of Fig.~\ref{fig-conv-e}(d).  For comparison with the discussion in
Sec.~\ref{sec-calc-en}, we note that the convergence rate at the
variational minimum ($\hbar\Omega=30\,\MeV$) is still $c\approx0.30$,
and the extrapolated energy is still approximately $-28.6\,\MeV$.

\section{Conclusion}
\label{sec-concl}

Although the conventional oscillator basis has definite advantages for
\textit{ab initio} nuclear many-body calculations with the NCCI approach,
namely, the potential for exact center-of-mass factorization of
eigenstates and the simplicity of the Moshinsky transformation for
Hamiltonian matrix elements, it also presents the disadvantage of
nonphysical Gaussian asymptotics at large distances, \textit{i.e.},
the oscillator wave functions satisfy the wrong boundary conditions at
infinity for use with bound states of nuclei.  The Coulomb-Sturmian
functions retain the advantages of forming a complete, discrete set of
square-integrable functions while also exhibiting realistic
exponential asymptotics.  We have seen that the technical and physical
challenges of carrying out NCCI calculations with a Coulomb-Sturmian
basis are tractable.  To briefly summarize the computational
framework, the many-body calculation has the standard structure for an
$nlj$ single-particle basis, the interaction matrix elements are
transformed from the harmonic-oscillator basis, and relative kinetic
energy matrix elements are calculated separably.  In the initial
exploratory calculations considered here, it is found that the
convergence rates for energies are competitive with those obtained
with an oscillator basis, the convergence rate for the RMS radius is
superior, and spurious center-of-mass excitations can be successfully
managed.  Many of the considerations addressed in this work could be
relevant to NCCI calculations with other possible radial bases as
well,
\textit{e.g.}, transformed harmonic oscillator
bases~\cite{stoitsov1998:tho-basis}.

The importance of the asymptotic properties of the basis functions may
be expected to vary depending upon the physical properties of the nucleus and state under
consideration.  A basis such as the Coulomb-Sturmian basis might well
be particularly appropriate to halo nuclei, where the mismatch with
the oscillator functions at large distances is particularly severe.
Another case of interest would be states involving clusters with
significant spatial separation.  The
importance of reproducing the large-$r$ properties of the nuclear
eigenstates may also be expected to depend upon the observable under
consideration, depending upon how
heavily large-$r$ contributions are weighted by that observable.  Thus,
\textit{e.g.}, the difference between Gaussian and exponential
asymptotics may be expected to be more important for the RMS radius or
$E2$ observables than for $M1$ observables.
Asymptotic properties also play a significant role in scattering problems.
The extent to which a Coulomb-Sturmian basis may be successfully used
in \textit{ab initio} scattering calculations, \textit{e.g.}, through
a generalization of the no-core shell model resonating group method~\cite{quaglioni2009:ncsm-rgm}, will
depend critically upon the details of the center-of-mass factorization properties.

To more fully ascertain the relative advantages or disadvantages of
the Coulomb-Sturmian basis for NCCI calculations, extensive and
systematic calculations are required, into both the convergence
properties of the basis and the robustness of extrapolations.  Most
obviously, these need to be carried to high $\Nmax$, for a variety of
nuclei and interactions.  However, there is also considerable room for
optimization within the method itself, which must be explored.  The
prescription for the $l$-dependence of the length parameter within the
single-particle basis (Sec.~\ref{sec-cs-scale}) and the many-body
truncation scheme (Sec.~\ref{sec-calc-over}), in which the present
oscillator-like $N=2n+l$ ``energy'' weighting is dictated purely by
convenience, are notable areas of possible improvement.  Although the
two-body JISP16 interaction was used in the illustrative calculations,
the transformation procedure (Sec.~\ref{sec-xform}) carries over
readily to three-body interactions, so convergence properties with,
\textit{e.g.}, chiral effective field theory interactions with similarity
renormalization group evolution~\cite{bogner2007:srg-nucleon}, can be investigated.


\begin{acknowledgments}
We thank M.~Pervin and W.~N.~Polyzou for pointing out the relevance of
the Coulomb-Sturmian basis.  We also thank T.~Dytrych for assistance
in the validation process, Ch.~Constantinou and A.~E.~McCoy for
comments on the manuscript, and T.~Papenbrock and S.~Quaglioni for
valuable discussions.  This work was supported by the Research
Corporation for Science Advancement through the Cottrell Scholar
program, by the US Department of Energy under Grants
No.~DE-FG02-95ER-40934, DE-FC02-09ER41582 (SciDAC/UNEDF), and
DE-FG02-87ER40371, and by the US National Science Foundation under
Grant No.~0904782. Computational resources were provided by the
National Energy Research Supercomputer Center (NERSC), which is
supported by the Office of Science of the U.S. Department of Energy
under Contract No.~DE-AC02-05CH11231.
\end{acknowledgments}


\ifproofpre{}{\vfil}

\appendix

\section{\boldmath Center-of-mass decomposition of $r^2$ and $k^2$}
\label{app-r2k2}

The one-body operators $r^2=\sum_i r_i^2$ and $k^2=\sum_i k_i^2$, for
the $A$-body system, may be decomposed into separate parts depending
only upon the center-of-mass coordinate (or momentum) and on the
relative coordinates (or momenta), respectively.  The decompositions
of the kinetic energy operator $T$ and noninteracting harmonic
oscillator potential $U$ into center-of-mass and relative parts follow
immediately.  In this appendix, we summarize the relations among
relative and center-of-mass operators, both for reference in the
present discussion and to establish a uniform notation for the
description of coordinate-space and momentum-space matrix elements in
Sec.~\ref{sec-sep}.

Recall that the
center-of-mass coordinate and momentum vectors are
\begin{equation}
\label{eqn-RCM-PCM}
\vec{R}=\frac{1}{A}\sum_i \vec{r}_i 
\qquad
\vec{P}=\sum_i \vec{p}_i.
\end{equation}
In the following, we let $\vec{p}_i=\hbar\vec{k}_i$, $\vec{P}=\hbar
\vec{K}$, \textit{etc.}

First, consider the one-body $r^2$ operator, defined by
\begin{equation}
\label{eqn-r2}
r^2=\sum_i r_i^2.
\end{equation}
Comparing the sum on the right hand side of~(\ref{eqn-r2}) with those
in the operators\footnote{We include the factors of $A^2$ on the left hand side
of~(\ref{eqn-decomp-r2}) as compensation
for the factor of $1/A$ appearing in the
definition~(\ref{eqn-RCM-PCM}) of $\vec{R}$, so as to simplify the
right hand side.
In particular, this maintains the parallel with the
decomposition of momentum space operators in~(\ref{eqn-decomp-k2})}
\begin{equation}
\label{eqn-decomp-r2}
\begin{aligned}
A^2R^2&=\biggl(\sum_i \vec{r}_i\biggr)^2
&&=\sum_ir_i^2+{\sumprime_{ij}}\vec{r}_i\cdot\vec{r}_j,
\\
A^2 \rrel^2&=\tfrac{1}{2}{\sumprime_{ij}}(\vec{r}_i-\vec{r}_j)^2
&&=(A-1)\sum_ir_i^2-{\sumprime_{ij}} \vec{r}_i\cdot\vec{r}_j.
\end{aligned}
\end{equation}
demonstrates that
\begin{equation}
\label{eqn-r2-addition}
Ar^2=A^2R^2+A^2 \rrel^2.
\end{equation}
Multiplying
by $(m\Omega^2)/(2A)$ gives the decomposition of the harmonic
oscillator potential energy operator $U^\Omega$ into center-of-mass
and relative contributions, 
$U^\Omega=\Ucm[\Omega]+\Urel[\Omega]$,
where
\begin{equation}
\begin{gathered}
\Ucm[\Omega]=\frac{m\Omega^2}{2A}(A^2R^2),
\quad
\Urel[\Omega]=\frac{m\Omega^2}{2A}(A^2\rrel^2),
\ifproofpre{\\}{\quad}
U^\Omega=\frac{m\Omega^2}{2A}(Ar^2).
\end{gathered}
\end{equation}
The quantity $\rrel^2$ has the geometric significance that it is
the mean square
radius relative to the center of mass, \textit{i.e.},
\begin{equation}
\rrel^2=\frac{1}{A}\sum_i(\vec{r}_i-\vec{R})^2.
\end{equation}
The square root of the expectation value of this operator,
$\tbracket{\rrel^2}^{1/2}$, is the point-nucleon RMS radius.

Similarly, consider the one-body $k^2$ operator, defined by
\begin{equation}
\label{eqn-k2}
k^2=\sum_i k_i^2.
\end{equation}
Comparison with the sums in
\begin{equation}
\label{eqn-decomp-k2}
\begin{aligned}
K^2&=\biggl(\sum_i \vec{k}_i\biggr)^2
&&=\sum_i k_i^2+{\sumprime_{ij}}\vec{k}_i\cdot\vec{k}_j,
\\
\krel^2&=\tfrac{1}{2}{\sumprime_{ij}}(\vec{k}_i-\vec{k}_j)^2
&&=(A-1)\sum_i k_i^2-{\sumprime_{ij}} \vec{k}_i\cdot\vec{k}_j,
\end{aligned}
\end{equation}
demonstrates that
\begin{equation}
\label{eqn-p2-addition}
Ak^2=K^2+\krel^2.
\end{equation}
Multiplying 
 by
$\hbar^2/(2Am_N)$ gives us the decomposition
 of the kinetic
 energy operator $T$
into center-of-mass and relative contributions,
$T=\Tcm+\Trel$, where
\begin{equation}
\begin{gathered}
\Tcm=\frac{(\hbar^2 K^2)}{2Am_N},
\quad
\Trel=\frac{(\hbar^2 \krel^2)}{2Am_N},
\ifproofpre{\\}{\quad}
T=\frac{(A\hbar^2 k^2)}{2Am_N}.
\end{gathered}
\end{equation}

\section{Zeros of generalized Laguerre and Jacobi polynomials}
\label{app-zeros}

Numerically robust evaluation of the radial integrals which arise in
evaluation of the overlaps between harmonic oscillator and
Coulomb-Sturmian bases (Sec.~\ref{sec-xform}) and the radial matrix
elements for the Coulomb-Sturmian basis (Sec.~\ref{sec-sep}) requires
accurate knowledge of the zeros of the integrands
in~(\ref{eqn-ol-S-R})--(\ref{eqn-ol-St-Rt}),
(\ref{eqn-me-r2})--(\ref{eqn-me-k2}), and
(\ref{eqn-me-r})--(\ref{eqn-me-k}), thus of generalized Laguerre
polynomials $\LaguerreL{n}{\alpha}(x)$ and Jacobi polynomials
$J_n^{(\alpha,\beta)}(x)$.  Although, in principle, generic numerical
rootfinding algorithms may be used, it is preferable to determine the
zeros according to a more reliable and efficient approach specific to
orthogonal polynomials, such as the Golub-Welch
algorithm~\cite{golub1969:gauss-quadrature}.  This method requires
recurrence coefficients for the relevant \textit{monic} polynomials, \textit{i.e.}, such that the highest-order
coefficient is unity,
as summarized in this appendix.

The Golub-Welch algorithm is specifically formulated for monic
polynomials.  Consider a family of polynomials $p_n(x)=\sum_{m=0}^n
c_m x^m$ ($n=0$, $1$, $\ldots$), orthogonal under weight function
$w(x)$ on the interval $[a,b]$, and with $c_n=1$.  Suppose these
polynomials satisfy the recurrence relation
\begin{equation}
\label{eqn-poly-recurrence}
p_{n+1}(x)+(B_n-x)p_n(x)+A_np_{n-1}(x)=0,
\end{equation}
characterized by recurrence coefficients $A_n$ and $B_n$.
Then, to find the zeros $p_n$ via the Golub-Welch
algorithm~\cite{golub1969:gauss-quadrature}, one must construct the
corresponding
Jacobi matrix $J$.  This is the $n\times n$  tridiagonal matrix
consisting of entries $J_{i,i}=B_{i-1}$ on the main diagonal and
$J_{i-1,i}=J_{i,i-1}=(A_{i-1})^{1/2}$ on the adjacent diagonals.  As a
tridiagonal matrix, $J$ is easily diagonalized.  The
eigenvalues $x_i$, for $i=1$, $2$, $\ldots$, $n$, are then the zeros of
$p_n$.

The generalized Laguerre polynomials $\LaguerreL{n}{\alpha}$ are not
monic, having $c_n=(-)^n n!$~\cite{olver2010:handbook}.  We must
therefore instead consider the \textit{monic} generalized Laguerre
polynomials $\LaguerreLhat{n}{\alpha}$, defined by
$\LaguerreLhat{n}{\alpha}(x)=(-)^n n! 
\LaguerreL{n}{\alpha}(x)$~\cite{cuyt2008:continued-fractions}.  These satisfy
a recurrence relation of the form~(\ref{eqn-poly-recurrence}), with
recurrence coefficients
\begin{equation}
\label{eqn-Lhat-recurrence}
\begin{aligned}
A_n&=n(n+\alpha)\\
B_n&=2n+\alpha+1.  
\end{aligned}
\end{equation}

The Jacobi polynomials $P_n^{(\alpha,\beta)}$ are likewise
not monic, having
$c_n=2^{-n}\binom{2n+\alpha+\beta}{n}$~\cite{olver2010:handbook}.
We must therefore instead consider the \textit{monic} Jacobi
polynomials $\Phat_n^{(\alpha,\beta)}$, defined by
$\Phat_n^{(\alpha,\beta)}(x)=2^{n}\binom{2n+\alpha+\beta}{n}^{-1}P_n^{(\alpha,\beta)}(x)$~\cite{cuyt2008:continued-fractions}.
These satisfy a recurrence relation of the
form~(\ref{eqn-poly-recurrence}), now with
\begin{equation}
\label{eqn-Jhat-recurrence}
\begin{aligned}
A_n&=\frac{4n(n+\alpha)(n+\beta)(n+\alpha+\beta)}{(2n+\alpha+\beta)^2(2n+\alpha+\beta+1)(2n+\alpha+\beta-1)}
\\
B_n&=\frac{\beta^2-\alpha^2}{(2n+\alpha+\beta)(2n+\alpha+\beta+2)}.
\end{aligned}
\end{equation}

The Golub-Welch algorithm also yields the weights $w_i$ appearing in the $n$-point Gaussian integration
formula associated with this family of polynomials, $\int_{a}^{b} f(x) w(x)\,
dx\approx \sum_{i=1}^n f(x_i) w_i$, which are obtained from
the eigenvectors of
$J$, as detailed in Ref.~\cite{golub1969:gauss-quadrature}.
The zeros and
weights appearing in $n$-point
Gauss-Legendre quadrature formulas used in evaluating the radial integrals of
Sec.~\ref{sec-cs} are widely
tabulated~\cite{olver2010:handbook}.  However, it is convenient to
note that the Jacobi matrix required in obtaining these may also be
obtained using the recurrence coefficients of~(\ref{eqn-Jhat-recurrence}). It is necessary
to consider the monic Legendre polynomials
$\Phat_n(x)=[2^n(n!)^2/(2n)!]P_n(x)$~\cite{cuyt2008:continued-fractions},
which constitute a special case of the monic Jacobi polynomials,
$\Phat_n(x)=\Phat_n^{(0,0)}(x)$, described by~(\ref{eqn-Jhat-recurrence}) with $\alpha=\beta=0$.

\section{Two-body states}
\label{app-tb}

In this appendix, the notation is established for the
\textit{antisymmetrized} (AS)
and \textit{normalized antisymmetrized} (NAS)
two-particle states, with angular momentum
coupling.  These definitions are required for discussion of like-particle two-body
matrix elements in Sec.~\ref{sec-cs}.

We first define angular momentum coupled states
\begin{equation}
\label{eqn-state-product}
\tpket{ab;J}=\sum_{m_am_b}\tcg{j_a}{m_a}{j_b}{m_b}{J}{M}\tket{am_a}\tket{bm_b}.
\end{equation}
for two \textit{distinguishable} particles, that is,
particle $1$ is in orbital $a$, and particle $2$ is in orbital $b$.
We denote such distinguishable-particle states by using parentheses
rather than angle brackets, following the conventions of Ref.~\cite{negele1988:many-particle}.  Such states may be used directly
for the case of one proton and one neutron, \textit{i.e.},
\begin{equation}
\label{eqn-state-pn}
\tket{ab;J}_\pn=\sum_{m_am_b}\tcg{j_a}{m_a}{j_b}{m_b}{J}{M}\tket{am_a}_p\tket{bm_b}_n.
\end{equation}
However, for two \textit{like}
fermions, \textit{antisymmetrized} states are then obtained as
\begin{equation}
\label{eqn-state-as}
\begin{aligned}
\tket{ab;JM}_\as&\equiv(\cd_a\times\cd_b)^J_M\tket{}
\\
&=\frac{1}{\sqrt{2}}\bigl[\tpket{ab;JM}-(-)^{J-j_a-j_b}\tpket{ba;JM}\bigr].
\end{aligned}
\end{equation}
These states have the basic symmetry property
\begin{equation}
\label{eqn-state-as-symm}
\tket{ab;JM}_\as=-(-)^{J-j_a-j_b}\tket{ba;JM}_\as.
\end{equation}
Therefore, if the orbitals $a$ and $b$ are identical, only states with
$J$ even may be obtained.  The states defined in~(\ref{eqn-state-as}) are antisymmetrized
but not strictly normalized, in that a further factor of $1/\sqrt{2}$ 
is required for normalization in the special case in which both particles
occupy the same orbital.  Strict normalization, even in this special case, is
obtained by taking \textit{normalized antisymmetrized} states
\begin{equation}
\label{eqn-state-nas}
\tket{ab;JM}_\nas=(1+\delta_{ab})^{-1/2}\tket{ab;JM}_\as.
\end{equation}

Two-body matrix elements may be represented in either the AS scheme or
NAS scheme, with the relation
\begin{multline}
\label{eqn-me-nas-as}
\tme{cd;J}{V}{ab;J}_\nas
\ifproofpre{\\}{}
=(1+\delta_{cd})^{-1/2}(1+\delta_{ab})^{-1/2}\tme{cd;J}{V}{ab;J}_\as,
\end{multline}
shown here for matrix elements of a scalar
operator $V$ within a
single $J$-space.
Both schemes are in common use for representing interaction matrix
elements.  The AS scheme may yield simpler expressions than the NAS scheme,
\textit{e.g.}, as seen comparing the change of basis
relation~(\ref{eqn-tbme-xform-as}) with~(\ref{eqn-tbme-xform-nas}).  

\section{Rescaling of separable matrix elements}
\label{app-rescaling}

For the separable calculation of matrix elements described in
Sec.~\ref{sec-sep}, the relations~(\ref{eqn-r2-k2-me}) provide
$A$-dependent expressions for the two-body matrix elements of $R^2$,
$\rrel^2$, $K^2$, and $\krel^2$ in terms of the $A$-independent two-body
matrix elements 
$\tme{cd;J}{V_{r^2}}{ab;J}$,
$\tme{cd;J}{V_{\vec{r}_1\cdot\vec{r}_2}}{ab;J}$,
$\tme{cd;J}{V_{k^2}}{ab;J}$, and
$\tme{cd;J}{V_{\vec{k}_1\cdot\vec{k}_2}}{ab;J}$.
These matrix elements still depend on the length parameter chosen for
the basis.  However, the operators $V_{r^2}$ and
$V_{\vec{r}_1\cdot\vec{r}_2}$ are homogeneous of order $2$ in the
coordinates, \textit{i.e.}, their matrix elements scale with the
length parameter as $b^2$, and the operators $V_{k^2}$ and
$V_{\vec{k}_1\cdot\vec{k}_2}$ are homogeneous of order $-2$,
\textit{i.e.}, their matrix elements scale as $b^{-2}$.
Recall that, under the prescription of Sec.~\ref{sec-cs-scale}, the
length parameters $b_l$ appearing in all Coulomb-Sturmian functions
are proportional to a common length parameter $\bho$ (this common
proportionality is a general property to be expected of any
prescription for the $b_l$).  Therefore, these matrix elements of
$V_{r^2}$, $V_{\vec{r}_1\cdot\vec{r}_2}$, $V_{k^2}$, and
$V_{\vec{k}_1\cdot\vec{k}_2}$ need only be calculated once, at some
particular reference value for the length scale, and then may be
transformed to the actual length scale, or $\hbar\Omega$ value, of the
many-body calculation by simple multiplication.  
For evaluation of the radial integrals appearing in~(\ref{eqn-me-r2}), (\ref{eqn-me-k2}), (\ref{eqn-me-r}),
and~(\ref{eqn-me-k}), it is natural to adopt a dimensionless reference
scale $\bho=1$.
Thus it is only necessary to evaluate matrix elements
$\tme{cd;J}{V_{r^2}}{ab;J}_0$,
$\tme{cd;J}{V_{\vec{r}_1\cdot\vec{r}_2}}{ab;J}_0$,
$\tme{cd;J}{V_{k^2}}{ab;J}_0$, and
$\tme{cd;J}{V_{\vec{k}_1\cdot\vec{k}_2}}{ab;J}_0$, by which we denote
matrix elements evaluated for $\bho=1$.  In this appendix, we give
explicit expressions for matrix elements of physically relevant
operators, for a given basis $\hbar\Omega$, in terms of these reference matrix elements and
dimensional scale factors.

Two-body matrix elements of the relative kinetic energy are given by
\begin{multline}
\label{eqn-tbme-dim-Trel}
\tme{cd;J}{\Trel}{ab;J}
=
\biggl(\frac{\hbar\Omega}{2A}\biggr)\tme{cd;J}{V_{k^2}}{ab;J}_0
\ifproofpre{\\}{}
-2\biggl(\frac{\hbar\Omega}{2A}\biggr)\tme{cd;J}{V_{\vec{k}_1\cdot\vec{k}_2}}{ab;J}_0
\end{multline}
and those of the $\rrel^2$ observable by
\begin{multline}
\label{eqn-tbme-dim-rrel2}
\tme{cd;J}{\rrel^2}{ab;J}
=
\biggl(\frac{\bho^2}{A^2}\biggr)\tme{cd;J}{V_{r^2}}{ab;J}_0
\ifproofpre{\\}{}
-2\biggl(\frac{\bho^2}{A^2}\biggr)\tme{cd;J}{V_{\vec{r}_1\cdot\vec{r}_2}}{ab;J}_0.
\end{multline}
For present purposes, it is most convenient to reexpress $\bho$ in terms of
$\hbar\Omega$ using combinations of physical constants chosen so as to
only involve energy and length units, as
\begin{equation}
\label{eqn-beta-omega-dim}
\bho=\frac{(\hbar c)}{[(m_Nc^2)(\hbar\Omega)]^{1/2}},
\end{equation}
where $m_Nc^2\approx938.92\,\mathrm{MeV}$ and $\hbar
c\approx197.327\,\mathrm{MeV\,fm}$.  

In the investigation of center-of-mass
separation and in the Lawson term as applied to NCCI calculations with the Coulomb-Sturmian basis
(Sec.~\ref{sec-calc-cm}), we consider the center-of-mass
oscillator number operator $\Ncm[\Omegat]$ of~(\ref{eqn-ncm}),
involving an arbitrary
oscillator energy $\hbar\Omegat$,
in general different from the basis $\hbar\Omega$.  This operator
has two-body matrix elements
\begin{multline}
\label{eqn-tbme-dim-Ncm}
\tme{cd;J}{\Ncm[\Omegat]}{ab;J}
=
\ifproofpre{\\}{}
\frac{1}{2A}\frac{(\hbar\Omega)}{(\hbar\Omegat)}
\Bigl[\frac{1}{A-1}\tme{cd;J}{V_{k^2}}{ab;J}_0
+2\tme{cd;J}{V_{\vec{k}_1\cdot\vec{k}_2}}{ab;J}_0\Bigr]
\\
+\frac{1}{2A}\frac{(\hbar\Omegat)}{(\hbar\Omega)}
\Bigl[\frac{1}{A-1}\tme{cd;J}{V_{r^2}}{ab;J}_0
+2\tme{cd;J}{V_{\vec{r}_1\cdot\vec{r}_2}}{ab;J}_0\Bigr]
\\
-\frac{3}{A(A-1)}\tme{cd;J}{\unity_{2b}}{ab;J},
\end{multline}
where $\unity_{2b}$ is the identity operator on the two-body space.
If one is evaluating $\tbracket{\Ncm[\Omegat]}$ for several values of $\hbar\Omegat$,
as in Fig.~\ref{fig-Ncm-distrib}, it suffices to 
calculate the expectation values of just the two operators
$P^2$ and $R^2$ for the many-body state, since these two numerical
values may then be combined arithmetically 
by~(\ref{eqn-ncm}) to deduce
$\tbracket{\Ncm[\Omegat]}$ for any value of $\hbar\Omegat$.  More simply, in terms of the expectation values of the dimensionless operators
$K^2_0=P^2/(m_N \hbar \Omega)$ and $A^2R^2_0=(m_N
\Omega A^2/\hbar)R^2$, corresponding to the two-body matrix elements 
appearing in
brackets, respectively, in~(\ref{eqn-tbme-dim-Ncm}),  we have 
\begin{equation}
\label{eqn-Ncm-ev}
\tbracket{\Ncm[\Omegat]}=\frac{1}{2A}\frac{(\hbar\Omega)}{(\hbar\Omegat)}\tbracket{K^2_0}+\frac{1}{2A}\frac{(\hbar\Omegat)}{(\hbar\Omega)}\tbracket{A^2R^2_0}-\frac32.
\end{equation}

The number operator $N^\Omegat$ for the one-body harmonic oscillator
Hamiltonian $H^\Omegat$, though not used in the present work, can
also be of interest.  For instance, if a many-body state has been
obtained from an NCCI calculation using the Coulomb-Sturmian basis,
$\tbracket{N^\Omega}$ provides an estimate of
the number of quanta which would be
required to represent this same state in a space spanned by a
conventional harmonic-oscillator basis of oscillator energy $\hbar\Omega$,
or a Hamiltonian term proportional to $N^\Omega$ may be used for calculations
involving an external harmonic oscillator trapping field.  The two-body matrix elements
are given by
\begin{multline}
\label{eqn-tbme-dim-Nob}
\tme{cd;J}{N^{\Omegat}}{ab;J}
=
\frac{1}{2(A-1)}\frac{(\hbar\Omega)}{(\hbar\Omegat)}
\tme{cd;J}{V_{k^2}}{ab;J}_0
\\
+\frac{1}{2(A-1)}\frac{(\hbar\Omegat)}{(\hbar\Omega)}
\tme{cd;J}{V_{r^2}}{ab;J}_0
\ifproofpre{\\}{}
-\frac{3}{A-1}\tme{cd;J}{\unity_{2b}}{ab;J}.
\end{multline}

\ifproofpre{\vfill}{}


\providecommand{\APSLONG}{}
\providecommand{\ELSEVIER}{}




\begin{thebibliography}{44}
\expandafter\ifx\csname natexlab\endcsname\relax\def\natexlab#1{#1}\fi
\expandafter\ifx\csname bibnamefont\endcsname\relax
  \def\bibnamefont#1{#1}\fi
\expandafter\ifx\csname bibfnamefont\endcsname\relax
  \def\bibfnamefont#1{#1}\fi
\expandafter\ifx\csname citenamefont\endcsname\relax
  \def\citenamefont#1{#1}\fi
\expandafter\ifx\csname url\endcsname\relax
  \def\url#1{\texttt{#1}}\fi
\expandafter\ifx\csname urlprefix\endcsname\relax\def\urlprefix{URL }\fi
\providecommand{\bibinfo}[2]{#2}
\providecommand{\eprint}[2][arXiv]{\url{#1:#2}}
\renewcommand{\eprint}[2][arXiv]{\url{#1:#2}}

\bibitem{entem2003:chiral-nn-potl}
\bibinfo{author}{\bibfnamefont{D.~R.} \bibnamefont{Entem}} \bibnamefont{and}
  \bibinfo{author}{\bibfnamefont{R.}~\bibnamefont{Machleidt}},
  \bibinfo{journal}{Phys. Rev. C} \textbf{\bibinfo{volume}{68}},
  \bibinfo{pages}{041001} (\bibinfo{year}{2003}).

\bibitem{epelbaum2009:nuclear-forces}
\bibinfo{author}{\bibfnamefont{E.}~\bibnamefont{Epelbaum}},
  \bibinfo{author}{\bibfnamefont{H.-W.} \bibnamefont{Hammer}},
  \bibnamefont{and} \bibinfo{author}{\bibfnamefont{U.-G.}
  \bibnamefont{Mei\ss{}ner}}, \bibinfo{journal}{Rev. Mod. Phys.}
  \textbf{\bibinfo{volume}{81}}, \bibinfo{pages}{1773} (\bibinfo{year}{2009}).

\bibitem{navratil2000:12c-ncsm-COMBO}
\bibinfo{author}{\bibfnamefont{P.}~\bibnamefont{Navr\'{a}til}},
  \bibinfo{author}{\bibfnamefont{J.~P.} \bibnamefont{Vary}}, \bibnamefont{and}
  \bibinfo{author}{\bibfnamefont{B.~R.} \bibnamefont{Barrett}},
  \bibinfo{journal}{Phys. Rev. Lett.} \textbf{\bibinfo{volume}{84}},
  \bibinfo{pages}{5728} (\bibinfo{year}{2000}); Phys. Rev. C \textbf{62},
  054311 (2000).

\bibitem{rotenberg1962:sturmian-scatt}
\bibinfo{author}{\bibfnamefont{M.}~\bibnamefont{Rotenberg}},
  \bibinfo{journal}{Ann. Phys. (N.Y.)} \textbf{\bibinfo{volume}{19}},
  \bibinfo{pages}{262} (\bibinfo{year}{1962}).

\bibitem{weniger1985:fourier-plane-wave}
\bibinfo{author}{\bibfnamefont{E.~J.} \bibnamefont{Weniger}},
  \bibinfo{journal}{J. Math. Phys.} \textbf{\bibinfo{volume}{26}},
  \bibinfo{pages}{276} (\bibinfo{year}{1985}).

\bibitem{hylleraas1928:helium-sturmian}
\bibinfo{author}{\bibfnamefont{E.~A.} \bibnamefont{Hylleraas}},
  \bibinfo{journal}{Z. Phys.} \textbf{\bibinfo{volume}{48}},
  \bibinfo{pages}{469} (\bibinfo{year}{1928}).

\bibitem{loewdin1956:natural-orbital}
\bibinfo{author}{\bibfnamefont{P.-O.} \bibnamefont{L{\"o}wdin}}
  \bibnamefont{and} \bibinfo{author}{\bibfnamefont{H.}~\bibnamefont{Shull}},
  \bibinfo{journal}{Phys. Rev.} \textbf{\bibinfo{volume}{101}},
  \bibinfo{pages}{1730} (\bibinfo{year}{1956}).

\bibitem{rotenberg1970:sturmian-scatt}
\bibinfo{author}{\bibfnamefont{M.}~\bibnamefont{Rotenberg}},
  \bibinfo{journal}{Adv. At. Mol. Phys.} \textbf{\bibinfo{volume}{6}},
  \bibinfo{pages}{233} (\bibinfo{year}{1970}).

\bibitem{jacobs1986:heavy-quark-sturmian}
\bibinfo{author}{\bibfnamefont{S.}~\bibnamefont{Jacobs}},
  \bibinfo{author}{\bibfnamefont{M.~G.} \bibnamefont{Olsson}},
  \bibnamefont{and} \bibinfo{author}{\bibfnamefont{C.}~\bibnamefont{Suchyta},
  \bibfnamefont{III}}, \bibinfo{journal}{Phys. Rev. D}
  \textbf{\bibinfo{volume}{33}}, \bibinfo{pages}{3338} (\bibinfo{year}{1986}).

\bibitem{fulcher1993:quarkonium-sturmian}
\bibinfo{author}{\bibfnamefont{L.~P.} \bibnamefont{Fulcher}},
  \bibinfo{author}{\bibfnamefont{Z.}~\bibnamefont{Chen}}, \bibnamefont{and}
  \bibinfo{author}{\bibfnamefont{K.~C.} \bibnamefont{Yeong}},
  \bibinfo{journal}{Phys. Rev. D} \textbf{\bibinfo{volume}{47}},
  \bibinfo{pages}{4122} (\bibinfo{year}{1993}).

\bibitem{keister1997:on-basis}
\bibinfo{author}{\bibfnamefont{B.~D.} \bibnamefont{Keister}} \bibnamefont{and}
  \bibinfo{author}{\bibfnamefont{W.~N.} \bibnamefont{Polyzou}},
  \bibinfo{journal}{J. Comput. Phys.} \textbf{\bibinfo{volume}{134}},
  \bibinfo{pages}{231} (\bibinfo{year}{1997}).

\bibitem{pervin2005:diss}
\bibinfo{author}{\bibfnamefont{M.}~\bibnamefont{Pervin}}, Ph.D. thesis,
  \bibinfo{school}{Florida State University} (\bibinfo{year}{2005}).

\bibitem{forssen2008:ncsm-sequences}
\bibinfo{author}{\bibfnamefont{C.}~\bibnamefont{Forssen}},
  \bibinfo{author}{\bibfnamefont{J.~P.} \bibnamefont{Vary}},
  \bibinfo{author}{\bibfnamefont{E.}~\bibnamefont{Caurier}}, \bibnamefont{and}
  \bibinfo{author}{\bibfnamefont{P.}~\bibnamefont{Navratil}},
  \bibinfo{journal}{Phys. Rev. C} \textbf{\bibinfo{volume}{77}},
  \bibinfo{pages}{024301} (\bibinfo{year}{2008}).

\bibitem{maris2009:ncfc}
\bibinfo{author}{\bibfnamefont{P.}~\bibnamefont{Maris}},
  \bibinfo{author}{\bibfnamefont{J.~P.} \bibnamefont{Vary}}, \bibnamefont{and}
  \bibinfo{author}{\bibfnamefont{A.~M.} \bibnamefont{Shirokov}},
  \bibinfo{journal}{Phys. Rev. C} \textbf{\bibinfo{volume}{79}},
  \bibinfo{pages}{014308} (\bibinfo{year}{2009}).

\bibitem{coon2012:nscm-ho-regulator}
\bibinfo{author}{\bibfnamefont{S.~A.} \bibnamefont{Coon}},
  \bibinfo{author}{\bibfnamefont{M.~I.} \bibnamefont{Avetian}},
  \bibinfo{author}{\bibfnamefont{M.~K.~G.} \bibnamefont{Kruse}},
  \bibinfo{author}{\bibfnamefont{U.}~\bibnamefont{van Kolck}},
  \bibinfo{author}{\bibfnamefont{P.}~\bibnamefont{Maris}}, \bibnamefont{and}
  \bibinfo{author}{\bibfnamefont{J.~P.} \bibnamefont{Vary}},
  \bibinfo{journal}{Phys. Rev. C} \textbf{\bibinfo{volume}{86}},
  \bibinfo{pages}{054002} (\bibinfo{year}{2012}).

\bibitem{moshinsky1996:oscillator}
\bibinfo{author}{\bibfnamefont{M.}~\bibnamefont{Moshinsky}} \bibnamefont{and}
  \bibinfo{author}{\bibfnamefont{Y.~F.} \bibnamefont{Smirnov}},
  \emph{\bibinfo{title}{The Harmonic Oscillator in Modern Physics}}
  (\bibinfo{publisher}{Harwood Academic Publishers},
  \bibinfo{address}{Amsterdam}, \bibinfo{year}{1996}).

\bibitem{davies1966:hartree-fock}
\bibinfo{author}{\bibfnamefont{K.~T.~R.} \bibnamefont{Davies}},
  \bibinfo{author}{\bibfnamefont{S.~J.} \bibnamefont{Krieger}},
  \bibnamefont{and} \bibinfo{author}{\bibfnamefont{M.}~\bibnamefont{Baranger}},
  \bibinfo{journal}{Nucl. Phys.} \textbf{\bibinfo{volume}{84}},
  \bibinfo{pages}{545} (\bibinfo{year}{1966}).

\bibitem{caprio2012:csbasis-hites12}
\bibinfo{author}{\bibfnamefont{M.~A.} \bibnamefont{Caprio}},
  \bibinfo{author}{\bibfnamefont{P.}~\bibnamefont{Maris}}, \bibnamefont{and}
  \bibinfo{author}{\bibfnamefont{J.~P.} \bibnamefont{Vary}},
  \bibinfo{journal}{J. Phys. Conf. Ser.} \textbf{\bibinfo{volume}{403}},
  \bibinfo{pages}{012014} (\bibinfo{year}{2012}).

\bibitem{fluegge1971:qm}
\bibinfo{author}{\bibfnamefont{S.}~\bibnamefont{Fl{\"u}gge}},
  \emph{\bibinfo{title}{Practical Quantum Mechanics I}}, \bibinfo{series}{{\it
  {G}rundlehren der mathematischen {W}issenschaften}} Vol.
  \bibinfo{volume}{177} (\bibinfo{publisher}{Springer-Verlag},
  \bibinfo{address}{Berlin}, \bibinfo{year}{1971}).

\bibitem{gloeckner1974:spurious-com}
\bibinfo{author}{\bibfnamefont{D.~H.} \bibnamefont{Gloeckner}}
  \bibnamefont{and} \bibinfo{author}{\bibfnamefont{R.~D.}
  \bibnamefont{Lawson}}, \bibinfo{journal}{Phys. Lett. B}
  \textbf{\bibinfo{volume}{53}}, \bibinfo{pages}{313} (\bibinfo{year}{1974}).

\bibitem{elliott1955:com-shell}
\bibinfo{author}{\bibfnamefont{J.~P.} \bibnamefont{Elliott}} \bibnamefont{and}
  \bibinfo{author}{\bibfnamefont{T.~H.~R.} \bibnamefont{Skyrme}},
  \bibinfo{journal}{Proc. R. Soc. London A} \textbf{\bibinfo{volume}{232}},
  \bibinfo{pages}{561} (\bibinfo{year}{1955}).

\bibitem{lanczos1950:algorithm}
\bibinfo{author}{\bibfnamefont{C.}~\bibnamefont{Lanczos}}, \bibinfo{journal}{J.
  Res. Natl. Bur. Stand.} \textbf{\bibinfo{volume}{45}}, \bibinfo{pages}{255}
  (\bibinfo{year}{1950}).

\bibitem{messiah1999:qm}
\bibinfo{author}{\bibfnamefont{A.}~\bibnamefont{Messiah}},
  \emph{\bibinfo{title}{Quantum Mechanics}} (\bibinfo{publisher}{Dover},
  \bibinfo{address}{Mineola, New York}, \bibinfo{year}{1999}).

\bibitem{suhonen2007:nucleons-nucleus}
\bibinfo{author}{\bibfnamefont{J.}~\bibnamefont{Suhonen}},
  \emph{\bibinfo{title}{From Nucleons to Nucleus}}
  (\bibinfo{publisher}{Springer-Verlag}, \bibinfo{address}{Berlin},
  \bibinfo{year}{2007}).

\bibitem{arfken1995:mathmethods}
\bibinfo{author}{\bibfnamefont{G.~B.} \bibnamefont{Arfken}} \bibnamefont{and}
  \bibinfo{author}{\bibfnamefont{H.~J.} \bibnamefont{Weber}},
  \emph{\bibinfo{title}{Mathematical Methods for Physicists}},
  \bibinfo{edition}{4th} ed. (\bibinfo{publisher}{Academic Press},
  \bibinfo{address}{San Diego}, \bibinfo{year}{1995}).

\bibitem{filter1980:sturmian-translation}
\bibinfo{author}{\bibfnamefont{E.}~\bibnamefont{Filter}} \bibnamefont{and}
  \bibinfo{author}{\bibfnamefont{E.~O.} \bibnamefont{Steinborn}},
  \bibinfo{journal}{J. Math. Phys.} \textbf{\bibinfo{volume}{21}},
  \bibinfo{pages}{2725} (\bibinfo{year}{1980}).

\bibitem{klahn1977:rayleigh-convergence-part2-special}
\bibinfo{author}{\bibfnamefont{B.}~\bibnamefont{Klahn}} \bibnamefont{and}
  \bibinfo{author}{\bibfnamefont{W.~A.} \bibnamefont{Bingel}},
  \bibinfo{journal}{Theor. Chim. Acta} \textbf{\bibinfo{volume}{44}},
  \bibinfo{pages}{27} (\bibinfo{year}{1977}).

\bibitem{olver2010:handbook}
\bibinfo{editor}{\bibfnamefont{F.~W.~J.} \bibnamefont{Olver}},
  \bibinfo{editor}{\bibfnamefont{D.~W.} \bibnamefont{Lozier}},
  \bibinfo{editor}{\bibfnamefont{R.~F.} \bibnamefont{Boisvert}},
  \bibnamefont{and} \bibinfo{editor}{\bibfnamefont{C.~W.} \bibnamefont{Clark}},
  eds., \emph{\bibinfo{title}{NIST Handbook of Mathematical Functions}}
  (\bibinfo{publisher}{Cambridge University Press},
  \bibinfo{address}{Cambridge}, \bibinfo{year}{2010}).

\bibitem{negele1988:many-particle}
\bibinfo{author}{\bibfnamefont{J.~W.} \bibnamefont{Negele}} \bibnamefont{and}
  \bibinfo{author}{\bibfnamefont{H.}~\bibnamefont{Orland}},
  \emph{\bibinfo{title}{Quantum Many-Particle Systems}}
  (\bibinfo{publisher}{Addison-Wesley}, \bibinfo{address}{Redwood City, CA},
  \bibinfo{year}{1988}).

\bibitem{edmonds1960:am}
\bibinfo{author}{\bibfnamefont{A.~R.} \bibnamefont{Edmonds}},
  \emph{\bibinfo{title}{Angular Momentum in Quantum Mechanics}},
  \bibinfo{edition}{2nd} ed., \bibinfo{series}{Investigations in Physics}
  No.~\bibinfo{number}{4} (\bibinfo{publisher}{Princeton University Press},
  \bibinfo{address}{Princeton, New Jersey}, \bibinfo{year}{1960}).

\bibitem{sternberg2008:ncsm-mfdn-sc08}
\bibinfo{author}{\bibfnamefont{P.}~\bibnamefont{Sternberg}},
  \bibinfo{author}{\bibfnamefont{E.~G.} \bibnamefont{Ng}},
  \bibinfo{author}{\bibfnamefont{C.}~\bibnamefont{Yang}},
  \bibinfo{author}{\bibfnamefont{P.}~\bibnamefont{Maris}},
  \bibinfo{author}{\bibfnamefont{J.~P.} \bibnamefont{Vary}},
  \bibinfo{author}{\bibfnamefont{M.}~\bibnamefont{Sosonkina}},
  \bibnamefont{and} \bibinfo{author}{\bibfnamefont{H.~V.} \bibnamefont{Le}}, in
  \emph{\bibinfo{booktitle}{SC '08: Proceedings of the 2008 ACM/IEEE Conference
  on Supercomputing}} (\bibinfo{publisher}{IEEE Press},
  \bibinfo{address}{Piscataway, NJ}, \bibinfo{year}{2008}),
  \bibinfo{note}{{A}rticle No. 15}.

\bibitem{vary2009:ncsm-mfdn-scidac09}
\bibinfo{author}{\bibfnamefont{J.~P.} \bibnamefont{Vary}},
  \bibinfo{author}{\bibfnamefont{P.}~\bibnamefont{Maris}},
  \bibinfo{author}{\bibfnamefont{E.}~\bibnamefont{Ng}},
  \bibinfo{author}{\bibfnamefont{C.}~\bibnamefont{Yang}}, \bibnamefont{and}
  \bibinfo{author}{\bibfnamefont{M.}~\bibnamefont{Sosonkina}},
  \bibinfo{journal}{J. Phys. Conf. Ser.} \textbf{\bibinfo{volume}{180}},
  \bibinfo{pages}{012083} (\bibinfo{year}{2009}).

\bibitem{maris2010:ncsm-mfdn-iccs10}
\bibinfo{author}{\bibfnamefont{P.}~\bibnamefont{Maris}},
  \bibinfo{author}{\bibfnamefont{M.}~\bibnamefont{Sosonkina}},
  \bibinfo{author}{\bibfnamefont{J.~P.} \bibnamefont{Vary}},
  \bibinfo{author}{\bibfnamefont{E.}~\bibnamefont{Ng}}, \bibnamefont{and}
  \bibinfo{author}{\bibfnamefont{C.}~\bibnamefont{Yang}},
  \bibinfo{journal}{Procedia Comput. Sci.} \textbf{\bibinfo{volume}{1}},
  \bibinfo{pages}{97} (\bibinfo{year}{2010}).

\bibitem{shirokov2007:nn-jisp16}
\bibinfo{author}{\bibfnamefont{A.~M.} \bibnamefont{Shirokov}},
  \bibinfo{author}{\bibfnamefont{J.~P.} \bibnamefont{Vary}},
  \bibinfo{author}{\bibfnamefont{A.~I.} \bibnamefont{Mazur}}, \bibnamefont{and}
  \bibinfo{author}{\bibfnamefont{T.~A.} \bibnamefont{Weber}},
  \bibinfo{journal}{Phys. Lett. B} \textbf{\bibinfo{volume}{644}},
  \bibinfo{pages}{33} (\bibinfo{year}{2007}).

\bibitem{cockrell2012:li-ncfc}
\bibinfo{author}{\bibfnamefont{C.}~\bibnamefont{Cockrell}},
  \bibinfo{author}{\bibfnamefont{J.~P.} \bibnamefont{Vary}}, \bibnamefont{and}
  \bibinfo{author}{\bibfnamefont{P.}~\bibnamefont{Maris}},
  \bibinfo{journal}{Phys. Rev. C} \textbf{\bibinfo{volume}{86}},
  \bibinfo{pages}{034325} (\bibinfo{year}{2012}).

\bibitem{bogner2008:ncsm-converg-2N}
\bibinfo{author}{\bibfnamefont{S.~K.} \bibnamefont{Bogner}},
  \bibinfo{author}{\bibfnamefont{R.~J.} \bibnamefont{Furnstahl}},
  \bibinfo{author}{\bibfnamefont{P.}~\bibnamefont{Maris}},
  \bibinfo{author}{\bibfnamefont{R.~J.} \bibnamefont{Perry}},
  \bibinfo{author}{\bibfnamefont{A.}~\bibnamefont{Schwenk}}, \bibnamefont{and}
  \bibinfo{author}{\bibfnamefont{J.}~\bibnamefont{Vary}},
  \bibinfo{journal}{Nucl. Phys. A} \textbf{\bibinfo{volume}{801}},
  \bibinfo{pages}{21} (\bibinfo{year}{2008}).

\bibitem{lipkin1958:com-shell}
\bibinfo{author}{\bibfnamefont{H.~J.} \bibnamefont{Lipkin}},
  \bibinfo{journal}{Phys. Rev.} \textbf{\bibinfo{volume}{110}},
  \bibinfo{pages}{1395} (\bibinfo{year}{1958}).

\bibitem{hagen2009:coupled-cluster-com-COMBO}
\bibinfo{author}{\bibfnamefont{G.}~\bibnamefont{Hagen}},
  \bibinfo{author}{\bibfnamefont{T.}~\bibnamefont{Papenbrock}},
  \bibnamefont{and} \bibinfo{author}{\bibfnamefont{D.~J.} \bibnamefont{Dean}},
  \bibinfo{journal}{Phys. Rev. Lett.} \textbf{\bibinfo{volume}{103}},
  \bibinfo{pages}{062503} (\bibinfo{year}{2009});
  \bibinfo{author}{\bibfnamefont{G.}~\bibnamefont{Hagen}},
  \bibinfo{author}{\bibfnamefont{T.}~\bibnamefont{Papenbrock}},
  \bibinfo{author}{\bibfnamefont{D.~J.} \bibnamefont{Dean}}, \bibnamefont{and}
  \bibinfo{author}{\bibfnamefont{M.}~\bibnamefont{Hjorth-Jensen}}, Phys. Rev. C
  \textbf{82}, 034330 (2010).

\bibitem{mcgrory1975:spurious-com}
\bibinfo{author}{\bibfnamefont{J.~B.} \bibnamefont{McGrory}} \bibnamefont{and}
  \bibinfo{author}{\bibfnamefont{B.~H.} \bibnamefont{Wildenthal}},
  \bibinfo{journal}{Phys. Lett. B} \textbf{\bibinfo{volume}{60}},
  \bibinfo{pages}{5} (\bibinfo{year}{1975}).

\bibitem{stoitsov1998:tho-basis}
\bibinfo{author}{\bibfnamefont{M.~V.} \bibnamefont{Stoitsov}},
  \bibinfo{author}{\bibfnamefont{W.}~\bibnamefont{Nazarewicz}},
  \bibnamefont{and} \bibinfo{author}{\bibfnamefont{S.}~\bibnamefont{Pittel}},
  \bibinfo{journal}{Phys. Rev. C} \textbf{\bibinfo{volume}{58}},
  \bibinfo{pages}{2092} (\bibinfo{year}{1998}).

\bibitem{quaglioni2009:ncsm-rgm}
\bibinfo{author}{\bibfnamefont{S.}~\bibnamefont{Quaglioni}} \bibnamefont{and}
  \bibinfo{author}{\bibfnamefont{P.}~\bibnamefont{Navr{\'a}til}},
  \bibinfo{journal}{Phys. Rev. C} \textbf{\bibinfo{volume}{79}},
  \bibinfo{pages}{044606} (\bibinfo{year}{2009}).

\bibitem{bogner2007:srg-nucleon}
\bibinfo{author}{\bibfnamefont{S.~K.} \bibnamefont{Bogner}},
  \bibinfo{author}{\bibfnamefont{R.~J.} \bibnamefont{Furnstahl}},
  \bibnamefont{and} \bibinfo{author}{\bibfnamefont{R.~J.} \bibnamefont{Perry}},
  \bibinfo{journal}{Phys. Rev. C} \textbf{\bibinfo{volume}{75}},
  \bibinfo{pages}{061001} (\bibinfo{year}{2007}).

\bibitem{golub1969:gauss-quadrature}
\bibinfo{author}{\bibfnamefont{G.~H.} \bibnamefont{Golub}} \bibnamefont{and}
  \bibinfo{author}{\bibfnamefont{J.~H.} \bibnamefont{Welsch}},
  \bibinfo{journal}{Math. Comput.} \textbf{\bibinfo{volume}{23}},
  \bibinfo{pages}{221} (\bibinfo{year}{1969}).

\bibitem{cuyt2008:continued-fractions}
\bibinfo{author}{\bibfnamefont{A.}~\bibnamefont{Cuyt}},
  \bibinfo{author}{\bibfnamefont{V.~B.} \bibnamefont{Petersen}},
  \bibinfo{author}{\bibfnamefont{B.}~\bibnamefont{Verdonk}},
  \bibinfo{author}{\bibfnamefont{H.}~\bibnamefont{Waadeland}},
  \bibnamefont{and} \bibinfo{author}{\bibfnamefont{W.~B.} \bibnamefont{Jones}},
  \emph{\bibinfo{title}{Handbook of Continued Fractions for Special Functions}}
  (\bibinfo{publisher}{Springer-Verlag}, \bibinfo{address}{Berlin},
  \bibinfo{year}{2008}).

\end{thebibliography}
\end{document}